\documentclass[pdflatex,sn-mathphys]{sn-jnl}
\usepackage{adjustbox}

\jyear{2023}%

\theoremstyle{thmstyleone}%
\theoremstyle{thmstyletwo}%

\theoremstyle{thmstylethree}%

\begin{document}

\title[Understanding User Intent Modeling for CRS: An SLR]{Understanding User Intent Modeling for Conversational Recommender Systems: A Systematic Literature Review}

\author*[1]{\small \fnm{Siamak} \sur{Farshidi}}\email{s.farshidi@uu.nl}

\author[2]{\small \fnm{Kiyan} \sur{Rezaee}}
\equalcont{\small These authors contributed equally to this work.}

\author[2]{\small \fnm{Sara} \sur{Mazaheri}}
\equalcont{\small These authors contributed equally to this work.}

\author[2]{\small \fnm{Amir Hossein} \sur{Rahimi}}
\equalcont{\small These authors contributed equally to this work.}

\author[2]{\small \fnm{Ali} \sur{Dadashzadeh}}

\author[2]{\small \fnm{Morteza} \sur{Ziabakhsh}}

\author*[2]{\small \fnm{Sadegh} \sur{Eskandari}}\email{eskandari@guilan.ac.ir}
\equalcont{\small These authors contributed equally to this work.}

\author*[1,3]{\small \fnm{Slinger} \sur{Jansen}}\email{slinger.jansen@uu.nl}

\affil*[1]{\small \orgdiv{Department of Information and Computer Science}, \orgname{Utrecht University}, \state{Utrecht}, \country{The Netherlands}}
\affil[2]{\small \orgdiv{Department of Computer Science}, \orgname{University of Guilan}, \city{Rasht},\country{Iran}}
\affil[3]{\small \orgname{Lappeenranta University of Technology},\city{Lappeenranta}, \country{Finland}}

\abstract{ 

\textbf{Context:} User intent modeling is a crucial process in Natural Language Processing that aims to identify the underlying purpose behind a user's request, enabling personalized responses. With a vast array of approaches introduced in the literature (over 13,000 papers in the last decade), understanding the related concepts and commonly used models in AI-based systems is essential. 

\textbf{Method:} We conducted a systematic literature review to gather data on models typically employed in designing conversational recommender systems. From the collected data, we developed a decision model to assist researchers in selecting the most suitable models for their systems. Additionally, we performed two case studies to evaluate the effectiveness of our proposed decision model.

\textbf{Results:} Our study analyzed 59 distinct models and identified 74 commonly used features. We provided insights into potential model combinations, trends in model selection, quality concerns, evaluation measures, and frequently used datasets for training and evaluating these models.

\textbf{Contribution:} Our study contributes practical insights and a comprehensive understanding of user intent modeling, empowering the development of more effective and personalized conversational recommender systems. With the Conversational Recommender System, researchers can perform a more systematic and efficient assessment of fitting intent modeling frameworks.

}

\maketitle

\section{Introduction}
User intent modeling is a fundamental process in Natural Language Processing (NLP) that aims to discern the underlying purpose or objective of a user's request~\cite{carmel2020future}. By leveraging machine learning algorithms to analyze various aspects of user input, such as words, phrases, and context, user intent modeling enables accurate identification of desired outcomes in conversational recommender systems~\cite{khilji2023multimodal}. Consequently, this approach leads to the delivery of personalized and precise responses~\cite{ge2018personalizing}.

Understanding and predicting user goals and motivations through user intent modeling play a vital role in optimizing search engines and recommender systems~\cite{zhang2019deep}. Aligning the user experience and search results with users' preferences and needs allows designers and developers to enhance user satisfaction and engagement~\cite{oulasvirta2008motivations}. This personalized approach results in providing relevant and tailored results~\cite{konishi2016extracting,bendersky2017learning}. For example, ChatGPT is a state-of-the-art generative language model that has garnered substantial interest for its potential applications in search engines and recommender systems~\cite{cao2023comprehensive}. It can comprehend user intentions and engage in meaningful interactions with them.

User intent modeling finds diverse practical applications in several domains, from e-commerce and healthcare to education, social media, and virtual assistants. In e-commerce, it plays a pivotal role in delivering personalized product recommendations, thereby enhancing the overall shopping experience for users~\cite{tanjim2020attentive, wang2020next, guo2020edgedipn}. Moreover, user intent modeling contributes to the detection of fake product reviews, which is a critical issue in e-commerce platforms~\cite{paul2021fake}. By identifying and filtering out fraudulent reviews, it helps build trust among customers and ensures more reliable product evaluations, ultimately benefiting both consumers and businesses. The healthcare domain benefits from user intent modeling by utilizing it to provide personalized health recommendations and interventions based on individual patients' health goals and motivations~\cite{zhang2016mining, wang2022recognizing}.

Similarly, in education, user intent modeling supports personalized learning experiences tailored to the specific goals and preferences of students~\cite{liu2021intent, bhaskaran2019efficient}. In the realm of social media, it enables a comprehensive understanding of user interests, preferences, and behaviors, which, in turn, drives the delivery of personalized content and advertising~\cite{ding2015mining, wang2019context}.

User intent modeling proves to be a valuable asset for virtual assistants as it assists them in comprehending user queries and providing relevant and personalized responses~\cite{penha2020does, hashemi2018measuring}. Moreover, its application extends to advertising targeting and personalization across various domains, benefiting businesses and users alike~\cite{gharibshah2020deep, bilenko2011predictive, yamamoto2012wisdom}.

User intent modeling finds utility in other contexts, such as chatbots, where it enhances the user experience by providing more human-like interactions~\cite{rapp2021human}. Recommender systems rely on user intent modeling to make more accurate and personalized suggestions~\cite{villegas2018characterizing}. In software applications, it contributes to a better understanding of user behavior and improving user interfaces~\cite{auch2020similarity}. Additionally, user intent modeling significantly optimizes web services and enhances user interactions~\cite{obidallah2020clustering}. 

The field of user intent modeling encompasses various machine learning models, including Support Vector Machines (SVM)~\cite{xia2018zero, hu2017deep}, Latent Dirichlet Allocation (LDA)~\cite{chen2013wt, weismayer2017identifying}, Naive Bayes~\cite{HuREc,GuRec}, and deep learning models like Bidirectional Encoder Representations from Transformers (BERT)~\cite{yao2022reprbert}, Word2vec~\cite{AminuRec,YeRec}, and Multilayer Perceptron (MLP)~\cite{XU2022102545,Qu7837964}. A thorough examination of these models and their characteristics provides a comprehensive understanding of their advantages and limitations, offering valuable insights for future research and development.

The process of selecting the most suitable machine learning model for user intent modeling in recommender systems can be challenging due to the wide array of models and approaches available~\cite{zhang2019deep,ricci2015recommender}. The lack of a clear classification scheme further complicates the model selection process~\cite{portugal2018use}. Researchers and developers often struggle to navigate the multitude of available models, leading to uncertainty and a lack of confidence in selecting the optimal model for their specific requirements~\cite{allamanis2018survey,hill2016trials}. Overcoming these challenges is crucial for developing effective solutions in user intent modeling and recommendation tasks, underscoring the need for continued research to enhance model selection and development processes.

While user intent modeling and its application in conversational recommender systems have gained significant attention, existing research in this field is often scattered across diverse sources, hindering comprehensive understanding. Moreover, the multitude of machine learning models, concepts, datasets, and evaluation measures utilized in this research can be overwhelming. To address these issues, we conducted a systematic literature review following the guidelines of Kitchenham~\cite{kitchenham2009systematic}, Xiao~\cite{xiao2019guidance}, and Okoli~\cite{okoli2015guide} to consolidate and analyze the information, providing a more comprehensive understanding of the field. Additionally, we developed a decision model based on the data collected from the literature review, serving as a valuable tool for selecting intent modeling approaches. To evaluate the effectiveness of the decision model, we conducted two academic case studies following the guidelines outlined by Yin~\cite{yin1981case}.

This study presents a Systematic Literature Review (SLR) on user intent modeling within conversational recommender systems. Additionally, it proposes a decision model based on the collected data to guide research modelers in making informed decisions. Section~\ref{ResearchApproach} defines the problem statement and research questions and outlines the research methods employed, including systematic literature study and case study research. Section~\ref{SLR-Methodology} outlines the methodology used in the SLR, covering the review protocol, paper collection procedures, inclusion/exclusion criteria, quality assessment techniques, data extraction methods, synthesis processes, and systematic search approach. In Section~\ref{SLR-results}, the findings and analysis of the SLR are presented, exploring various aspects of user intent modeling, such as models and their characteristics, feature engineering techniques, model combinations, emerging trends, quality evaluation measures, and available datasets. Section~\ref{Decisionmakingprocess} focuses on the practical utilization of the collected data, addressing project-specific concerns through the introduced decision model. This meta-model serves as a framework for effective decision-making, particularly in model selection. Section~\ref{CaseStudies} includes insightful academic case studies that provide practical insights and validate the conducted research to enrich the evaluation of findings. Section~\ref{Discussion} critically examines the outcomes of the SLR, discussing lessons learned, implications of the findings, and addressing potential threats to the study's validity. Section~\ref{RelatedStudies} situates our study and the decision model within the broader landscape of related research studies, establishing their unique contributions and relevance. Finally, in Section~\ref{Conclusion-FutureWork}, the paper summarizes the study's contributions and highlights avenues for future research, providing a cohesive closure to the research on user intent modeling in conversational recommender systems.
\section{Research Approach}\label{ResearchApproach}

This study adopted a systematic research approach, combining SLR and Case Study Research to investigate user intent modeling approaches. The SLR enabled us to gather and analyze relevant information from existing literature, while the case studies allowed us to assess the practical applicability of our findings.

\subsection{Problem Statement}\label{ProblemStatement}
Developing effective search engines and recommendation systems relies on accurately identifying and understanding user intent~\cite{ye2016starrysky, wang2021learning}. However, user intent modeling lacks consensus and comprehensive analysis of optimal approaches~\cite{portugal2018use}. This scattered knowledge makes it challenging for researchers to choose suitable models for specific scenarios~\cite{nguyen2004capturing}. Additionally, combining models to enhance conversational recommender systems' accuracy presents a formidable challenge~\cite{hernandez2019comparative}. Understanding prevailing trends, emerging patterns, and appropriate evaluation measures for intent modeling approaches further complicate the development of effective systems~\cite{chen2015recommender,jordan2015machine, telikani2021evolutionary, singh2016review}. Furthermore, selecting representative datasets for training and evaluation is not straightforward~\cite{zaib2022conversational}. Consequently, in the realm of intent modeling approaches, the following research challenges have been identified:

\noindent\textbf{Scattered knowledge:} The concepts, models, and characteristics of intent modeling approaches are dispersed across diverse academic literature~\cite{portugal2018use}, hindering informed decision-making for developing conversational recommender systems. Systematically consolidating and categorizing existing approaches is demanding. Researchers need a comprehensive landscape of intent modeling techniques to make better choices.

\noindent\textbf{Model combinations and integration:} Combining and integrating models in user intent modeling is challenging~\cite{von2020combining}. Finding effective model combinations to improve conversational recommender systems' accuracy requires investigating compatibility and synergy between models.

\noindent\textbf{Trends and emerging patterns:} Understanding prevailing trends and emerging patterns in user intent modeling approaches is crucial. Researchers need to analyze a large volume of research papers to identify such patterns and tailor their efforts accordingly~\cite{chen2015recommender,jordan2015machine}.

\noindent\textbf{Selecting assessment criteria:} Choosing appropriate evaluation measures and quality attributes for assessing intent modeling approaches is challenging. Researchers must identify measures tailored to each approach to evaluate their performance accurately~\cite{telikani2021evolutionary, singh2016review}.

\noindent\textbf{Selecting datasets:} Selecting suitable datasets for training and evaluating intent modeling approaches is complex. Researchers must analyze and choose representative datasets encompassing various intents and user behaviors to develop robust intent models~\cite{yuan2020expert}.

\noindent\textbf{Decision-making process:} A comprehensive decision model encompassing various intent modeling concepts and guidelines for selecting model combinations and conducting systematic evaluations is missing from the existing literature~\cite{farshidi2020capturing,farshidi2020multi}. Such a model would aid researchers in navigating the complexities of intent modeling and streamlining their decision-making processes.

\subsection{Research Questions}\label{ResearchQuestions}
Based on the identified research challenges in intent modeling approaches, the following research questions are formulated:

\noindent$RQ_1$: What types and categories of models have researchers commonly used in the literature, following best practices, for developing decision-making in conversational recommender systems?

\noindent$RQ_2$: What are the essential features that models in the context of conversational recommender systems must possess to address the requirements of researchers effectively?

\noindent$RQ_3$: Are there any discernible trends in using models to develop conversational recommender systems?

\noindent$RQ_4$: What evaluation measures and quality attributes are most suitable for accurately assessing the performance of user intent modeling approaches?

\noindent$RQ_5$: How can researchers identify and select representative datasets that accurately depict real-world scenarios, enabling effective training and evaluation of intent modeling approaches?

\noindent$RQ_6$: How can we develop a comprehensive decision model to guide researchers in making informed decisions while developing user intent modeling approaches?

\subsection{Research Methods}

We utilized a mixed research method~\cite{jansen2009applied,johnson2004mixed} to tackle the research questions, combining SLR and Case Study Research. The SLR allowed us to gain a comprehensive understanding of user intent modeling approaches, and the case studies assessed the practical applicability of the proposed decision model in real-world scenarios.

The SLR followed guidelines by Kitchenham~\cite{kitchenham2009systematic}, Xiao~\cite{xiao2019guidance}, and Okoli~\cite{okoli2015guide} to identify models, their definitions, model combinations, supported features, potential evaluation measures, and relevant concepts from existing literature. Based on the SLR findings, we developed a decision model, drawing from our previous studies on multi-criteria decision-making in software engineering~\cite{farshidi2020multi}.

To evaluate the practical applicability of the decision model, we conducted two case studies, following the guidelines of Yin~\cite{yin2017case}. These case studies assessed if the proposed decision model effectively assisted research modelers in selecting models for their projects.

We addressed the research questions by employing this mixed research method, including SLR and case studies, contributing meaningful insights and practical solutions to advance intent modeling and improve conversational recommender systems.
\section{Systematic Literature Review Methodology}\label{SLR-Methodology}

In this study, we followed the procedures and guidelines outlined by Kitchenham~\cite{kitchenham2009systematic}, Xiao~\cite{xiao2019guidance}, and Okoli~\cite{okoli2015guide} to address the research question highlighted in Section~\ref{ResearchQuestions}. Accordingly, we adopted the following review protocol (see Figure~\ref{fig:6step_DSR}) to systematically collect and extract data from relevant studies. The following steps were taken to conduct the SLR:

\begin{figure*}[!ht]
\centering
\includegraphics[trim=31 15 90 152,clip,width=1.0\textwidth]{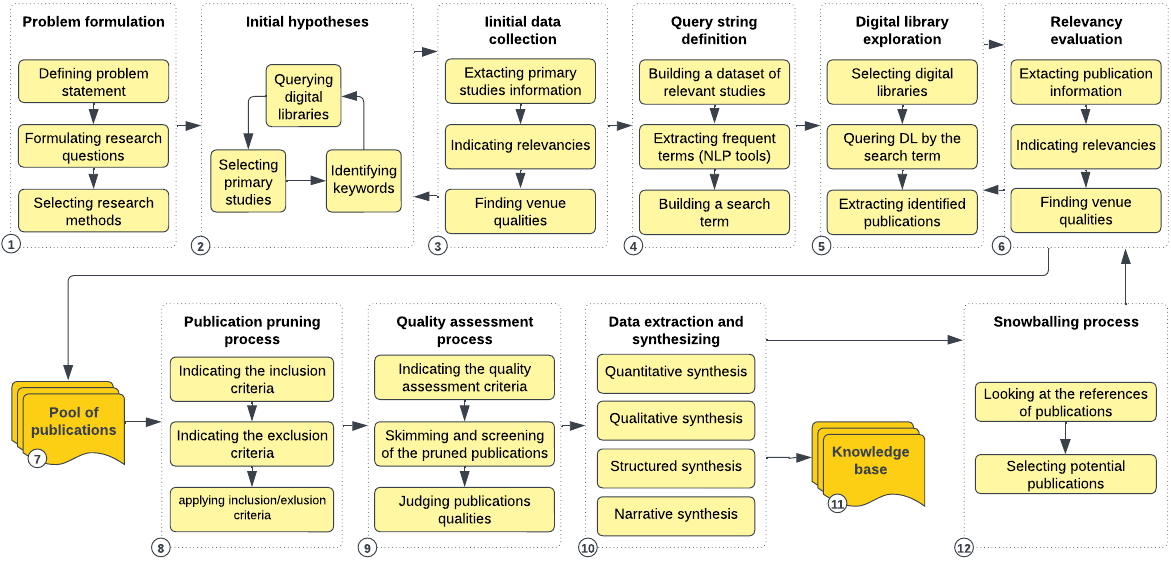}
\caption{ illustrates the review protocol employed in this study, following the prescribed procedures and guidelines outlined by Kitchenham~\cite{kitchenham2009systematic}, Xiao~\cite{xiao2019guidance}, and Okoli~\cite{okoli2015guide}. The review protocol consists of 12 elements systematically executed to collect and extract data from relevant studies. These steps ensure a rigorous investigation and adherence to scientific standards in the research process.}
\label{fig:6step_DSR}
\end{figure*}

\noindent\textbf{(1) Problem formulation:} In this research phase, we followed the prescribed procedures and guidelines of Xiao~\cite{xiao2019guidance} to define the problem statement and research questions. By identifying the research methods, including using an SLR, we ensured that our study addressed a subset of research questions suitable for an SLR. This systematic approach allowed us to conduct a rigorous investigation.

\noindent\textbf{(2) Initial hypotheses:} During the initial stage, we considered a set of keywords to search for primary studies that could address our research questions. These keywords formed the basis for identifying potential seed papers, which served as the starting point for our literature review. This method enabled us to explore relevant publications systematically.

\noindent\textbf{(3) Initial data collection:} We manually collected a comprehensive set of characteristics for primary studies, including source, URL, title, keywords, abstract, venue, venue quality, type of publication, number of citations, publication year, relevancy level. This meticulous process ensured that our review focused on essential information and facilitated the establishment of inclusion/exclusion criteria.

\noindent\textbf{(4) Query string definition:} By analyzing primary studies' keywords, abstracts, and titles, we constructed a search query based on frequent terms found in highly relevant and high-quality papers. This approach helped refine our search and ensure the inclusion of relevant publications.

\noindent\textbf{(5) Digital library exploration:} We thoroughly explored digital libraries such as ACM, ScienceDirect, and Elsevier, using the generated search query to query these databases. This systematic exploration of reputable sources ensured the comprehensive coverage of relevant publications.

\noindent\textbf{(6) Relevancy Evaluation:} We assessed the characteristics of the resulting publications and added them to our collection while estimating their relevancy based on their alignment with our research questions and challenges. This evaluation process ensured the inclusion of highly relevant publications in our review.

\noindent\textbf{(7) The pool of publications:} The collected papers and their associated characteristics formed the pool of publications that served as the foundation for our subsequent review. This pool was continuously expanded during the snowballing process, ensuring a comprehensive examination of the literature.

\noindent\textbf{(8) Publication pruning process:} We rigorously applied inclusion/exclusion criteria to evaluate the pool of publications, eliminating irrelevant material and refining the selection to include the most relevant and high-quality studies. This process enhanced the quality and focus of our review.

\noindent\textbf{(9) Quality assessment process:} We assessed the quality of the remaining publications based on established criteria, including the clarity of research questions and findings. This evaluation ensured that only high-quality studies were included in our review, enhancing the reliability of our findings.

\noindent\textbf{(10) Data extraction and synthesizing:} Through systematic data extraction, we obtained relevant information from the selected publications, synthesizing the findings to identify key insights. This rigorous process facilitated the identification and summarization of critical information.

\noindent\textbf{(11) Knowledge base:} The final set of selected highly relevant and high-quality publications, along with their characteristics, formed a comprehensive knowledge base. Additionally, the extracted data provided a mapping that connected specific findings to their respective sources. This knowledge base is a valuable resource for future research, offering a consolidated summary of essential findings and enabling further analysis.

\noindent\textbf{(12) Snowballing process:} By reviewing the references of selected publications, we identified additional relevant papers that may have been initially overlooked. This snowballing process ensured our review's comprehensiveness and enriched our findings.

By meticulously following this systematic review protocol, we adhered to rigorous and scientific standards in collecting and analyzing the relevant literature on user intent modeling approaches. This approach ensured the validity and reliability of our study, allowing us to address the research questions identified in our study effectively.

\subsection{Review protocol}
This section explains how we followed the review protocol presented in Figure~\ref{fig:6step_DSR} to conduct our SLR.
\subsubsection{Paper collection}
During the \textit{automatic search} phase of our systematic literature review, we implemented a robust search strategy to retrieve pertinent and high-quality publications from scientific search engines. To formulate our search query, we extracted keywords from an initial set of publications obtained through the \textit{manual search} process. These keywords were identified based on the frequent terms used by researchers in highly relevant and high-quality papers. We further refined the keyword selection using a topic modeling tool, Sketch Engine~\cite{kilgarriff2014sketch}, which helped identify additional relevant terms. In total, we identified 314 highly relevant and high-quality publications during the initial part of this phase of the SLR.

The search query was carefully constructed to target publications that specifically addressed user intent modeling in the context of search engines and recommender systems. It aimed to cover various topics such as intent detection, intent prediction, interactive intent modeling, conversational search, intent classification, and user behavior modeling. The query was formulated using logical operators "AND" and "OR" to combine the selected keywords. The search query in this SLR is as follows.

\textit{\scriptsize("user intent" OR "user intent modeling" OR "topic model" OR "user intent detection" OR "user intent prediction" OR "interactive intent modeling" OR "conversational search" OR "intent classification" OR "intent mining" OR "conversational recommender system" OR "user response prediction" OR "user behavior modeling" OR "interactive user intent" OR "intent detection" OR "concept discovery") AND ("search engine" OR "recommender system")}

The search query was employed during the \textit{automatic search} phase, and the resulting publications (a total of 3,828 out of 13,168 results considered in the pool of publications) underwent a rigorous screening process based on our predefined inclusion/exclusion criteria. This ensured that only relevant and high-quality publications were included in our data extraction and analysis. The effectiveness of the search query was assessed by comparing the search results with those obtained from the \textit{manual search} to ensure consistency and comprehensiveness. The search query used in our study was derived from previous research and validated to retrieve publications relevant to user intent modeling in search engines and recommender systems.

\subsubsection{Inclusion/exclusion criteria}

Inclusion/exclusion criteria are essential guidelines used to determine the relevance and eligibility of studies for inclusion in a systematic literature review or meta-analysis. These criteria are crucial in ensuring that the selected studies are high quality and directly address the research question under investigation. Inclusion criteria specify the characteristics or attributes a study must possess to be considered for inclusion in the review.

We employed rigorous inclusion and exclusion criteria during this study phase to filter out irrelevant and low-quality publications. Our criteria encompassed several factors, including the quality of the publication venue, the publication year, the number of citations, and the relevancy of the publication to our research topic. These criteria were carefully defined and consistently applied to ensure that only high-quality and relevant publications were included in our review. By adhering to these criteria, we evaluated publications that provided valuable insights and contributed significantly to our research topic. After applying our predefined inclusion/exclusion criteria, we identified and selected 1,067 publications out of the initial pool of 3,828 publications.

\subsubsection{Quality assessment}

During the SLR, we comprehensively assessed the quality of the selected publications after applying the inclusion/exclusion criteria. Several factors were taken into consideration to evaluate the quality and suitability of the publications for our research:

\noindent\textbf{Research Method:} We evaluated whether the chosen research method was appropriate for addressing the research question. The clarity and transparency of the research methodology were also assessed.

\noindent\textbf{Research Type:} We considered whether the publication presented original research, a review article, a case study, or a meta-analysis. The relevance and scope of the research in the field of machine learning were also taken into account.

\noindent\textbf{Data Collection Method:} We evaluated the appropriateness of the data collection method in relation to the research question. The adequacy and clarity of the reported data collection process were also assessed.

\noindent\textbf{Evaluation Method:} We assessed whether the chosen evaluation method was suitable for addressing the research question. The transparency and statistical significance of the reported results were considered.

\noindent\textbf{Problem Statement:} We evaluated whether the publication identified the research problem and provided sufficient background information. The clarity and definition of the research question were also taken into account.

\noindent\textbf{Research Questions:} We assessed the relevance, clarity, and definition of the research questions in relation to the research problem.

\noindent\textbf{Research Challenges:} We considered whether the publication identified and acknowledged the challenges and limitations associated with the research.

\noindent\textbf{Statement of Findings:} We evaluated whether the publication reported the research results and whether the findings were relevant to the research problem and questions.

\noindent\textbf{Real-World Use Cases:} We assessed whether the publication provided real-world use cases or applications for the proposed method or model.

Based on the aforementioned factors' assessment, a team of five researchers involved in the SLR evaluated the publications' quality. Each researcher independently assessed the publications based on the established criteria. In cases where there were discrepancies or differences in evaluating a publication's quality, the researchers engaged in discussions to reach a consensus and ensure a consistent assessment.

Through this collaborative evaluation process, a final selection of 791 publications was made from the initial pool of 1,067 publications. These selected publications demonstrated high quality and relevance to our research question, meeting the predefined inclusion/exclusion criteria. The consensus reached by the research team ensured a rigorous and reliable selection of publications for further analysis and data extraction in the SLR.

\subsubsection{Data extraction and synthesizing}
During the data extraction and synthesis phase of the SLR, our primary objective was to address the identified research questions and gain insights into the foundational models commonly employed by researchers in their intent modeling approaches. We aimed to understand the features of these models, the associated quality attributes, and the evaluation measures utilized by research modelers to assess their approaches. Furthermore, we explored the potential combinations of models that researchers incorporated into their research papers.

We extracted relevant data from the papers included in our review to achieve these objectives. In our perspective, evaluation measures encompassed a range of measurements and key performance indicators (KPIs) used to evaluate the performance of the models. Quality attributes represent the characteristics of models that are not easily quantifiable and are typically assigned values using Likert scales or similar approaches. For example, authors may assess the performance of a model as high or low compared to other models. On the other hand, features encompassed any characteristics of models that authors highlighted to demonstrate specific functionalities. These features played a role in the selection of models by research modelers. Examples of features include ranking and prediction capabilities.

In this context, "models" refer to mathematical, algorithmic models or processes that can be applied in various domains. For instance, Support Vector Machines (SVM)~\cite{xia2018zero, hu2017deep} and Bayesian Personalized Ranking (BPR)~\cite{ni2021effective, wang2018streaming} are examples of models commonly utilized in intent modeling.

By extracting and analyzing this data, we aimed to comprehensively understand the existing literature, including popular open-access datasets used for training and evaluating the models. This knowledge empowered us to contribute insights and recommendations to the academic community, supporting them in selecting appropriate models and approaches for their intent modeling research endeavors.

\subsection{Search process}

In this study, we followed the review protocol presented in this section (see Figure~\ref{fig:6step_DSR}) to gather relevant studies. The search process involved an automated search phase, which utilized renowned digital libraries such as ACM DL, IEEE Xplore, ScienceDirect, and Springer. However, Google Scholar was excluded from the automated search due to its tendency to generate numerous irrelevant studies. Furthermore, Google Scholar significantly overlaps the other digital libraries considered in this SLR. Table~\ref{table:SearchProcess} provides a comprehensive overview of the sequential phases of the search process, outlining the number of studies encompassed within each stage.

\begin{table}[!ht]
\centering
\caption{\scriptsize presents an overview of the systematic search process employed to identify relevant publications on user intent modeling. The search process involved both manual and automatic searches, incorporating specific inclusion/exclusion criteria to ensure the retrieval of high-quality results. The search query used in the automatic search was carefully designed to retrieve relevant publications from scientific search engines, while the manual search involved screening articles from selected venues. The final set of articles obtained from both searches was then subjected to comprehensive analysis and synthesis to provide valuable insights into the current state of research on user intent modeling.}
\begin{tabular}{rrrrrr}
\scriptsize

               & \textbf{\#hits} & \textbf{Phase 1} &\textbf{ Phase 2 }& \textbf{Phase 3 }&\textbf{ Phase 4} \\ \hline
Google Scholar & 3,940   & 314     & 96      & 96      & 68      \\
ACM DL         & 2,152   & 586     & 311     & 311     & 243     \\
IEEE Xplore    & 89     & 82      & 9       & 9       & 7       \\
ScienceDirect  & 1,528   & 921     & 246     & 246     & 190     \\
Springer       & 5,459   & 1,896    & 379     & 379     & 263     \\
Snowballing    & N/A    & 29      & 26      & 26      & 20      \\\hline
               & 13,168  & 3,828    & 1,067    & 1,067    & \textbf{791}    
\end{tabular}
\label{table:SearchProcess}
\end{table}

Table~\ref{table:SearchProcess} provides insights into the search process conducted in four phases: Phase 1, Phase 2, Phase 3, and Phase 4.

\noindent\textbf{Phase 1 (Pool of Publications):} We initially performed a manual search, resulting in 314 relevant publications from Google Scholar. Additionally, automated searches from ACM DL, IEEE Xplore, ScienceDirect, and Springer contributed to the pool of publications with 586, 82, 921, and 1896 relevant papers, respectively.

\noindent\textbf{Phase 2 (Publication pruning process):} In this phase, the inclusion/exclusion criteria were applied to the collected publications, ensuring the selection of high-quality and relevant studies. The numbers were reduced to 96 in ACM DL, 9 in IEEE Xplore, 246 in ScienceDirect, and 379 in Springer.

\noindent\textbf{Phase 3 (Quality assessment process):} Quality assessment was conducted for the publications based on several criteria, resulting in a final selection of 1067 studies from all sources.

\noindent\textbf{Phase 4 (Data extraction and synthesizing + Snowballing process):} During this phase, data extraction and synthesis were performed to gain insights into foundational intent modeling models, quality attributes, evaluation measures, and potential combinations of models used by researchers. Additionally, snowballing, involving reviewing references of selected publications, led to an additional 20 relevant papers. By carefully applying the review protocol and snowballing, we retrieved 791 high-quality studies for our comprehensive analysis and synthesis in this systematic literature review.

\section{Findings and Analysis}\label{SLR-results}
In this section, we present the SLR results and provide an overview of the collected data, which were analyzed to address the research questions identified in our study. 

\subsection{Models}
The SLR conducted in our study has revealed a diverse array of models employed in user intent modeling. These models encompass a range of approaches, each characterized by unique characteristics and methodologies. For a comprehensive understanding of these models, including their definitions and descriptions, please refer to the appendix (Appendix~\ref{Appendix_Models}).

We have examined their underlying principles and methodologies to categorize these models effectively. The appendix (Appendix~\ref{Appendix_CategoriesOfModels}) provides detailed definitions and explanations of the identified categories, offering comprehensive insights into each category and its specific characteristics.

Among the identified categories, prominent ones include Classification~\cite{qu2019user, zhang2016mining} and Clustering~\cite{zhang2021discovering, agarwal2020evaluation} models, Convolutional Neural Network (CNN)\cite{wang2020next, zhang2016mining}, Deep Belief Networks (DBN)\cite{zhang2018discrete, hu2017deep}, and Graph Neural Networks (GNN)~\cite{yu2022graph, lin2021go}, among others. These categories encompass a broad range of modeling techniques applied in user intent modeling research. However, it is important to note that these categories represent only a subset of the diverse range of models identified in our SLR.

\begin{table}[!ht]
\scriptsize
\caption{shows the mapping of 59 models to their respective categories in user intent modeling. The table showcases the models that appear in at least six publications. However, for access to the complete list of models, please refer to the supplementary materials available on \textit{Mendeley Data}~\cite{Farshidi_Rezaee_2023}. }
\centering
\includegraphics[trim=10 10 5 10  ,clip,width=1\textwidth]{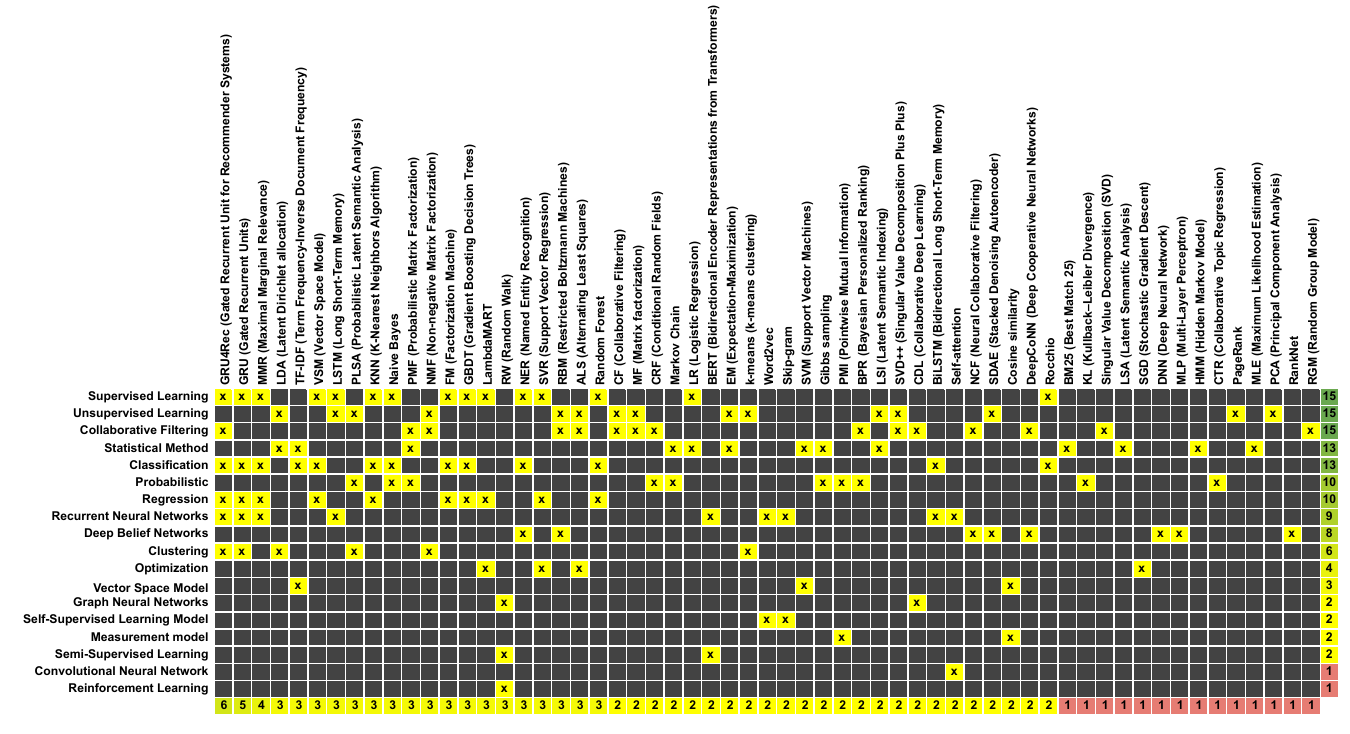}
\label{table:ModelCategories}
\end{table}

Table~\ref{table:ModelCategories} presents an overview of the 59 most frequently mentioned models in the SLR on user intent modeling. The table showcases the models appearing in at least six publications (columns) and their corresponding 18 categories (rows). Each model in user intent modeling can often be categorized into multiple categories, highlighting their versatility and diverse functionalities. For example, GRU4Rec, a widely recognized model in the field (cited in 10 publications included in our review), exhibits characteristics that align with various categories. GRU4Rec falls under Supervised Learning, as it uses labeled examples during training to predict user intent. Additionally, it incorporates Collaborative Filtering techniques by analyzing user behavior and preferences to generate personalized recommendations, associating it with the Collaborative Filtering category~\cite{latifi2021session}. Moreover, GRU4Rec can be classified as a Classification model as it categorizes input data into specific classes or categories to predict user intent~\cite{park2020click}. It also demonstrates traits of Regression models by estimating and predicting user preferences or ratings based on the available data. Considering its reliance on recurrent connections, GRU4Rec can be associated with the Recurrent Neural Networks (RNN) category, enabling it to process sequential data and capture temporal dependencies~\cite{ludewig2018evaluation}. Lastly, GRU4Rec's ability to cluster similar users or items based on their behavior and preferences places it within the Clustering category. This clustering capability provides valuable insights and recommendations to users based on their respective clusters.

\subsection{Features}
Our study conducted a comprehensive investigation of the features supported by models in user intent modeling, emphasizing their significance in the field. We identified a total of 74 distinct features that were consistently mentioned in at least six publications\footnote{For access to the complete list of features, please refer to the supplementary materials available on \textit{Mendeley Data}~\cite{Farshidi_Rezaee_2023}.}, highlighting their relevance and impact in intent modeling research. For a comprehensive understanding of these features, please consult Appendix~\ref{Appendix_Features}, where detailed definitions and explanations are provided.

\begin{table}[!ht]
\scriptsize
\caption{illustrates a mapping of features to models in user intent modeling. The table presents the comprehensive mapping of 74 distinct features to the corresponding models, grouped into 20 categories. For detailed definitions and explanations of the features, please refer to Appendix~\ref{Appendix_Features}.}
\centering
\includegraphics[trim=0 220 0 0  ,clip,width=1\textwidth]{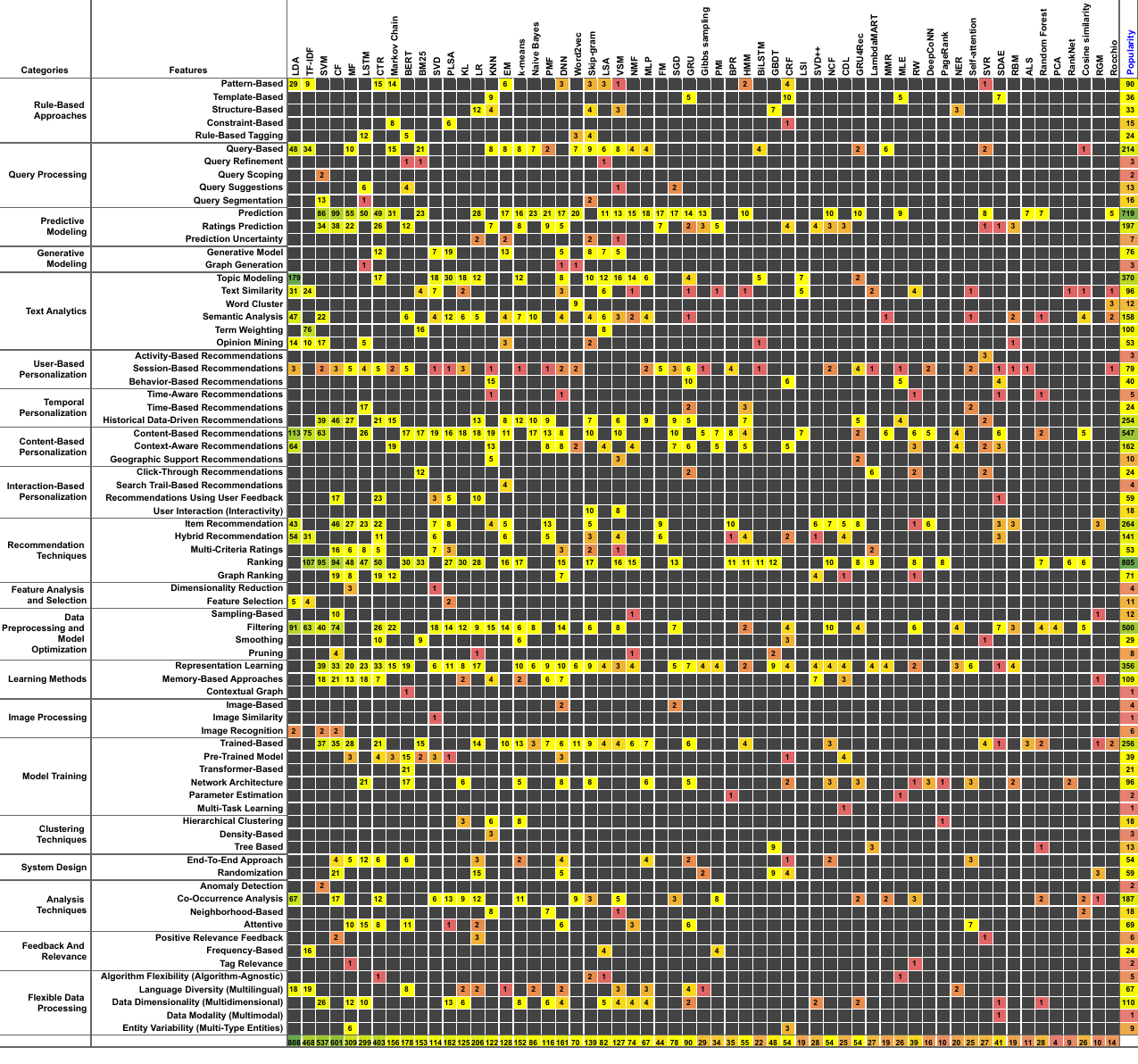}
\label{table:ModelsToFeatures}
\end{table}

To effectively organize and comprehend these features, we categorized them into 20 categories based on their context, domain, and applications. Machine learning models possess the versatility to support a wide range of features, each tailored to specific use cases and applications. Some common features include historical data~\cite{zhou2020leveraging,white2013enhancing,zou2022improving}, enabling models to learn from past experiences and predict future outcomes. Algorithm-agnostic models~\cite{zhou2019real,musto2019linked,mandayam2017intent} provide the flexibility to select the most suitable algorithm for a particular task. Model-based~\cite{ding2022tdtmf,pradhan2021claver,yu2018pave} features leverage statistical methods~\cite{schlaefer2011statistical,kim2017deep} and semantic analysis~\cite{zhang2016improving,xu2015topic} to offer predictions based on specific models.

Table~\ref{table:ModelsToFeatures} illustrates the mapping of features to models in user intent modeling, highlighting the frequency of explicit mentions in relevant publications. Each cell represents the number of publications that specifically refer to the corresponding feature in relation to the associated models. Authors of these papers have emphasized their feature requirements as a pivotal factor in selecting particular models. The color-coded cells indicate the range of publication counts, ranging from low to high, reflecting the level of support for each feature by the models. It is noteworthy that gray cells indicate the absence of evidence supporting the feature's compatibility with a specific model, based on the comprehensive review of 791 papers conducted in this study. For example, among the analyzed publications, we identified 29 instances where LDA (Latent Dirichlet Allocation) was mentioned as being applicable in pattern-based approaches within the context of rule-based methods~\cite{tang2010combination, li2014identifying}. This implies that researchers and authors found LDA to be relevant and applicable in scenarios where patterns are analyzed, and rules are used to extract meaningful information or make decisions.

This mapping process involves determining which models are most suitable for addressing specific features in a given problem. It necessitates a comprehensive analysis of the problem's characteristics and an understanding of the capabilities, strengths, and weaknesses of the available models. For instance, in the domain of image classification, deep learning models such as Convolutional Neural Networks (CNNs) and Recurrent Neural Networks (RNNs) have proven to be effective in handling features related to image recognition and processing. Conversely, for time series forecasting problems, models like ARIMA, LSTM, or GRU may be more suitable choices.

\subsection{Model combinations}
Following the completion of the data extraction and synthesis phase of the SLR, a total of 59 models were identified, each mentioned in at least six publications. It became evident that some of these models were integrated to address the considerations of research modelers, including feature requirements, quality attributes, and evaluation measures (see Figure~\ref{fig:meta-model}). The selected publications proposed viable combinations of models based on the authors' research and assessed the outcomes resulting from these combinations.

\begin{figure}[!ht]
\centering
\includegraphics[trim=0 310 0 0,clip,width=1\textwidth]{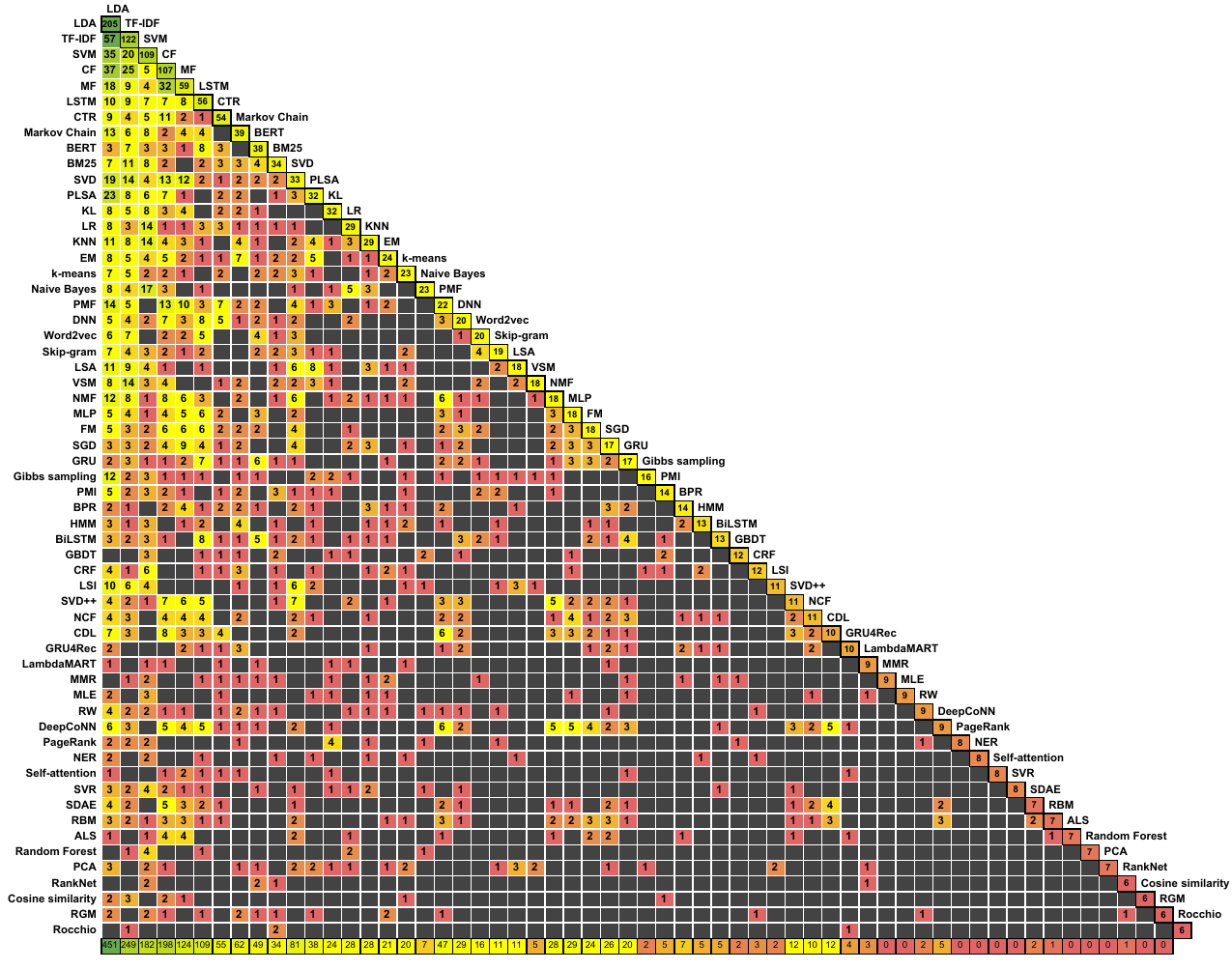}
\scriptsize
\caption{shows a matrix representation of model combinations in user intent modeling research. The matrix illustrates the combinations of 59 selected models, where each cell indicates the number of publications discussing the corresponding model combination. The diagonal cells represent the number of publications discussing each model individually. Green cells indicate a higher research volume, while yellow and red cells indicate lower volumes. Gray cells represent areas where no evidence was found for valid combinations. The last row of the matrix represents the frequency of publications in which the models on the diagonal cells were considered in combination with others. For instance, we identified 451 publications that mentioned LDA as one of their design decisions in combination with other models. The combination matrix provides insights into the frequency and popularity of model combinations, aiding researchers in identifying existing combinations and areas for further investigation.}
\label{fig:ModelCombinations}
\end{figure}

To thoroughly examine the various model combinations, a matrix resembling a symmetric adjacency matrix was constructed, treating the models as nodes and the combinations as edges in a graph representation. The upper or lower triangular matrix was utilized to depict unique combinations. Figure~\ref{fig:ModelCombinations} visually presents this combination matrix, encompassing the 59 selected models. The diagonal cells of the matrix indicate the number of publications discussing each model independently. For instance, our analysis identified 205 papers concerning LDA~\cite{chen2013wt, weismayer2017identifying} and 122 papers focusing on TF-IDF~\cite{binkley2018need, izadi2022predicting}.

Within the matrix, the cells represent the number of papers discussing the combinations of the corresponding columns and rows. For example, there were 57 papers discussing the combination of LDA and TF-IDF~\cite{venkateswara2022societal}, while 35 papers delved into the combination of SVM and LDA~\cite{yu2015combining}.

The color coding in the matrix indicates the number of research articles associated with each combination. Green cells signify a higher volume of research conducted in the literature, while yellow and red cells denote lower volumes. Additionally, gray cells indicate areas without evidence regarding valid combinations based on the authors' perspectives. However, it is crucial to note that these gray cells represent potential areas warranting further investigation, offering researchers opportunities to explore the feasibility of such combinations.

Overall, the combination matrix serves as an extensive overview of the model combinations in user intent modeling research, shedding light on the frequency of their occurrence in the literature. It can be considered a valuable resource for researchers and practitioners seeking to identify existing combinations and areas requiring further exploration.

\subsection{Model trends}
In recent studies, machine learning models have witnessed significant advancements across various fields, leading to notable trends in their development and application. However, it is worth noting that our study goes beyond recent years. By using the term "models," we refer to a wide range of models that research modelers can employ in user intent modeling.

\begin{table}[!ht]
\scriptsize
\caption{illustrates the trend of models mentioned in user intent modeling research over publication years, highlighting the popularity and emergence of various models. }
\centering
\includegraphics[trim=0 600 630 0  ,clip,width=1\textwidth]{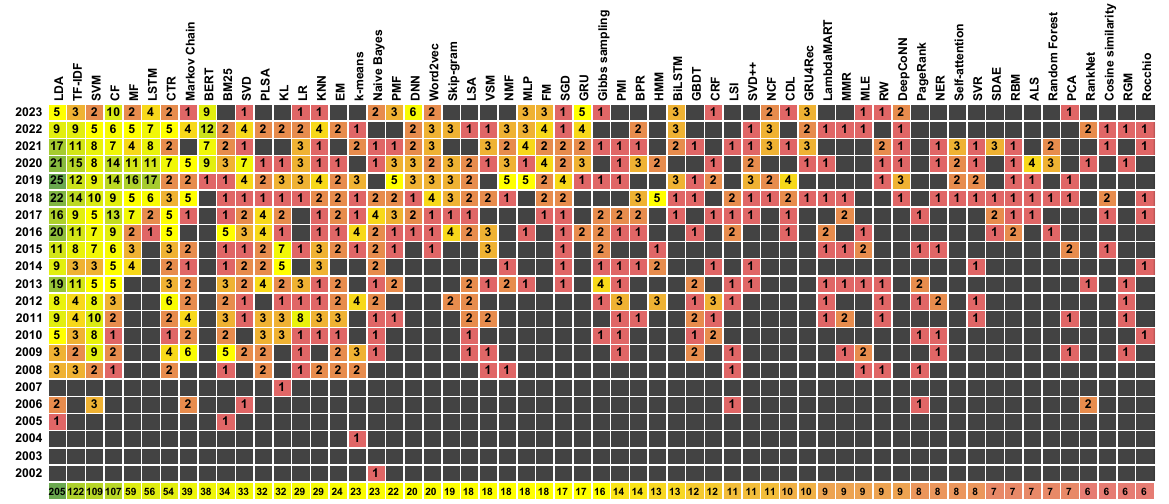}
\label{table:ModelTrends}
\end{table}

To gain insights into the usage patterns of these models, we organized the 59 selected models (mentioned in at least six publications\footnote{For access to the complete list of model, please refer to the supplementary materials available on \textit{Mendeley Data}\cite{Farshidi_Rezaee_2023}.}) based on the publication years of the studies that referenced them. The span of these publications ranges from 2002 to 2023. Table~\ref{table:ModelTrends} provides an overview of these trends.

Among the selected models, LDA, TF-IDF, SVM, CF, and MF emerged as the top five most frequently mentioned models, appearing in over 500 papers. It is important to note that while some recently gained substantial attention, such as BERT~\cite{yao2022reprbert}, CF~\cite{yadav2022clus}, LSTM~\cite{xu2022deep, gozuacik2023technological}, DNN~\cite{yengikand2023dhsirs}, and GRU~\cite{chen2020handling, elfaik2023leveraging}, our study encompasses models from various time periods. 

These trends shed light on the popularity and usage patterns of different models in user intent modeling. By identifying frequently mentioned models and observing shifts in their prevalence over time, researchers and practitioners can stay informed about the evolving landscape of user intent modeling and make informed decisions when selecting models for their specific applications~\cite{zaib2022conversational, ittoo2016text}.

\subsection{Quality models and evaluation measures}

In AI-based projects, high-quality models and comprehensive evaluation measures are crucial. Quality attributes refer to a set of metrics that assess the performance of a model~\cite{de2020intelligent,hernandez2019comparative}, while evaluation measures quantitatively gauge the quality of model outputs~\cite{zaib2022conversational}. These attributes and measures play a critical role in ensuring the generation of accurate and reliable results~\cite{pan2022test,pu2012evaluating,hernandez2019comparative}.

\begin{table}[!ht]
\scriptsize
\caption{shows an overview of quality models and evaluation measures used in machine learning, including performance metrics such as accuracy, precision, recall, F1 score, and AUC-ROC, as well as other evaluation techniques such as cross-validation, holdout validation, and confusion matrices.}
\centering
\includegraphics[trim=-2 460 315 -2,clip,width=0.8\textwidth]{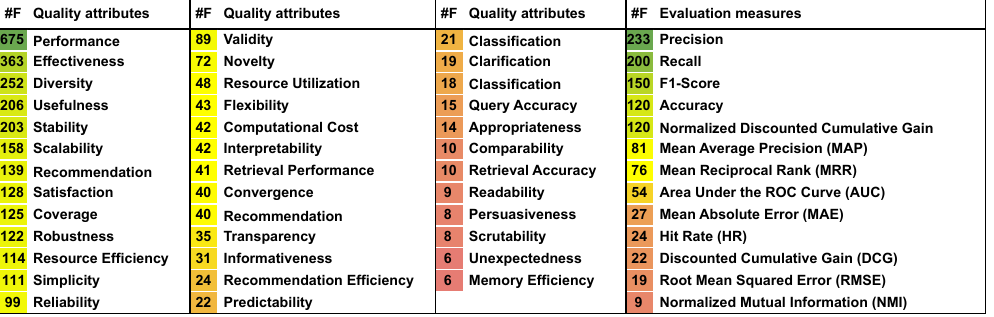}
\label{table:QualitiesEvaluationMeasures}
\end{table}

While accuracy is a commonly employed evaluation measure, it may not adequately represent the model's performance, especially in imbalanced classes. Alternative measures such as precision~\cite{salle2022cosearcher, baykan2011comprehensive}, recall~\cite{wang2022causal, phan2010hidden}, and F1-score~\cite{yu2019adaptive, ashkan2009classifying} are used to evaluate model performance, particularly when dealing with imbalanced data. Additionally, evaluation measures like the area under the curve (AUC)\cite{xu2016spatio, liu2022multi} and receiver operating characteristic (ROC)\cite{wu2019context,wang2020next} curve are frequently used to assess binary classifiers. These measures provide insights into the model's ability to differentiate between positive and negative instances, particularly when the costs associated with false positives and false negatives differ.

For ranking problems, evaluation measures such as mean average precision (MAP)\cite{mao2019multiobjective, ni2012user} and normalized discounted cumulative gain (NDCG)\cite{liu2020dynamic, kaptein2013exploiting} are commonly employed. These measures evaluate the quality of the ranked lists generated by the model and estimate its effectiveness in predicting relevant instances.

When evaluating regression models, measures such as root mean squared error (RMSE)\cite{cai2014object,colace2015collaborative} and mean absolute error (MAE)\cite{yao2017version, yadav2022clus} are used to quantify the discrepancy between predicted values and actual values of the target variable.

The selection of appropriate evaluation measures is crucial to ensure the accuracy and reliability of machine learning models. The suitable measure(s) choice depends on the specific problem domain, data type, and project objectives. These factors are pivotal in selecting the most appropriate quality attributes and evaluation measures. Table~\ref{table:QualitiesEvaluationMeasures} presents the quality attributes and evaluation measures identified in at least six publications\footnote{For access to the complete list of quality attributes and evaluation measures, please refer to the supplementary materials available on \textit{Mendeley Data}\cite{Farshidi_Rezaee_2023}.}. Performance, Effectiveness, Diversity, Usefulness, and Stability are among the top five quality attributes. Precision, Recall, F1-Score, Accuracy, and NDCG are among the top five evaluation measures identified in the SLR. For detailed explanations of the identified quality attributes and evaluation measures, please refer to Appendix~\ref{Appendix_QA_EvaluationMeasures}.

\subsection{Datasets}

Datasets are fundamental to machine learning and data science research, as they provide the raw material for training and testing models and enable the development of solutions to complex problems. They come in various forms and sizes, ranging from small, well-curated collections to large, messy datasets with millions of records. The quality of datasets is crucial~\cite{pan2022test}, as high-quality data ensures the accuracy and reliability of models, while poor-quality data can introduce biases and inaccuracies. Data quality encompasses completeness, accuracy, consistency, and relevance, and ensuring data quality involves cleaning, normalization, transformation, and validation.

\begin{table}[!ht]
\scriptsize

\caption{shows datasets commonly used for user intent modeling approaches. The table includes the names of the datasets and their corresponding URLs.}
\centering
\includegraphics[trim=0 460 300 0  ,clip,width=1\textwidth]{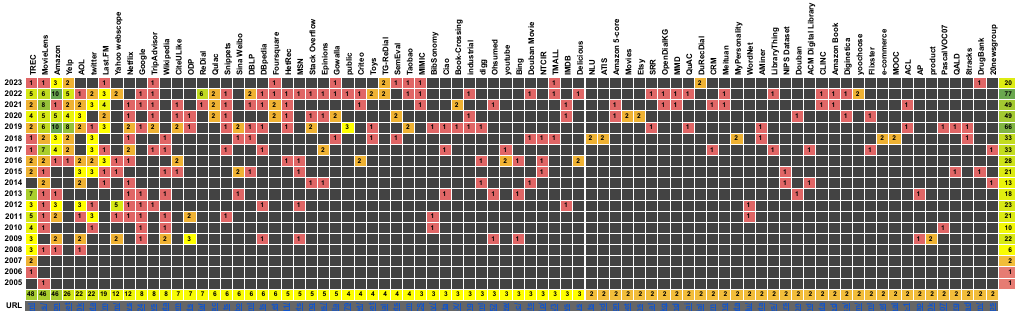}
\label{table:Datasets}
\end{table}

The size and complexity of datasets pose challenges in terms of storage, processing, and analysis. Big datasets require specialized tools and infrastructure to handle the volume and velocity of data. On the other hand, complex datasets, such as graphs, images, and text, may require specialized techniques and models for extracting meaningful information and patterns.

Furthermore, the availability of datasets is a vital consideration in advancing machine learning research and applications. Open datasets that are freely accessible and well-documented foster collaboration and innovation, while proprietary datasets may restrict access and impede progress~\cite{zhang2016mining, teevan2008personalize, ittoo2016text}. Data sharing and ethical considerations in data use are increasingly recognized, leading to efforts to promote open access and responsible data practices.

In this study, we identified 80 datasets that researchers have utilized in the context of intent modeling approaches, and these datasets were mentioned in at least two publications\footnote{For access to the complete list of datasets, please refer to the supplementary materials available on \textit{Mendeley Data}\cite{Farshidi_Rezaee_2023}.}. Table~\ref{table:Datasets} provides an overview of these datasets and their frequency of usage from 2005 to 2023. Notably, TREC, MovieLens, Amazon, Yelp, and AOL emerged as the top five datasets commonly used in evaluating intent modeling approaches for recommender systems~\cite{wang2021adapting, papadimitriou2012generalized, wang2020next} and search engines~\cite{fan2022modeling, liu2022category, konishi2016extracting}. These datasets have been utilized in over 200 publications, highlighting their significance and wide adoption in the field.

\section{Decision making process}\label{Decisionmakingprocess}

This section describes how researchers make decisions when selecting intent modeling approaches. It illustrates a systematic approach to choosing intent modeling methods based on academic literature.

\subsection{Decision meta-model}

Research modelers face the challenge of selecting the most suitable combination of models to develop an intent modeling approach for a conversational recommender system. In this section, we present a meta-model for the decision-making process in the context of intent modeling. The adoption of this meta-model is based on the principles outlined in the ISO/IEC/IEEE standard 42010~\cite{iso420102011iec}, which provides a framework for conceptual modeling of Architecture Description. This process requires a systematic approach to ensure that the chosen models effectively capture and understand users' intentions. Let's consider a scenario where research modelers encounter this challenge and go through the decision-making process:

\noindent\textbf{Goal and Concerns:} The research modelers aim to build an intent modeling approach for a conversational recommender system. Their goal is to accurately determine the underlying purposes or goals behind users' requests, enabling personalized and precise responses. The modelers have concerns regarding quality attributes and functional requirements, and they aim to achieve an acceptable level of quality based on their evaluation measures.

\noindent\textbf{Identification of Models and Features:} To address this problem, the modelers consider various models that can capture users' intentions in the conversational context. They identify essential features, such as \textit{user intent prediction} or \textit{context analysis} based on their concerns. They explore the available models and techniques, such as \textit{Supervised Learning}, \textit{Unsupervised Learning}, \textit{Recurrent Neural Networks}, \textit{Deep Belief Networks}, \textit{Clustering}, and \textit{Self-Supervised Learning Models}. The modelers also consider the recent trends in employing models for intent modeling.

\noindent\textbf{Evaluation of Models:} The modelers review the descriptions and capabilities of several models that align with capturing users' intentions in conversational interactions. They analyze each model's strengths, limitations, and applicability to the intent modeling problem. They consider factors such as the models' ability to handle natural language input, understand context, and predict user intents accurately. This evaluation allows them to shortlist a set of candidate models that have the potential to address the intent modeling challenge effectively.

\noindent\textbf{In-depth Analysis:} The research modelers conduct a more detailed analysis of the shortlisted models. They examine the associated techniques for each model to ensure their suitability in the conversational recommender system. They assess factors such as training data requirements, model complexity, interpretability, and scalability. Additionally, they explore the possibility of combining models to identify compatible combinations or evaluate the existing literature on such combinations. If necessary, further study may be conducted to assess the feasibility of model combinations. This step helps them identify the optimal combination of models that best capture users' intentions in the conversational setting and address their concerns.

\begin{figure*}[!ht]
\centering
\includegraphics[trim=15 25 15 25,clip,width=0.9\textwidth]{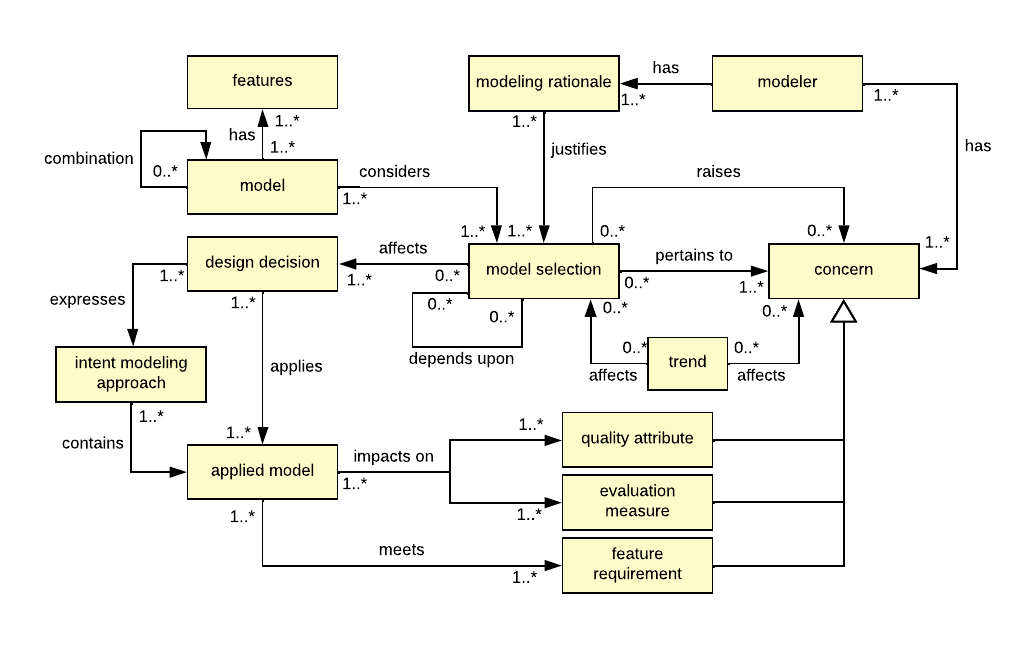}
\caption{illustrates the decision-making process researchers employ in selecting intent modeling approaches within the academic literature.}
\label{fig:meta-model}
\end{figure*}

\subsection{A decision model for intent modeling selection}\label{DecisionModel}

Decision theories have wide-ranging applications in various fields, including e-learning \cite{garg2018madm} and software production \cite{xu2007concepts,fitzgerald2014continuous,rus2003supporting}. In the literature, \textit{decision-making} is commonly defined as a process involving problem identification, data collection, defining alternatives, and selecting feasible solutions with ranked preferences \cite{fitzgerald2017differences,kaufmann2012rationality,garg2020mcdm,garg2017mcdm,sandhya2018computational,garg2019parametric}. However, decision-makers approach decision problems differently, as they have their own priorities, tacit knowledge, and decision-making policies \cite{doumpos2013multicriteria}. These differences in judgment necessitate addressing them in decision models, which is a primary focus in the field of multiple-criteria decision-making (MCDM).

MCDM problems involve evaluating a set of alternatives and considering decision criteria \cite{farshidi2020multi}. The challenge lies in selecting the most suitable alternatives based on decision-makers' preferences and requirements \cite{majumder2015multi}. It is important to note that MCDM problems do not have a single optimal solution, and decision-makers' preferences play a vital role in differentiating between solutions \cite{majumder2015multi}. In this study, we approach the problem of model selection as an MCDM problem within the context of intent modeling approaches for conversational recommender systems.

Let $Models={m_1,m_2, \dots, m_{\| Models\|}}$ be a set of models found in the literature (decision space), such as \textit{LDA}, \textit{SVM}, and \textit{BERT}. Let $Features={f_1,f_2, \dots, f_{\| Features\|}}$ be a set of features associated with the models, such as ranking, prediction, and recommendation. Each model $m \in Models$ supports a subset of the set $Features$ and satisfies a set of evaluation measures ($Measures={e_1,e_2, \dots, e_{\| Measures\|}}$) and quality attributes ($Qualities={q_1,q_2, \dots, q_{\| Qualities\|}}$). The objective is to identify the most suitable models, or a combination of models, represented by the set $Solutions \subset Models$, that address the concerns of researchers denoted as \textit{Concerns}, where $Concerns \subseteq \{ Features \cup Measures \cup Qualities \}$. Accordingly, research modelers can adopt a systematic strategy to select combinations of models by employing an MCDM approach. This approach involves taking $Models$ and their associated $Features$ as input and applying a weighting method to prioritize the $Features$ based on the preferences of decision-makers. Subsequently, the defined $Concerns$ are considered, and an aggregation method is utilized to rank the $Models$ and propose fitting $Solutions$. Consequently, the MCDM approach can be formally expressed as follows:

\begin{equation*}
\begin{aligned}
MCDM: Models \times Features \times Concerns \to Solutions
\end{aligned}
\end{equation*}

The decision model developed for intent modeling, using MCDM theory and depicted in Figure~\ref{fig:meta-model}, is a valuable tool for researchers working on conversational recommender systems. This approach helps researchers explore options systematically, consider important factors for conversational interactions, and choose the best combination of models to create an effective intent modeling approach. The decision model suggests five steps for selecting a combination of models for conversational recommender systems:

\noindent(1) \textbf{Models:} In this phase, researchers should gain insights into best practices and well-known models employed by other researchers in designing conversational recommender systems. Appendix~\ref{Appendix_Models} can be used to understand the definitions of models, while Appendix~\ref{Appendix_CategoriesOfModels} can help in becoming familiar with the categories used to classify these models. Table~\ref{table:ModelCategories} illustrates the categorization of models in this study, and Table~\ref{table:ModelTrends} presents the trends observed among research modelers in utilizing models to build their conversational recommender systems.

\noindent(2) \textbf{Feature Requirements Elicitation:} In this step, researchers need to fully understand the core aspects of the intent modeling problem they are studying. They should carefully analyze their specific scenario to identify the key characteristics required in the models they are seeking, which may involve using a combination of models. For instance, researchers might consider prediction, ranking, and recommendation as essential feature requirements for their conversational recommender systems. Researchers can refer to Appendix~\ref{Appendix_Features} to gain a better understanding of feature definitions and model characteristics, which will help them select the most suitable features for their intent modeling project. 

\noindent(3) \textbf{Finding Feasible Solutions:} In this step, researchers should identify models that can feasibly fulfill all of their feature requirements. Table~\ref{table:ModelsToFeatures} can be used to determine which models support specific features. For example, the table shows that 99 publications explicitly mentioned Collaborative Filtering (CF) as a suitable model for applications requiring predictions, and 94 publications indicated CF's applicability for ranking. Moreover, 46 studies employed CF for item recommendation. Based on these findings, if a conversational recommender system requires these three feature requirements, CF could be selected as one of the potential solutions. If the number of feature requirements increases, the selection problem can be converted into a set covering problem \cite{caprara2000algorithms} to identify the smallest sub-collection of models that collectively satisfy all feature requirements.

\noindent(4) \textbf{Selecting Feasible Combinations:} In this phase, researchers need to assess whether the identified models can be integrated or combined. Figure~\ref{fig:ModelCombinations} provides information on the feasibility of combining models based on the reviewed articles in this study. If the table does not indicate a potential combination, it does not necessarily imply that the combination is impossible. It simply means no evidence supports its feasibility, and researchers should investigate the combination independently.

\noindent(5) \textbf{Performance Analysis:} After identifying a set of feasible combinations, researchers should address their remaining concerns regarding quality attributes and evaluation measures. Table~\ref{table:QualitiesEvaluationMeasures} and Appendix~\ref{Appendix_QA_EvaluationMeasures} can be used to understand the typical concerns other researchers in the field employ. Additionally, Table~\ref{table:Datasets} provides insights into frequently used datasets across domains and applications. Researchers can then utilize off-the-shelf models from various libraries, such as TensorFlow and scikit-learn, to build their own solutions (pipelines). These solutions can be evaluated using desired datasets to assess whether they meet all the specified concerns. This phase of the decision model differs from the previous four phases, as it requires significant ad-hoc efforts in developing, training, and evaluating the models.

By employing this decision-making process, research modelers can develop an intent modeling approach that accurately captures and understands users' intentions in the conversational recommender system. This enables personalized and precise responses, enhancing the overall user experience and satisfaction.

\section{Evaluation of Findings: Case Studies}\label{CaseStudies}
In this section, we present an evaluation of the proposed decision model (refer to Section~\ref{Decisionmakingprocess}) through two scientific case studies conducted by eight researchers from the University of California San Diego in the United States and the University of Klagenfurt in Austria. 
The primary objective of the case studies was to understand the applicability of the decision model to the participants' projects and gain insights into their decision-making processes. The participants emphasized their specific feature requirements throughout the case studies, which we diligently documented in Table~\ref{table:ModelsToFeatures}. Drawing from this information, we identified feasible models based on the comprehensive data presented in Table~\ref{table:ModelCategories} and Table~\ref{table:ModelsToFeatures}. We further explored the viable combinations of these models, as outlined in Figure~\ref{fig:ModelCombinations}. To assess the attention and recognition received by the selected models in the academic literature, we conducted a thorough analysis, referring to Table~\ref{table:ModelTrends}. This analysis provided valuable insights into the popularity and relevance of the models over time among researchers. Finally, the prominent and trending feasible combinations were shared with the case study participants. Figure~\ref{fig:meta-model} offers an overview of the typical decision-making process employed by researchers when selecting intent modeling models. 

In Table~\ref{table:CaseStudy}, we have provided a comprehensive overview of the case studies conducted in this research. The table includes details about the specific contexts of each case study, the feature requirements identified by the case study participants, the design decisions (model selection) made by the researchers based on those requirements, and the outcomes of our decision model for each case study. Subsequent sections of this paper provide an in-depth exploration of the case studies, covering the addressed concerns, the outcomes obtained through utilizing the decision model, and the implications derived from our rigorous analysis. 

\begin{table}[!ht]
\scriptsize

\caption{ provides an overview of the feature requirements considered by the case study participants\cite{MANZOOR-CaseStudy,Mehrab-caseStudy} during their decision-making process for developing their conversational recommender systems. The selected feature requirements were instrumental in guiding the participants' model selection. We employed the decision model based on the defined feature requirements to identify feasible combinations of models. The results of the decision model are also presented in this table, showcasing how it aligns with the participants' choices, validating the effectiveness of the decision-making process for developing innovative and effective conversational recommender systems.}
\centering
\includegraphics[trim=0 0 0 0  ,clip,width=0.85\textwidth]{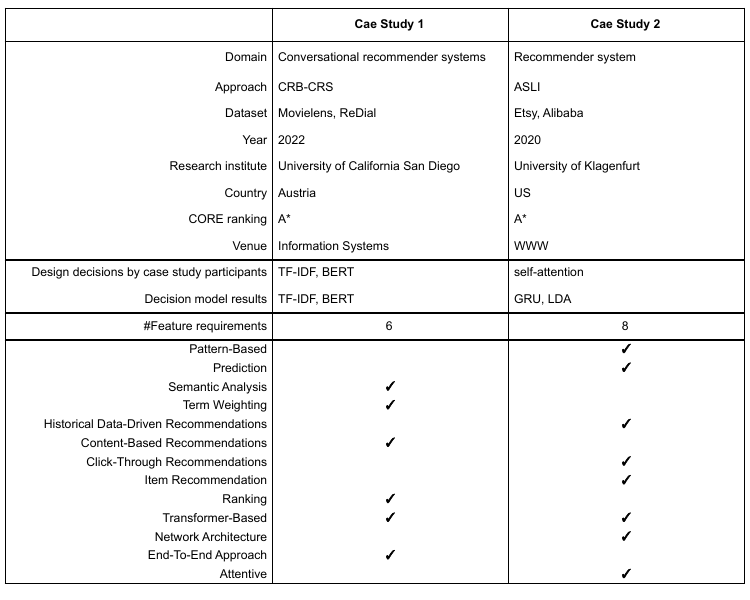}
\label{table:CaseStudy}
\end{table}

\subsection{Case Study Method}
Case study research is an empirical research method~\cite{jansen2009applied} that investigates a phenomenon within a particular context in the domain of interest~\cite{yin2017case}. Case studies can be employed to describe, explain, and evaluate a hypothesis. They involve collecting data regarding a specific phenomenon and applying a tool to evaluate its efficiency and effectiveness, often through interviews. In our study, we followed the guidelines outlined by Yin~\cite{yin1981case} to conduct and plan the case studies.

\noindent\textbf{Objective:} The main aim of this research was to conduct case studies to evaluate the effectiveness of the decision model and its applicability in the academic setting for supporting research modelers in selecting appropriate models for their intent modeling approaches.

\noindent\textbf{The cases:} We conducted two case studies within the academic domain to assess the practicality and usefulness of the proposed decision model. The case studies aimed to evaluate the decision model's effectiveness in assisting research modelers and researchers in selecting models for their intent modeling tasks.

\noindent\textbf{Methods:} For the case studies, we engaged with research modelers and researchers actively involved in intent modeling approaches. We collected data through expert interviews and discussions to gain a comprehensive understanding of their specific requirements, preferences, and challenges when selecting models. The case study participants provided valuable insights into the decision-making process and offered feedback on the suitability of the decision model for their intent modeling needs. 

\noindent\textbf{Selection strategy:} In line with our research objective, we employed a multiple case study approach~\cite{yin1981case} to capture a diverse range of perspectives and scenarios within the academic domain. This selection strategy aimed to ensure the credibility and reliability of our findings. We deliberately selected two publications from highly regarded communities with an A* CORE rank. We verified the expertise of the authors, who actively engage in selecting and implementing intent modeling models. Their knowledge and experience allowed us to consider various factors in different application contexts, including quality attributes, evaluation measures, and feature requirements.

By conducting these case studies, our research aimed to validate the practicality of the decision model and demonstrate its value in supporting research modelers and researchers in their intent modeling endeavors. The insights gained from the case studies provided valuable feedback for refining the decision model and contributed to advancing the intent modeling field within the academic community.

\subsection{Case Study 1:}
The first case study presented in our paper revolves around a research project conducted at the University of Klagenfurt in Austria. The study focused on investigating a retrieval-based approach for conversational recommender systems (CRS)~\cite{MANZOOR-CaseStudy}. The primary objective of the researchers was to assess the effectiveness of this approach as an alternative or complement to language generation methods in CRS. They conducted user studies and carefully analyzed the results to understand the potential benefits of retrieval-based approaches in enhancing user intent modeling for conversational recommender systems.

Throughout the project, the case study participants made two important design decisions (models), TF-IDF and BERT, to develop the CRS. They evaluated their approach using Movielens and ReDial datasets to measure its performance.

By applying the decision model presented in our paper (in Section~\ref{DecisionModel}), the case study participants identified six essential features that were crucial in guiding their decision-making process for selecting the most suitable models and datasets. These features provided valuable insights into designing and implementing an effective retrieval-based approach for conversational recommender systems, contributing to improving user intent modeling in this context.

\subsubsection{Feature requirements:}
In this section, we outline the feature requirements that the case study participants considered during their decision-making process for the research project. Each feature requirement was carefully chosen based on its relevance and potential to enhance the retrieval-based approach for CRS. Below are the feature requirements and their rationale for selection:

\noindent\textbf{Semantic Analysis:} The case study participants recognized the importance of analyzing the meaning and context of words and phrases in natural language data. Semantic analysis helps the model understand user intents more accurately, leading to more relevant and contextually appropriate recommendations.

\noindent\textbf{Term Weighting:} Assigning numerical weights to terms or words in a document or dataset helps the machine learning model comprehend the significance of different terms in the data. The participants adopted term weighting to improve the model's ability to identify relevant features and make better recommendations.

\noindent\textbf{Content-Based Recommendations:} This feature involves utilizing item characteristics or features to recommend similar items to users. The participants valued this approach, allowing the system to tailor recommendations based on users' past interactions and preferences.

\noindent\textbf{Ranking:} The case study participants sought a model capable of ranking items or entities based on their relevance to specific queries or users. By incorporating ranking, the system ensures that the most relevant recommendations appear at the top, enhancing user satisfaction.

\noindent\textbf{Transformer-Based:} Transformer-based models, such as neural networks, excel at learning contextual relationships in sequential data like natural language. The participants chose this approach to effectively leverage the model's ability to understand and process conversational context.

\noindent\textbf{End-To-End Approach:} The case study participants preferred an end-to-end modeling strategy, where a single model directly learns complex tasks from raw data inputs to desired outputs. By avoiding intermediate stages and hand-crafted features, the participants aimed to simplify the model and improve its performance in CRS tasks.

\subsubsection{Results and analysis:}
During the expert interview session with the case study participants, we systematically followed the decision model presented in Section~\ref{DecisionModel} to identify appropriate combinations of models that align with the defined feature requirements for their conversational recommender systems. In the initial steps (Steps 1 and 2), we collaboratively established the essential feature requirements for their CRS, carefully considering the critical aspects that would enhance their system's performance. Subsequently, we referred to Table~\ref{table:ModelsToFeatures} (Steps 3 and 4) to evaluate which models could fulfill these specific feature requirements.

Upon analyzing the table, both the case study participants and we discovered that BERT offered support for Semantic Analysis, Content-Based Recommendations, Ranking, Transformer-Based, and End-To-End Approaches. Additionally, TF-IDF was found to be supportive of Term Weighting, Content-Based Recommendations, and Ranking. This insightful information made us realize that combining these two models would adequately address all six feature requirements for their CRS. Consequently, the case study participants confirmed that combining BERT and TF-IDF would be a suitable choice to fulfill their CRS needs. This combination was validated as a compatible and valid option, consistent with the guidance provided by the decision model.

The data presented in Table~\ref{table:ModelTrends} further reinforces the popularity and relevance of BERT and TF-IDF as widely used models for conversational recommender systems. The case study participants were well-aware of these trends and acknowledged that their model choices aligned with prevailing practices. This alignment provides additional validation to their model selections, demonstrating their dedication to adopting the latest technologies in their research project to create an effective CRS.

Furthermore, Table~\ref{table:Datasets} provided valuable insights into the popularity and significance of various datasets, including Movielens and ReDial. These datasets have been cited and utilized in over 50 publications, underscoring their recognition within the research community. The case study participants acknowledged the widespread use of these datasets by other researchers, reflecting an interesting trend in dataset selection. This awareness further highlights their commitment to utilizing well-established and reputable datasets in their research, contributing to the credibility and reliability of their study findings.

\subsection{Case Study 2:}
The second case study presented in our paper focuses on a research project conducted at the University of California San Diego in the United States~\cite{Mehrab-caseStudy}. The study introduces the Attentive Sequential model of Latent Intent (ASLI) to enhance recommender systems by capturing users' hidden intents from their interactions.

Understanding user intent is essential for delivering relevant recommendations in conventional recommender systems. However, user intents are often latent, meaning they are not directly observable from their interactions. ASLI addresses this challenge by uncovering and leveraging these latent user intents.

Using a self-attention layer, the researchers (case study participants) designed a model that initially learns item similarities based on users' interaction histories. They incorporated a Temporal Convolutional Network (TCN) layer to derive latent representations of user intent from their actions within specific categories. ASLI employs an attentive model guided by the latent intent representation to predict the next item for users. This enables ASLI to capture the dynamic behavior and preferences of users, resulting in state-of-the-art performance on two major e-commerce datasets from Etsy and Alibaba.

By utilizing the decision model presented in our paper (in Section~\ref{DecisionModel}), the case study participants identified eight essential features crucial in guiding their decision-making process for selecting the most suitable models and datasets. 

\subsubsection{Feature requirements:}
In this section, we present the feature requirements that were crucial considerations for the case study participants during their decision-making process for the research project. The following are the feature requirements and the reasons behind their selection:

\noindent\textbf{Pattern-Based:} In the case study, the researchers aimed to improve conversational recommender systems by capturing users' hidden intents from their interactions. By identifying user interactions and behavior patterns, the ASLI model can make informed guesses about users' intents and preferences, leading to more accurate and relevant recommendations.

\noindent\textbf{Prediction:} The ASLI model predicts the next item for users based on their latent intents derived from their historical interactions within specific categories. The model can deliver personalized and effective recommendations by predicting users' preferences and future actions.

\noindent\textbf{Historical Data-Driven Recommendations:} The researchers used previously collected data from users' interactions to train the ASLI model. By analyzing historical data, the model can identify patterns, relationships, and trends in users' behaviors, which inform its predictions and recommendations for future interactions.

\noindent\textbf{Click-Through Recommendations:} In the case study, the ASLI model considers users' clicks on items to understand their preferences and improve the relevance and ranking of future recommendations. The model can adapt and refine its recommendations by utilizing click-through data to meet users' needs better.

\noindent\textbf{Item Recommendation:} The ASLI model suggests items to users based on their previous interactions, enabling it to offer personalized recommendations tailored to individual users' preferences and behaviors.

\noindent\textbf{Transformer-Based:} ASLI is a neural network model based on the Transformer architecture. Transformers are well-suited for learning context and meaning from sequential data, making them suitable for capturing the dynamic behavior and preferences of users in conversational recommender systems.

\noindent\textbf{Network Architecture:} The ASLI model's network architecture is crucial in guiding information flow through the model's layers. By designing an effective network architecture, the researchers ensure that the model can capture and leverage users' latent intents to make accurate recommendations.

\noindent\textbf{Attentive:} ASLI utilizes attention mechanisms to focus on the most relevant parts of users' interactions and behaviors. The model can better understand users' intents and preferences by paying attention to critical information, leading to more attentive and accurate recommendations.

\subsubsection{Results and analysis:}
During the expert interview session with the case study participants, we used the decision model (outlined in Section~\ref{DecisionModel}) to identify suitable combinations of models that align with the defined feature requirements for their conversational recommender systems. In Steps 1 and 2, we collaboratively established the essential feature requirements for the ASLI, carefully considering critical aspects to enhance system performance. Then, in Steps 3 and 4, we referred to Table~\ref{table:ModelsToFeatures} to evaluate models that could fulfill these specific feature requirements.

According to the table,both the case study participants and ourselves found that the GRU model supports Prediction, Historical Data-Driven Recommendations, Click-Through Recommendations, Network Architecture, and Attentive features. Additionally, the LDA model supports Pattern-Based and Item Recommendation features. We also discovered that BERT is the only model in our list supporting Transformer-Based features, and the case study participants agreed with this combination, considering these models as the baseline of their approach. However, after performance analysis, they found that GRU's performance was unsatisfactory in their setting. Consequently, they chose to develop their own model from scratch, modifying the self-attentive model. It's worth noting that the Self-attentive model only supports Network Architecture and Attentive features, making it a suitable baseline in combination with other models for their solutions. The case study participants mentioned considering LDA and BERT as potential models for their upcoming research project due to their similar requirements, although they were not previously aware of this combination. As per Step 5 of the decision model, researchers should address any remaining concerns about quality attributes and evaluation measures after identifying feasible combinations. Thus, the decision model provided valid models in this case study, but in real-world scenarios, model combinations may be modified based on other researchers' concerns, such as quality attributes and evaluation measures.

The case study participants emphasized the value of the data presented in Table~\ref{table:ModelTrends} and their intention to incorporate it into their future design decisions. Understanding trends in model usage is crucial to identify models that may perform well in conversational recommender systems, taking into account similar concerns and requirements from other researchers.

Furthermore, Table~\ref{table:Datasets} indicates that Etsy and Alibaba datasets are not widely known in the context of user intent modeling, although the case study participants clarified that these datasets are well-known in e-commerce services, aligning with their project's specific domain of focus. Nonetheless, they expressed their intention to utilize the data presented in this table to explore potential datasets for evaluating their approach and comparing their work against other approaches in the literature.
\section{Discussion}\label{Discussion}

\subsection{SLR outcomes}
In our comprehensive review of 791 publications, only 68 of them (8.59\%) explicitly mentioned sharing their code repositories, such as GitHub. This finding highlights that a significant number of researchers do not openly share their code, which can create challenges in replicating experiments and hinder the progress of scientific research. Openly providing access to code is essential to promote transparency and ensure reproducibility in machine learning research~\cite{haefliger2008code}.

Throughout the systematic literature review, we collected 600 models, out of which 352 were singletons, representing 58.66\% of the total models. This observation indicates that many researchers develop and use unique models tailored to their research questions. However, relying heavily on singletons can restrict the generalizability of research outcomes and impede meaningful comparisons between different approaches. Encouraging the adoption of common models or establishing standards for model evaluation could significantly enhance the reproducibility and comparability of machine learning research~\cite{amershi2019software}.

In some instances, the methodology for combining models was not clearly described in the publications. This lack of transparency challenges understanding the underlying techniques used and evaluating their effectiveness. Explicitly providing descriptions of model combination techniques and the reasons behind their selection is crucial to increase transparency and facilitate the replication and extension of research findings~\cite{kuwajima2020engineering}.

Our analysis revealed a substantial number of variations in the collected models, including BERT4Rec~\cite{chen2022intent}, SBERT~\cite{garcia2021topic}, BERT-NeuQS~\cite{hashemi2020guided}, BioBERT~\cite{carvallo2020automatic}, ELBERT~\cite{gao2022search}, and RoBERTa~\cite{wu2021exploration}, among others. These variations are derived from BERT~\cite{devlin2018bert}, a well-known language model, and we found 35 publications related to this model based on Figure~\ref{fig:ModelCombinations}. Researchers often leverage diverse model variations to address different research questions and tasks. However, the extensive use of multiple variations can make comparisons with other models complex and hinder the replication of experiments. Developing standardized categories and taxonomies for model variations would be beneficial to address this challenge. Such categorization would greatly assist researchers in understanding the differences and similarities between models, thereby promoting the sharing and reuse of models across various research domains. This standardized approach can enhance collaboration and facilitate advancements in machine learning research~\cite{sarker2021machine}.

Figure~\ref{fig:ModelCombinations} displays that LDA is the most prevalent model in the context of user intent modeling approaches (as indicated in Table~\ref{table:ModelTrends}). It is important to recognize that LDA and other traditional models have significantly influenced the field and inspired the development of newer models like BERT~\cite{devlin2018bert}. These traditional models have served as foundational building blocks, offering initial insights into various NLP tasks. While traditional models like LDA~\cite{blei2003latent} have been widely used and contributed to understanding natural language, their adoption may have gradually decreased over time for several reasons. One crucial factor is the advent of more sophisticated and advanced models like BERT. BERT's bidirectional contextual embeddings and transformer architecture have shown exceptional performance on various NLP tasks, setting new benchmarks and gaining substantial attention in the research community~\cite{raffel2020exploring}. Moreover, larger datasets and advancements in computational resources have facilitated the training and fine-tuning of complex models like BERT, making them more practical and feasible for real-world applications. Additionally, the interpretability and ease of use of traditional models like LDA have been balanced by the increased complexity and opaqueness of modern models like BERT. This trade-off between interpretability and performance has influenced researchers and practitioners in selecting the most suitable models for their specific tasks. Furthermore, the diversity of downstream NLP applications has also influenced the preference for modern models~\cite{ribeiro2016should}. While traditional models may perform well in specific tasks, BERT's ability to generalize and excel across a wide range of NLP benchmarks has made it a popular choice for various applications.

Regarding dataset usage, we observed that only 394 out of 791 publications (49.81\%) opted to utilize public and open-access datasets. This finding implies that more than half of the publications relied on proprietary datasets that were specifically generated for individual cases, rendering them inaccessible for reuse by other researchers. Our investigation further revealed the existence of 253 public open-access datasets that authors employed to evaluate and train their approaches. However, it is worth noting that 173 of these datasets (68.37\%) were mentioned in only one publication and were not subsequently reused by other researchers in the domain of user intent modeling. This observation highlights a potential deficiency in dataset-sharing and reuse practices within this research area, which could have significant consequences for the advancement and credibility of scientific endeavors in the field.

The limited availability of previous datasets presents researchers with challenges in reproducing and validating reported results, as access to such datasets is often restricted. Consequently, the ability to objectively compare and benchmark different models becomes hampered, impeding the identification of state-of-the-art techniques and areas for improvement~\cite{pujol2020fair}. Moreover, the absence of diverse and openly accessible datasets may result in biased model development and evaluation, limiting the generalizability of models to real-world scenarios and diverse user populations~\cite{bagdasaryan2019differential}. The consequences of this issue extend further, as the duplication of effort in collecting and preparing new datasets consumes valuable resources and consequently decelerates research progress. To mitigate these challenges, fostering a culture of openness and collaboration within the research community is essential.

\subsection{Case Study Participants}

The case study participants showed a careful and thorough approach to decision-making by conducting extensive research and literature reviews. This method allowed them to carefully select models for their research project carefully, showcasing the effectiveness of the decision model in helping researchers make well-informed and compatible model choices for developing conversational recommender systems.

Both case study participants emphasized the value of using the decision model and the knowledge gained during this study. They expressed their intention to use this information to make informed decisions when selecting the appropriate combinations of models for user intent modeling approaches.

Furthermore, the case study participants recognized that the decision model serves as a valuable tool for generating an initial list of models to develop their approaches. However, they acknowledged that Step 5 of the decision model highlights the importance of further analysis, such as performance testing, to identify the right combinations of models that work well for specific use cases. This recognition underscores the need for practical testing and validation to ensure the chosen model combinations are effective and suitable for their particular research goals.

The use of well-known datasets, such as Movielens and ReDial in the first case study, and Etsy and Alibaba datasets in the second case study, underlines the researchers' commitment to using credible data sources for evaluation. The decision model allowed researchers to consider dataset popularity and relevance, enhancing the credibility and reliability of their study findings.

The decision model provided valuable insights into the trends in model usage, as presented in Table~\ref{table:ModelTrends}. Both case study participants expressed interest in incorporating these trends into their future research decisions, ensuring they stay up-to-date with the latest advancements in intent modeling approaches.

Throughout the case studies, the discussion highlighted the dynamic nature of the decision-making process. While the decision model offered feasible model combinations based on feature requirements, the final choices were influenced by additional factors such as model performance, quality attributes, and evaluation measures. This adaptability showcased the decision model's flexibility in accommodating researchers' unique priorities and preferences.

Both case studies effectively demonstrated that the decision model offers a systematic approach to model selection and helps researchers explore various options and combinations of models. This exploratory nature allowed researchers to consider novel solutions and build upon existing models, creating innovative intent modeling approaches.

The success of the decision model in assisting researchers in their model selection process holds promising implications for the broader academic community. By providing a structured and comprehensive methodology, the decision model can streamline the development of conversational recommender systems with accurate intent modeling capabilities, ultimately enhancing user experience and satisfaction.

\subsection{Threat to Validity}
Validity evaluation is paramount in empirical studies, encompassing systematic literature reviews (SLRs) and case study research~\cite{zhou2016map}. This paper's validity assessment covers various dimensions, including Construct Validity, Internal Validity, External Validity, and Conclusion Validity. Although other types of validity, such as Theoretical Validity and Interpretive Validity, are relevant to intent modeling, they are not explicitly addressed in this context due to their relatively limited exploration.

\noindent\textbf{Construct Validity} pertains to the accuracy of operational measures or tests used to investigate concepts. In this research, we developed a meta-model (refer to Figure~\ref{fig:meta-model}) based on the ISO/IEC/IEEE standard 42010~\cite{iso420102011iec} to represent the decision-making process in intent modeling for conversational recommender systems. We formulated comprehensive research questions by utilizing the meta-model's essential elements, ensuring an exhaustive coverage of pertinent publications on intent modeling approaches.

\noindent\textbf{Internal Validity} concerns verifying cause-effect relationships within the study's scope and ensures the study's robustness. We employed a rigorous quasi-gold standard (QGS)~\cite{zhang2011identifying} to minimize selection bias in paper inclusion. Combining manual and automated search strategies, the QGS provided an accurate evaluation of sensitivity and precision. Our search spanned four major online digital libraries, widely regarded to encompass a substantial portion of high-quality publications relevant to intent modeling for conversational recommender systems. Additionally, we used snowballing to complement our search and mitigate the risk of missing essential publications. The review process involved a team of researchers, including three principal investigators and five research assistants. Furthermore, the findings were validated by real-world researchers in intent modeling to ensure their practicality and effectiveness.

\noindent\textbf{External Validity} pertains to the generalizability of research findings to real-world applications. This study considered publications discussing intent modeling approaches across multiple years. Although some exclusions and inaccessibility of studies may impact the generalizability of SLR and case study results, the proportion of inaccessible studies (less than 2\%) is not expected to affect the overall findings significantly. The knowledge extracted from this research can be applied to support the development of new theories and methods for future intent modeling challenges, benefiting both academia and practitioners in this field.

\noindent\textbf{Conclusion Validity} ensures that the study's methods, including data collection and analysis, can be replicated to yield consistent results. We extracted knowledge from selected publications, encompassing various aspects such as \textit{Models}, \textit{Datasets}, \textit{Evaluation Metrics}, \textit{Quality Attributes}, \textit{Combinations}, and \textit{Trends} in intent modeling approaches. The accuracy of the extracted knowledge was safeguarded through a well-defined protocol governing the knowledge extraction strategy and format. The authors proposed and reviewed the review protocol, establishing a clear and consistent approach to knowledge extraction. A data extraction form was employed to ensure uniform extraction of relevant knowledge, and the acquired knowledge was validated against the research questions. All authors independently determined quality assessment criteria, and crosschecking was conducted among reviewers, with at least three researchers independently extracting data, thus enhancing the reliability of the results.

\section{Related work}\label{RelatedStudies}
This section contextualizes our study within the broader landscape of systematic literature reviews (SLRs) focused on intent modeling approaches.

\begin{table}[!ht]
\scriptsize

\caption{situates our study within the existing body of literature on user intent modeling, as identified through our SLR.}
\resizebox{\textwidth}{!}{
\begin{tabular}{|c|cccrcccccc|rrrr|rrrr|r|}
\hline
Ref. & Year & Type & RM & \#Pub & DM & Tr. & DS & Cat & MC & F & \#QA & \#F & \#E & \#M & \#CQA & \#CF & \#CE & \#CM & Cov. (\%) \\ \hline
Our Study & 2023 & Aca & SLR/CS & 791 & Yes & Yes & Yes & Yes & Yes & Yes & 38 & 74 & 13 & 59 & 38 & 74 & 13 & 59 & 100 \\
\cite{de2020intelligent} & 2020 & Aca & SLR & 58 & No & No & No & Yes & No & No & 7 & 4 & 8 & 6 & 6 & 3 & 5 & 2 & 64.00 \\
\cite{rapp2021human} & 2021 & Aca & SLR & 83 & No & No & No & Yes & No & No & 8 & 4 & 5 & 9 & 6 & 5 & 3 & 2 & 61.54 \\
\cite{pan2022test} & 2022 & Aca & SLR & 29 & No & No & Yes & Yes & No & No & 7 & 6 & 7 & 14 & 4 & 3 & 5 & 7 & 55.88 \\
\cite{keyvan2022approach} & 2022 & Aca & Survey & N/A & No & No & Yes & No & No & No & 8 & 12 & 14 & 15 & 5 & 9 & 10 & 7 & 63.27 \\
\cite{zaib2022conversational} & 2022 & Aca & Survey & 88 & No & Yes & Yes & No & No & No & 5 & 8 & 9 & 18 & 4 & 7 & 5 & 5 & 52.50 \\
\cite{iovine2023virtual} & 2023 & Aca & Survey & 116 & No & No & No & No & No & No & 6 & 7 & 5 & 11 & 6 & 7 & 2 & 8 & 79.31 \\
\cite{saka2023conversational} & 2023 & Aca & Survey & 21 & No & No & No & No & No & No & 4 & 7 & 9 & 11 & 4 & 4 & 9 & 5 & 70.97 \\
\cite{pu2012evaluating} & 2012 & Aca & Survey & N/A & Yes & Yes & No & No & No & No & 8 & 7 & 5 & 11 & 8 & 7 & 5 & 11 & 100 \\
\cite{liu2022review} & 2022 & Aca & Survey & N/A & No & No & No & Yes & No & No & 2 & 7 & 3 & 8 & 2 & 7 & 1 & 4 & 70.00 \\
\cite{tamine2010evaluation} & 2010 & Aca & Survey & N/A & No & No & No & No & No & No & 5 & 4 & 10 & 8 & 3 & 4 & 4 & 4 & 55.56 \\
\cite{chen2015recommender} & 2015 & Aca & Survey & N/A & No & Yes & No & No & No & No & 7 & 8 & 10 & 19 & 5 & 5 & 9 & 9 & 63.64 \\
\cite{latifi2021session} & 2021 & Aca & Review & N/A & No & No & Yes & No & No & No & 3 & 5 & 10 & 14 & 2 & 3 & 7 & 8 & 62.50 \\
\cite{zhang2019deep} & 2019 & Aca & Survey & N/A & No & Yes & No & Yes & No & No & 6 & 8 & 6 & 15 & 6 & 7 & 3 & 8 & 68.57 \\
\cite{jiang2013mining} & 2013 & Gry & Survey & N/A & No & No & No & No & No & No & 6 & 7 & 7 & 13 & 5 & 5 & 5 & 10 & 75.76 \\
\cite{hernandez2019comparative} & 2019 & Gry & Review & N/A & No & No & Yes & Yes & Yes & No & 4 & 11 & 6 & 12 & 4 & 11 & 3 & 9 & 81.82 \\
\cite{jindal2014review} & 2014 & Aca & Review & 15 & No & No & Yes & Yes & No & No & 5 & 4 & 2 & 3 & 5 & 4 & 1 & 2 & 85.71 \\
\cite{yuan2020expert} & 2020 & Gry & Review & N/A & No & No & Yes & Yes & Yes & No & 3 & 8 & 1 & 15 & 2 & 5 & 1 & 8 & 59.26 \\
\hline
\end{tabular}
\label{tbl:SLRposition}
}
\end{table}

Table~\ref{tbl:SLRposition} provides a comprehensive overview of our study's position within the existing body of literature on user intent modeling, identified through our systematic literature review. Our review encompassed a substantial number of publications, totaling 791, making our investigation one of the most extensive in this domain. The table comprises various columns, each serving distinct purposes. Through the SLR process, we curated both academic literature (Aca) and gray literature (Gry) reviews that contributed to a well-rounded understanding of user intent modeling.

Academic literature reviews (Aca) were prevalent among the selected studies, accounting for over 80 percent of the reviewed literature. This choice aligns with our approach, which primarily focuses on academic sources. The research methods (RM) employed in the selected studies include SLR, Case Study (CS), Survey (Surv.), and Review (Rev.). Notably, none of the reviewed SLRs utilized case studies (CS) to evaluate their findings; instead, they solely reported on the outcomes of the SLR process. In contrast, our study took a more comprehensive approach by incorporating case studies within the research methods (RM), enabling a holistic perspective on decision-making in user intent modeling.

In comparison to the reviewed SLRs, our study stands out for its emphasis on decision-making processes and decision models (DM). While only one paper~\cite{pu2012evaluating} among the reviewed SLRs reported on this aspect, our study introduced a decision model based on the evidence extracted from the literature. This decision model serves as a valuable tool for research modelers, guiding informed decisions and identifying suitable individual models or combinations that address specific concerns.

Regarding observed trends (Tr.) within the models, four studies~\cite{zaib2022conversational, pu2012evaluating, chen2015recommender, zhang2019deep} (23.52\%) reported on this aspect. Additionally, seven studies~\cite{latifi2021session, yuan2020expert, jindal2014review, hernandez2019comparative, zaib2022conversational, keyvan2022approach, pan2022test} (41.17\%) provided valuable insights into open-access datasets (DS) suitable for training or evaluating the models, serving as valuable resources for the research community.

Furthermore, our study categorized (Cat) the models, in line with eight other SLRs~\cite{de2020intelligent,rapp2021human,pan2022test,liu2022review,zhang2019deep,hernandez2019comparative,jindal2014review,yuan2020expert} (47.05\%) in the field. However, we noted that only two publications~\cite{hernandez2019comparative,yuan2020expert} (11.76\%) reported on combinations of models (MC), making it challenging to ascertain which models are feasible to combine effectively.

The table underscores the rigorous analysis conducted, encompassing a significant number of models (\#M), evaluation measures (\#E), quality attributes (\#QA), and features (\#F), compared to other studies. Moreover, we identified common concepts among our results and the selected publications, presented in the last four columns, \#CQA, \#CF, \#CE, and \#CM, along with the percentage of coverage (Cov.). Additionally, the last five columns of the table indicate that our study covers almost 70 percent of the models, quality attributes, evaluation measures, and features reported in other SLRs, showcasing the relevance of our research to the broader literature on user intent modeling.

To maintain conciseness, we have focused on concepts mentioned in more than five publications in our report. For a comprehensive understanding and access to the complete set of data and references, we encourage readers to explore our repository on \textit{Mendeley Data}~\cite{Farshidi_Rezaee_2023}. Furthermore, our study's inclusion of both academic literature and gray literature reviews contributes to a comprehensive understanding of user intent modeling, incorporating insights from diverse sources.

The combination of SLR and case study methods offers a robust research design, allowing us to explore existing literature while also delving deeper into specific real-world scenarios. By examining decision-making processes and introducing a decision model, our study addresses a crucial aspect often overlooked in the reviewed SLRs, providing valuable guidance to researchers and practitioners. Moreover, our analysis reveals emerging trends within the models and the availability of open-access datasets, enhancing the visibility of valuable resources for the research community. Categorizing the models facilitates a structured taxonomy, aiding researchers in navigating the diverse landscape of user intent modeling approaches.
While only a limited number of publications explored combinations of models, our study highlights this as a potential avenue for further investigation. By shedding light on the relationships between models, our findings can inform the development of more robust and effective ensemble approaches.
The extensive coverage of models, evaluation measures, quality attributes, and features in our analysis offers a comprehensive view of user intent modeling, providing valuable insights for researchers seeking to refine their models and evaluation strategies.

\section{Conclusion and future work}\label{Conclusion-FutureWork}
In this paper, we have undertaken a comprehensive investigation of the decision-making process involved in intent modeling for conversational recommender systems. Our main objective was to address the challenge faced by research modelers in selecting the most effective combination of models for developing intent modeling approaches.

To ensure the credibility and reliability of our findings, we conducted a systematic literature review and carried out two academic case studies, meticulously examining various dimensions of validity, including Construct Validity, Internal Validity, External Validity, and Conclusion Validity.

Drawing inspiration from the ISO/IEC/IEEE standard 42010~\cite{iso420102011iec}, we devised a meta-model as the foundational framework for representing the decision-making process in intent modeling. By formulating comprehensive research questions, we ensured the inclusion of relevant studies and achieved an exhaustive coverage of pertinent publications.

Our study offers a holistic understanding of user intent modeling within the context of conversational recommender systems. The SLR analyzed over 13,000 papers from the last decade, identifying 59 distinct models and 74 commonly used features. These analyses provide valuable insights into the design and implementation of user intent modeling approaches, contributing significantly to the advancement of the field.

Building on the findings from the SLR, we proposed a decision model to guide researchers and practitioners in selecting the most suitable models for developing conversational recommender systems. The decision model takes into account essential factors such as model characteristics, evaluation measures, and dataset requirements, facilitating informed decision-making and enhancing the development of more effective and efficient intent modeling approaches.

We demonstrated the practical applicability of the decision model through two case studies, showcasing its usefulness in real-world scenarios. The decision model aids researchers in identifying initial model sets and considering essential quality attributes and functional requirements, streamlining the process and enhancing its reliability.

The significance of contributions in User Intent Modeling cannot be overstated in the current landscape of scientific research. Whether actively engaged in advancing the fundamentals or exploring its applications within their respective domains, scientists are undeniably conscious of this field. Amidst this crucial juncture, our study holds paramount importance as it contributes to the consolidation of the field's foundations. We envision our research to become an integral component of essential literature for newcomers, fostering the promotion of this vital field and streamlining researchers' efforts in selecting suitable models and techniques. By solidifying the understanding and relevance of User Intent Modeling, we aim to facilitate future advancements and innovation in this area of study.

To ensure the longevity and up-to-dateness of the knowledge base constructed from our SLR, we are enthusiastic about taking the necessary steps to maintain its relevance and value for future researchers embarking on similar projects. We plan to establish a collaborative platform or repository, inviting researchers to contribute their latest findings and studies pertaining to the addressed research challenges. By fostering a community-driven approach, we aim to create an engaging environment that encourages regular and meaningful contributions. To streamline the process, we intend to develop user-friendly interfaces and implement effective content moderation to ensure the knowledge base's scientific integrity.

Moreover, we are excited to explore implementing an automated data crawling mechanism, periodically and systematically searching reputable literature sources and academic databases. This technology will enable seamless integration of the latest research into the knowledge base. Additionally, we are committed to maintaining a meticulous record of changes and updates to the knowledge base, including precise timestamps and new information sources. This transparent documentation will empower future researchers to follow the knowledge base's evolution and confidently leverage it for their specific research needs.

By embracing these proactive measures, we envision establishing a continuously updated and robust knowledge base that serves as a valuable resource for researchers in the dynamic domain of user intent modeling and recommender systems.
\section*{Acknowledgement}
We extend our sincere gratitude to the domain experts who actively participated in and contributed to this research project. Their valuable insights and expertise have significantly enriched the quality of this study. We would like to express our appreciation to Sjaak Brinkkemper, Fabiano Dalpiaz, Gerard Wagenaar, Fernando Castor de Lima Filho, and Sergio Espana Cubillo for their invaluable feedback, which has helped us in presenting the results of this study more effectively.

We are also deeply thankful to all the participants of the case studies for their cooperation and willingness to share their valuable publications, which served as essential resources in evaluating and validating the proposed decision model. Their contributions have been pivotal in ensuring the practical applicability and effectiveness of the decision model in real-world scenarios.

Finally, we extend our appreciation to the journal editors and reviewers for their meticulous review of this manuscript and their constructive feedback. Their efforts have played a crucial role in enhancing the quality and clarity of this research, making it a more valuable contribution to the scientific community.
\newpage
\bibliography{References}


\begin{thebibliography}{166}
\ifx \bisbn   \undefined \def \bisbn  #1{ISBN #1}\fi
\ifx \binits  \undefined \def \binits#1{#1}\fi
\ifx \bauthor  \undefined \def \bauthor#1{#1}\fi
\ifx \batitle  \undefined \def \batitle#1{#1}\fi
\ifx \bjtitle  \undefined \def \bjtitle#1{#1}\fi
\ifx \bvolume  \undefined \def \bvolume#1{\textbf{#1}}\fi
\ifx \byear  \undefined \def \byear#1{#1}\fi
\ifx \bissue  \undefined \def \bissue#1{#1}\fi
\ifx \bfpage  \undefined \def \bfpage#1{#1}\fi
\ifx \blpage  \undefined \def \blpage #1{#1}\fi
\ifx \burl  \undefined \def \burl#1{\textsf{#1}}\fi
\ifx \doiurl  \undefined \def \doiurl#1{\url{https://doi.org/#1}}\fi
\ifx \betal  \undefined \def \betal{\textit{et al.}}\fi
\ifx \binstitute  \undefined \def \binstitute#1{#1}\fi
\ifx \binstitutionaled  \undefined \def \binstitutionaled#1{#1}\fi
\ifx \bctitle  \undefined \def \bctitle#1{#1}\fi
\ifx \beditor  \undefined \def \beditor#1{#1}\fi
\ifx \bpublisher  \undefined \def \bpublisher#1{#1}\fi
\ifx \bbtitle  \undefined \def \bbtitle#1{#1}\fi
\ifx \bedition  \undefined \def \bedition#1{#1}\fi
\ifx \bseriesno  \undefined \def \bseriesno#1{#1}\fi
\ifx \blocation  \undefined \def \blocation#1{#1}\fi
\ifx \bsertitle  \undefined \def \bsertitle#1{#1}\fi
\ifx \bsnm \undefined \def \bsnm#1{#1}\fi
\ifx \bsuffix \undefined \def \bsuffix#1{#1}\fi
\ifx \bparticle \undefined \def \bparticle#1{#1}\fi
\ifx \barticle \undefined \def \barticle#1{#1}\fi
\bibcommenthead
\ifx \bconfdate \undefined \def \bconfdate #1{#1}\fi
\ifx \botherref \undefined \def \botherref #1{#1}\fi
\ifx \url \undefined \def \url#1{\textsf{#1}}\fi
\ifx \bchapter \undefined \def \bchapter#1{#1}\fi
\ifx \bbook \undefined \def \bbook#1{#1}\fi
\ifx \bcomment \undefined \def \bcomment#1{#1}\fi
\ifx \oauthor \undefined \def \oauthor#1{#1}\fi
\ifx \citeauthoryear \undefined \def \citeauthoryear#1{#1}\fi
\ifx \endbibitem  \undefined \def \endbibitem {}\fi
\ifx \bconflocation  \undefined \def \bconflocation#1{#1}\fi
\ifx \arxivurl  \undefined \def \arxivurl#1{\textsf{#1}}\fi
\csname PreBibitemsHook\endcsname

\bibitem{carmel2020future}
\begin{botherref}
\oauthor{\bsnm{Carmel}, \binits{D.}},
\oauthor{\bsnm{Chang}, \binits{Y.}},
\oauthor{\bsnm{Deng}, \binits{H.}},
\oauthor{\bsnm{Nie}, \binits{J.-Y.}}:
Future directions of query understanding.
Query Understanding for Search Engines,
205--224
(2020)
\end{botherref}
\endbibitem

\bibitem{khilji2023multimodal}
\begin{barticle}
\bauthor{\bsnm{Khilji}, \binits{A.F.U.R.}},
\bauthor{\bsnm{Sinha}, \binits{U.}},
\bauthor{\bsnm{Singh}, \binits{P.}},
\bauthor{\bsnm{Ali}, \binits{A.}},
\bauthor{\bsnm{Dadure}, \binits{P.}},
\bauthor{\bsnm{Manna}, \binits{R.}},
\bauthor{\bsnm{Pakray}, \binits{P.}}:
\batitle{Multimodal recipe recommendation system using deep learning and
  rule-based approach}.
\bjtitle{SN Computer Science}
\bvolume{4}(\bissue{4}),
\bfpage{421}
(\byear{2023})
\end{barticle}
\endbibitem

\bibitem{ge2018personalizing}
\begin{bchapter}
\bauthor{\bsnm{Ge}, \binits{S.}},
\bauthor{\bsnm{Dou}, \binits{Z.}},
\bauthor{\bsnm{Jiang}, \binits{Z.}},
\bauthor{\bsnm{Nie}, \binits{J.-Y.}},
\bauthor{\bsnm{Wen}, \binits{J.-R.}}:
\bctitle{Personalizing search results using hierarchical rnn with query-aware
  attention}.
In: \bbtitle{Proceedings of the 27th ACM International Conference on
  Information and Knowledge Management},
pp. \bfpage{347}--\blpage{356}
(\byear{2018})
\end{bchapter}
\endbibitem

\bibitem{zhang2019deep}
\begin{barticle}
\bauthor{\bsnm{Zhang}, \binits{S.}},
\bauthor{\bsnm{Yao}, \binits{L.}},
\bauthor{\bsnm{Sun}, \binits{A.}},
\bauthor{\bsnm{Tay}, \binits{Y.}}:
\batitle{Deep learning based recommender system: A survey and new
  perspectives}.
\bjtitle{ACM computing surveys (CSUR)}
\bvolume{52}(\bissue{1}),
\bfpage{1}--\blpage{38}
(\byear{2019})
\end{barticle}
\endbibitem

\bibitem{oulasvirta2008motivations}
\begin{barticle}
\bauthor{\bsnm{Oulasvirta}, \binits{A.}},
\bauthor{\bsnm{Blom}, \binits{J.}}:
\batitle{Motivations in personalisation behaviour}.
\bjtitle{Interacting with computers}
\bvolume{20}(\bissue{1}),
\bfpage{1}--\blpage{16}
(\byear{2008})
\end{barticle}
\endbibitem

\bibitem{konishi2016extracting}
\begin{bchapter}
\bauthor{\bsnm{Konishi}, \binits{T.}},
\bauthor{\bsnm{Ohwa}, \binits{T.}},
\bauthor{\bsnm{Fujita}, \binits{S.}},
\bauthor{\bsnm{Ikeda}, \binits{K.}},
\bauthor{\bsnm{Hayashi}, \binits{K.}}:
\bctitle{Extracting search query patterns via the pairwise coupled topic
  model}.
In: \bbtitle{Proceedings of the Ninth ACM International Conference on Web
  Search and Data Mining},
pp. \bfpage{655}--\blpage{664}
(\byear{2016})
\end{bchapter}
\endbibitem

\bibitem{bendersky2017learning}
\begin{bchapter}
\bauthor{\bsnm{Bendersky}, \binits{M.}},
\bauthor{\bsnm{Wang}, \binits{X.}},
\bauthor{\bsnm{Metzler}, \binits{D.}},
\bauthor{\bsnm{Najork}, \binits{M.}}:
\bctitle{Learning from user interactions in personal search via attribute
  parameterization}.
In: \bbtitle{Proceedings of the Tenth ACM International Conference on Web
  Search and Data Mining},
pp. \bfpage{791}--\blpage{799}
(\byear{2017})
\end{bchapter}
\endbibitem

\bibitem{cao2023comprehensive}
\begin{botherref}
\oauthor{\bsnm{Cao}, \binits{Y.}},
\oauthor{\bsnm{Li}, \binits{S.}},
\oauthor{\bsnm{Liu}, \binits{Y.}},
\oauthor{\bsnm{Yan}, \binits{Z.}},
\oauthor{\bsnm{Dai}, \binits{Y.}},
\oauthor{\bsnm{Yu}, \binits{P.S.}},
\oauthor{\bsnm{Sun}, \binits{L.}}:
A comprehensive survey of ai-generated content (aigc): A history of generative
  ai from gan to chatgpt.
arXiv preprint arXiv:2303.04226
(2023)
\end{botherref}
\endbibitem

\bibitem{tanjim2020attentive}
\begin{bchapter}
\bauthor{\bsnm{Tanjim}, \binits{M.M.}},
\bauthor{\bsnm{Su}, \binits{C.}},
\bauthor{\bsnm{Benjamin}, \binits{E.}},
\bauthor{\bsnm{Hu}, \binits{D.}},
\bauthor{\bsnm{Hong}, \binits{L.}},
\bauthor{\bsnm{McAuley}, \binits{J.}}:
\bctitle{Attentive sequential models of latent intent for next item
  recommendation}.
In: \bbtitle{Proceedings of The Web Conference 2020},
pp. \bfpage{2528}--\blpage{2534}
(\byear{2020})
\end{bchapter}
\endbibitem

\bibitem{wang2020next}
\begin{bchapter}
\bauthor{\bsnm{Wang}, \binits{J.}},
\bauthor{\bsnm{Ding}, \binits{K.}},
\bauthor{\bsnm{Hong}, \binits{L.}},
\bauthor{\bsnm{Liu}, \binits{H.}},
\bauthor{\bsnm{Caverlee}, \binits{J.}}:
\bctitle{Next-item recommendation with sequential hypergraphs}.
In: \bbtitle{Proceedings of the 43rd International ACM SIGIR Conference on
  Research and Development in Information Retrieval},
pp. \bfpage{1101}--\blpage{1110}
(\byear{2020})
\end{bchapter}
\endbibitem

\bibitem{guo2020edgedipn}
\begin{barticle}
\bauthor{\bsnm{Guo}, \binits{L.}},
\bauthor{\bsnm{Hua}, \binits{L.}},
\bauthor{\bsnm{Jia}, \binits{R.}},
\bauthor{\bsnm{Fang}, \binits{F.}},
\bauthor{\bsnm{Zhao}, \binits{B.}},
\bauthor{\bsnm{Cui}, \binits{B.}}:
\batitle{Edgedipn: a unified deep intent prediction network deployed at the
  edge}.
\bjtitle{Proceedings of the VLDB Endowment}
\bvolume{14}(\bissue{3}),
\bfpage{320}--\blpage{328}
(\byear{2020})
\end{barticle}
\endbibitem

\bibitem{paul2021fake}
\begin{barticle}
\bauthor{\bsnm{Paul}, \binits{H.}},
\bauthor{\bsnm{Nikolaev}, \binits{A.}}:
\batitle{Fake review detection on online e-commerce platforms: a systematic
  literature review}.
\bjtitle{Data Mining and Knowledge Discovery}
\bvolume{35}(\bissue{5}),
\bfpage{1830}--\blpage{1881}
(\byear{2021})
\end{barticle}
\endbibitem

\bibitem{zhang2016mining}
\begin{bchapter}
\bauthor{\bsnm{Zhang}, \binits{C.}},
\bauthor{\bsnm{Fan}, \binits{W.}},
\bauthor{\bsnm{Du}, \binits{N.}},
\bauthor{\bsnm{Yu}, \binits{P.S.}}:
\bctitle{Mining user intentions from medical queries: A neural network based
  heterogeneous jointly modeling approach}.
In: \bbtitle{Proceedings of the 25th International Conference on World Wide
  Web},
pp. \bfpage{1373}--\blpage{1384}
(\byear{2016})
\end{bchapter}
\endbibitem

\bibitem{wang2022recognizing}
\begin{bchapter}
\bauthor{\bsnm{Wang}, \binits{Y.}},
\bauthor{\bsnm{Wang}, \binits{S.}},
\bauthor{\bsnm{Li}, \binits{Y.}},
\bauthor{\bsnm{Dou}, \binits{D.}}:
\bctitle{Recognizing medical search query intent by few-shot learning}.
In: \bbtitle{Proceedings of the 45th International ACM SIGIR Conference on
  Research and Development in Information Retrieval},
pp. \bfpage{502}--\blpage{512}
(\byear{2022})
\end{bchapter}
\endbibitem

\bibitem{liu2021intent}
\begin{bchapter}
\bauthor{\bsnm{Liu}, \binits{Z.}},
\bauthor{\bsnm{Chen}, \binits{H.}},
\bauthor{\bsnm{Sun}, \binits{F.}},
\bauthor{\bsnm{Xie}, \binits{X.}},
\bauthor{\bsnm{Gao}, \binits{J.}},
\bauthor{\bsnm{Ding}, \binits{B.}},
\bauthor{\bsnm{Shen}, \binits{Y.}}:
\bctitle{Intent preference decoupling for user representation on online
  recommender system}.
In: \bbtitle{Proceedings of the Twenty-Ninth International Conference on
  International Joint Conferences on Artificial Intelligence},
pp. \bfpage{2575}--\blpage{2582}
(\byear{2021})
\end{bchapter}
\endbibitem

\bibitem{bhaskaran2019efficient}
\begin{barticle}
\bauthor{\bsnm{Bhaskaran}, \binits{S.}},
\bauthor{\bsnm{Santhi}, \binits{B.}}:
\batitle{An efficient personalized trust based hybrid recommendation (tbhr)
  strategy for e-learning system in cloud computing}.
\bjtitle{Cluster Computing}
\bvolume{22},
\bfpage{1137}--\blpage{1149}
(\byear{2019})
\end{barticle}
\endbibitem

\bibitem{ding2015mining}
\begin{bchapter}
\bauthor{\bsnm{Ding}, \binits{X.}},
\bauthor{\bsnm{Liu}, \binits{T.}},
\bauthor{\bsnm{Duan}, \binits{J.}},
\bauthor{\bsnm{Nie}, \binits{J.-Y.}}:
\bctitle{Mining user consumption intention from social media using domain
  adaptive convolutional neural network}.
In: \bbtitle{Proceedings of the AAAI Conference on Artificial Intelligence},
vol. \bseriesno{29}
(\byear{2015})
\end{bchapter}
\endbibitem

\bibitem{wang2019context}
\begin{bchapter}
\bauthor{\bsnm{Wang}, \binits{W.}},
\bauthor{\bsnm{Hosseini}, \binits{S.}},
\bauthor{\bsnm{Awadallah}, \binits{A.H.}},
\bauthor{\bsnm{Bennett}, \binits{P.N.}},
\bauthor{\bsnm{Quirk}, \binits{C.}}:
\bctitle{Context-aware intent identification in email conversations}.
In: \bbtitle{Proceedings of the 42nd International ACM SIGIR Conference on
  Research and Development in Information Retrieval},
pp. \bfpage{585}--\blpage{594}
(\byear{2019})
\end{bchapter}
\endbibitem

\bibitem{penha2020does}
\begin{bchapter}
\bauthor{\bsnm{Penha}, \binits{G.}},
\bauthor{\bsnm{Hauff}, \binits{C.}}:
\bctitle{What does bert know about books, movies and music? probing bert for
  conversational recommendation}.
In: \bbtitle{Proceedings of the 14th ACM Conference on Recommender Systems},
pp. \bfpage{388}--\blpage{397}
(\byear{2020})
\end{bchapter}
\endbibitem

\bibitem{hashemi2018measuring}
\begin{bchapter}
\bauthor{\bsnm{Hashemi}, \binits{S.H.}},
\bauthor{\bsnm{Williams}, \binits{K.}},
\bauthor{\bsnm{El~Kholy}, \binits{A.}},
\bauthor{\bsnm{Zitouni}, \binits{I.}},
\bauthor{\bsnm{Crook}, \binits{P.A.}}:
\bctitle{Measuring user satisfaction on smart speaker intelligent assistants
  using intent sensitive query embeddings}.
In: \bbtitle{Proceedings of the 27th ACM International Conference on
  Information and Knowledge Management},
pp. \bfpage{1183}--\blpage{1192}
(\byear{2018})
\end{bchapter}
\endbibitem

\bibitem{gharibshah2020deep}
\begin{barticle}
\bauthor{\bsnm{Gharibshah}, \binits{Z.}},
\bauthor{\bsnm{Zhu}, \binits{X.}},
\bauthor{\bsnm{Hainline}, \binits{A.}},
\bauthor{\bsnm{Conway}, \binits{M.}}:
\batitle{Deep learning for user interest and response prediction in online
  display advertising}.
\bjtitle{Data Science and Engineering}
\bvolume{5}(\bissue{1}),
\bfpage{12}--\blpage{26}
(\byear{2020})
\end{barticle}
\endbibitem

\bibitem{bilenko2011predictive}
\begin{bchapter}
\bauthor{\bsnm{Bilenko}, \binits{M.}},
\bauthor{\bsnm{Richardson}, \binits{M.}}:
\bctitle{Predictive client-side profiles for personalized advertising}.
In: \bbtitle{Proceedings of the 17th ACM SIGKDD International Conference on
  Knowledge Discovery and Data Mining},
pp. \bfpage{413}--\blpage{421}
(\byear{2011})
\end{bchapter}
\endbibitem

\bibitem{yamamoto2012wisdom}
\begin{bchapter}
\bauthor{\bsnm{Yamamoto}, \binits{T.}},
\bauthor{\bsnm{Sakai}, \binits{T.}},
\bauthor{\bsnm{Iwata}, \binits{M.}},
\bauthor{\bsnm{Yu}, \binits{C.}},
\bauthor{\bsnm{Wen}, \binits{J.-R.}},
\bauthor{\bsnm{Tanaka}, \binits{K.}}:
\bctitle{The wisdom of advertisers: mining subgoals via query clustering}.
In: \bbtitle{Proceedings of the 21st ACM International Conference on
  Information and Knowledge Management},
pp. \bfpage{505}--\blpage{514}
(\byear{2012})
\end{bchapter}
\endbibitem

\bibitem{rapp2021human}
\begin{barticle}
\bauthor{\bsnm{Rapp}, \binits{A.}},
\bauthor{\bsnm{Curti}, \binits{L.}},
\bauthor{\bsnm{Boldi}, \binits{A.}}:
\batitle{The human side of human-chatbot interaction: A systematic literature
  review of ten years of research on text-based chatbots}.
\bjtitle{International Journal of Human-Computer Studies}
\bvolume{151},
\bfpage{102630}
(\byear{2021})
\end{barticle}
\endbibitem

\bibitem{villegas2018characterizing}
\begin{barticle}
\bauthor{\bsnm{Villegas}, \binits{N.M.}},
\bauthor{\bsnm{S{\'a}nchez}, \binits{C.}},
\bauthor{\bsnm{D{\'\i}az-Cely}, \binits{J.}},
\bauthor{\bsnm{Tamura}, \binits{G.}}:
\batitle{Characterizing context-aware recommender systems: A systematic
  literature review}.
\bjtitle{Knowledge-Based Systems}
\bvolume{140},
\bfpage{173}--\blpage{200}
(\byear{2018})
\end{barticle}
\endbibitem

\bibitem{auch2020similarity}
\begin{barticle}
\bauthor{\bsnm{Auch}, \binits{M.}},
\bauthor{\bsnm{Weber}, \binits{M.}},
\bauthor{\bsnm{Mandl}, \binits{P.}},
\bauthor{\bsnm{Wolff}, \binits{C.}}:
\batitle{Similarity-based analyses on software applications: A systematic
  literature review}.
\bjtitle{Journal of Systems and Software}
\bvolume{168},
\bfpage{110669}
(\byear{2020})
\end{barticle}
\endbibitem

\bibitem{obidallah2020clustering}
\begin{barticle}
\bauthor{\bsnm{Obidallah}, \binits{W.J.}},
\bauthor{\bsnm{Raahemi}, \binits{B.}},
\bauthor{\bsnm{Ruhi}, \binits{U.}}:
\batitle{Clustering and association rules for web service discovery and
  recommendation: a systematic literature review}.
\bjtitle{SN Computer Science}
\bvolume{1},
\bfpage{1}--\blpage{33}
(\byear{2020})
\end{barticle}
\endbibitem

\bibitem{xia2018zero}
\begin{botherref}
\oauthor{\bsnm{Xia}, \binits{C.}},
\oauthor{\bsnm{Zhang}, \binits{C.}},
\oauthor{\bsnm{Yan}, \binits{X.}},
\oauthor{\bsnm{Chang}, \binits{Y.}},
\oauthor{\bsnm{Yu}, \binits{P.S.}}:
Zero-shot user intent detection via capsule neural networks.
arXiv preprint arXiv:1809.00385
(2018)
\end{botherref}
\endbibitem

\bibitem{hu2017deep}
\begin{barticle}
\bauthor{\bsnm{Hu}, \binits{Z.}},
\bauthor{\bsnm{Zhang}, \binits{Z.}},
\bauthor{\bsnm{Yang}, \binits{H.}},
\bauthor{\bsnm{Chen}, \binits{Q.}},
\bauthor{\bsnm{Zuo}, \binits{D.}}:
\batitle{A deep learning approach for predicting the quality of online health
  expert question-answering services}.
\bjtitle{Journal of biomedical informatics}
\bvolume{71},
\bfpage{241}--\blpage{253}
(\byear{2017})
\end{barticle}
\endbibitem

\bibitem{chen2013wt}
\begin{bchapter}
\bauthor{\bsnm{Chen}, \binits{L.}},
\bauthor{\bsnm{Wang}, \binits{Y.}},
\bauthor{\bsnm{Yu}, \binits{Q.}},
\bauthor{\bsnm{Zheng}, \binits{Z.}},
\bauthor{\bsnm{Wu}, \binits{J.}}:
\bctitle{Wt-lda: user tagging augmented lda for web service clustering}.
In: \bbtitle{Service-Oriented Computing: 11th International Conference, ICSOC
  2013, Berlin, Germany, December 2-5, 2013, Proceedings 11},
pp. \bfpage{162}--\blpage{176}
(\byear{2013}).
\bcomment{Springer}
\end{bchapter}
\endbibitem

\bibitem{weismayer2017identifying}
\begin{barticle}
\bauthor{\bsnm{Weismayer}, \binits{C.}},
\bauthor{\bsnm{Pezenka}, \binits{I.}}:
\batitle{Identifying emerging research fields: a longitudinal latent semantic
  keyword analysis}.
\bjtitle{Scientometrics}
\bvolume{113}(\bissue{3}),
\bfpage{1757}--\blpage{1785}
(\byear{2017})
\end{barticle}
\endbibitem

\bibitem{HuREc}
\begin{bchapter}
\bauthor{\bsnm{Hu}, \binits{Y.}},
\bauthor{\bsnm{Da}, \binits{Q.}},
\bauthor{\bsnm{Zeng}, \binits{A.}},
\bauthor{\bsnm{Yu}, \binits{Y.}},
\bauthor{\bsnm{Xu}, \binits{Y.}}:
\bctitle{Reinforcement learning to rank in e-commerce search engine:
  Formalization, analysis, and application}.
In: \bbtitle{Proceedings of the 24th ACM SIGKDD International Conference on
  Knowledge Discovery \& Data Mining}.
\bsertitle{KDD '18},
pp. \bfpage{368}--\blpage{377}.
\bpublisher{Association for Computing Machinery},
\blocation{New York, NY, USA}
(\byear{2018}).
\doiurl{10.1145/3219819.3219846}.
\burl{https://doi.org/10.1145/3219819.3219846}
\end{bchapter}
\endbibitem

\bibitem{GuRec}
\begin{bchapter}
\bauthor{\bsnm{Gu}, \binits{Y.}},
\bauthor{\bsnm{Zhao}, \binits{B.}},
\bauthor{\bsnm{Hardtke}, \binits{D.}},
\bauthor{\bsnm{Sun}, \binits{Y.}}:
\bctitle{Learning global term weights for content-based recommender systems}.
In: \bbtitle{Proceedings of the 25th International Conference on World Wide
  Web}.
\bsertitle{WWW '16},
pp. \bfpage{391}--\blpage{400}.
\bpublisher{International World Wide Web Conferences Steering Committee},
\blocation{Republic and Canton of Geneva, CHE}
(\byear{2016}).
\doiurl{10.1145/2872427.2883069}.
\burl{https://doi.org/10.1145/2872427.2883069}
\end{bchapter}
\endbibitem

\bibitem{yao2022reprbert}
\begin{bchapter}
\bauthor{\bsnm{Yao}, \binits{S.}},
\bauthor{\bsnm{Tan}, \binits{J.}},
\bauthor{\bsnm{Chen}, \binits{X.}},
\bauthor{\bsnm{Zhang}, \binits{J.}},
\bauthor{\bsnm{Zeng}, \binits{X.}},
\bauthor{\bsnm{Yang}, \binits{K.}}:
\bctitle{Reprbert: Distilling bert to an efficient representation-based
  relevance model for e-commerce}.
In: \bbtitle{Proceedings of the 28th ACM SIGKDD Conference on Knowledge
  Discovery and Data Mining},
pp. \bfpage{4363}--\blpage{4371}
(\byear{2022})
\end{bchapter}
\endbibitem

\bibitem{AminuRec}
\begin{barticle}
\bauthor{\bsnm{Da’u}, \binits{A.}},
\bauthor{\bsnm{Salim}, \binits{N.}}:
\batitle{Sentiment-aware deep recommender system with neural attention
  networks}.
\bjtitle{IEEE Access}
\bvolume{7},
\bfpage{45472}--\blpage{45484}
(\byear{2019}).
\doiurl{10.1109/ACCESS.2019.2907729}
\end{barticle}
\endbibitem

\bibitem{YeRec}
\begin{bchapter}
\bauthor{\bsnm{Ye}, \binits{Q.}},
\bauthor{\bsnm{Wang}, \binits{F.}},
\bauthor{\bsnm{Li}, \binits{B.}}:
\bctitle{Starrysky: A practical system to track millions of high-precision
  query intents}.
In: \bbtitle{Proceedings of the 25th International Conference Companion on
  World Wide Web}.
\bsertitle{WWW '16 Companion},
pp. \bfpage{961}--\blpage{966}.
\bpublisher{International World Wide Web Conferences Steering Committee},
\blocation{Republic and Canton of Geneva, CHE}
(\byear{2016}).
\doiurl{10.1145/2872518.2890588}.
\burl{https://doi.org/10.1145/2872518.2890588}
\end{bchapter}
\endbibitem

\bibitem{XU2022102545}
\begin{barticle}
\bauthor{\bsnm{Xu}, \binits{H.}},
\bauthor{\bsnm{Ding}, \binits{W.}},
\bauthor{\bsnm{Shen}, \binits{W.}},
\bauthor{\bsnm{Wang}, \binits{J.}},
\bauthor{\bsnm{Yang}, \binits{Z.}}:
\batitle{Deep convolutional recurrent model for region recommendation with
  spatial and temporal contexts}.
\bjtitle{Ad Hoc Networks}
\bvolume{129},
\bfpage{102545}
(\byear{2022}).
\doiurl{10.1016/j.adhoc.2021.102545}
\end{barticle}
\endbibitem

\bibitem{Qu7837964}
\begin{bchapter}
\bauthor{\bsnm{Qu}, \binits{Y.}},
\bauthor{\bsnm{Cai}, \binits{H.}},
\bauthor{\bsnm{Ren}, \binits{K.}},
\bauthor{\bsnm{Zhang}, \binits{W.}},
\bauthor{\bsnm{Yu}, \binits{Y.}},
\bauthor{\bsnm{Wen}, \binits{Y.}},
\bauthor{\bsnm{Wang}, \binits{J.}}:
\bctitle{Product-based neural networks for user response prediction}.
In: \bbtitle{2016 IEEE 16th International Conference on Data Mining (ICDM)},
pp. \bfpage{1149}--\blpage{1154}
(\byear{2016}).
\doiurl{10.1109/ICDM.2016.0151}
\end{bchapter}
\endbibitem

\bibitem{ricci2015recommender}
\begin{botherref}
\oauthor{\bsnm{Ricci}, \binits{F.}},
\oauthor{\bsnm{Rokach}, \binits{L.}},
\oauthor{\bsnm{Shapira}, \binits{B.}}:
Recommender systems: introduction and challenges.
Recommender systems handbook,
1--34
(2015)
\end{botherref}
\endbibitem

\bibitem{portugal2018use}
\begin{barticle}
\bauthor{\bsnm{Portugal}, \binits{I.}},
\bauthor{\bsnm{Alencar}, \binits{P.}},
\bauthor{\bsnm{Cowan}, \binits{D.}}:
\batitle{The use of machine learning algorithms in recommender systems: A
  systematic review}.
\bjtitle{Expert Systems with Applications}
\bvolume{97},
\bfpage{205}--\blpage{227}
(\byear{2018})
\end{barticle}
\endbibitem

\bibitem{allamanis2018survey}
\begin{barticle}
\bauthor{\bsnm{Allamanis}, \binits{M.}},
\bauthor{\bsnm{Barr}, \binits{E.T.}},
\bauthor{\bsnm{Devanbu}, \binits{P.}},
\bauthor{\bsnm{Sutton}, \binits{C.}}:
\batitle{A survey of machine learning for big code and naturalness}.
\bjtitle{ACM Computing Surveys (CSUR)}
\bvolume{51}(\bissue{4}),
\bfpage{1}--\blpage{37}
(\byear{2018})
\end{barticle}
\endbibitem

\bibitem{hill2016trials}
\begin{bchapter}
\bauthor{\bsnm{Hill}, \binits{C.}},
\bauthor{\bsnm{Bellamy}, \binits{R.}},
\bauthor{\bsnm{Erickson}, \binits{T.}},
\bauthor{\bsnm{Burnett}, \binits{M.}}:
\bctitle{Trials and tribulations of developers of intelligent systems: A field
  study}.
In: \bbtitle{2016 IEEE Symposium on Visual Languages and Human-Centric
  Computing (VL/HCC)},
pp. \bfpage{162}--\blpage{170}
(\byear{2016}).
\bcomment{IEEE}
\end{bchapter}
\endbibitem

\bibitem{kitchenham2009systematic}
\begin{barticle}
\bauthor{\bsnm{Kitchenham}, \binits{B.}},
\bauthor{\bsnm{Brereton}, \binits{O.P.}},
\bauthor{\bsnm{Budgen}, \binits{D.}},
\bauthor{\bsnm{Turner}, \binits{M.}},
\bauthor{\bsnm{Bailey}, \binits{J.}},
\bauthor{\bsnm{Linkman}, \binits{S.}}:
\batitle{Systematic literature reviews in software engineering--a systematic
  literature review}.
\bjtitle{Information and software technology}
\bvolume{51}(\bissue{1}),
\bfpage{7}--\blpage{15}
(\byear{2009})
\end{barticle}
\endbibitem

\bibitem{xiao2019guidance}
\begin{barticle}
\bauthor{\bsnm{Xiao}, \binits{Y.}},
\bauthor{\bsnm{Watson}, \binits{M.}}:
\batitle{Guidance on conducting a systematic literature review}.
\bjtitle{Journal of planning education and research}
\bvolume{39}(\bissue{1}),
\bfpage{93}--\blpage{112}
(\byear{2019})
\end{barticle}
\endbibitem

\bibitem{okoli2015guide}
\begin{botherref}
\oauthor{\bsnm{Okoli}, \binits{C.}},
\oauthor{\bsnm{Schabram}, \binits{K.}}:
A guide to conducting a systematic literature review of information systems
  research
(2015)
\end{botherref}
\endbibitem

\bibitem{yin1981case}
\begin{bbook}
\bauthor{\bsnm{Yin}, \binits{R.K.}}:
\bbtitle{Case Study Research: Design and Methods}
vol. \bseriesno{5}.
\bpublisher{sage}, \blocation{???}
(\byear{2009})
\end{bbook}
\endbibitem

\bibitem{ye2016starrysky}
\begin{bchapter}
\bauthor{\bsnm{Ye}, \binits{Q.}},
\bauthor{\bsnm{Wang}, \binits{F.}},
\bauthor{\bsnm{Li}, \binits{B.}}:
\bctitle{Starrysky: A practical system to track millions of high-precision
  query intents}.
In: \bbtitle{Proceedings of the 25th International Conference Companion on
  World Wide Web},
pp. \bfpage{961}--\blpage{966}
(\byear{2016})
\end{bchapter}
\endbibitem

\bibitem{wang2021learning}
\begin{bchapter}
\bauthor{\bsnm{Wang}, \binits{X.}},
\bauthor{\bsnm{Huang}, \binits{T.}},
\bauthor{\bsnm{Wang}, \binits{D.}},
\bauthor{\bsnm{Yuan}, \binits{Y.}},
\bauthor{\bsnm{Liu}, \binits{Z.}},
\bauthor{\bsnm{He}, \binits{X.}},
\bauthor{\bsnm{Chua}, \binits{T.-S.}}:
\bctitle{Learning intents behind interactions with knowledge graph for
  recommendation}.
In: \bbtitle{Proceedings of the Web Conference 2021},
pp. \bfpage{878}--\blpage{887}
(\byear{2021})
\end{bchapter}
\endbibitem

\bibitem{nguyen2004capturing}
\begin{bchapter}
\bauthor{\bsnm{Nguyen}, \binits{H.}},
\bauthor{\bsnm{Santos~Jr}, \binits{E.}},
\bauthor{\bsnm{Zhao~Jr}, \binits{Q.}},
\bauthor{\bsnm{Wang~Jr}, \binits{H.}}:
\bctitle{Capturing user intent for information retrieval}.
In: \bbtitle{Proceedings of the Human Factors and Ergonomics Society Annual
  Meeting},
vol. \bseriesno{48},
pp. \bfpage{371}--\blpage{375}
(\byear{2004}).
\bcomment{SAGE Publications Sage CA: Los Angeles, CA}
\end{bchapter}
\endbibitem

\bibitem{hernandez2019comparative}
\begin{barticle}
\bauthor{\bsnm{Hern{\'a}ndez-Rubio}, \binits{M.}},
\bauthor{\bsnm{Cantador}, \binits{I.}},
\bauthor{\bsnm{Bellog{\'\i}n}, \binits{A.}}:
\batitle{A comparative analysis of recommender systems based on item aspect
  opinions extracted from user reviews}.
\bjtitle{User Modeling and User-Adapted Interaction}
\bvolume{29}(\bissue{2}),
\bfpage{381}--\blpage{441}
(\byear{2019})
\end{barticle}
\endbibitem

\bibitem{chen2015recommender}
\begin{barticle}
\bauthor{\bsnm{Chen}, \binits{L.}},
\bauthor{\bsnm{Chen}, \binits{G.}},
\bauthor{\bsnm{Wang}, \binits{F.}}:
\batitle{Recommender systems based on user reviews: the state of the art}.
\bjtitle{User Modeling and User-Adapted Interaction}
\bvolume{25},
\bfpage{99}--\blpage{154}
(\byear{2015})
\end{barticle}
\endbibitem

\bibitem{jordan2015machine}
\begin{barticle}
\bauthor{\bsnm{Jordan}, \binits{M.I.}},
\bauthor{\bsnm{Mitchell}, \binits{T.M.}}:
\batitle{Machine learning: Trends, perspectives, and prospects}.
\bjtitle{Science}
\bvolume{349}(\bissue{6245}),
\bfpage{255}--\blpage{260}
(\byear{2015})
\end{barticle}
\endbibitem

\bibitem{telikani2021evolutionary}
\begin{barticle}
\bauthor{\bsnm{Telikani}, \binits{A.}},
\bauthor{\bsnm{Tahmassebi}, \binits{A.}},
\bauthor{\bsnm{Banzhaf}, \binits{W.}},
\bauthor{\bsnm{Gandomi}, \binits{A.H.}}:
\batitle{Evolutionary machine learning: A survey}.
\bjtitle{ACM Computing Surveys (CSUR)}
\bvolume{54}(\bissue{8}),
\bfpage{1}--\blpage{35}
(\byear{2021})
\end{barticle}
\endbibitem

\bibitem{singh2016review}
\begin{bchapter}
\bauthor{\bsnm{Singh}, \binits{A.}},
\bauthor{\bsnm{Thakur}, \binits{N.}},
\bauthor{\bsnm{Sharma}, \binits{A.}}:
\bctitle{A review of supervised machine learning algorithms}.
In: \bbtitle{2016 3rd International Conference on Computing for Sustainable
  Global Development (INDIACom)},
pp. \bfpage{1310}--\blpage{1315}
(\byear{2016}).
\bcomment{Ieee}
\end{bchapter}
\endbibitem

\bibitem{zaib2022conversational}
\begin{barticle}
\bauthor{\bsnm{Zaib}, \binits{M.}},
\bauthor{\bsnm{Zhang}, \binits{W.E.}},
\bauthor{\bsnm{Sheng}, \binits{Q.Z.}},
\bauthor{\bsnm{Mahmood}, \binits{A.}},
\bauthor{\bsnm{Zhang}, \binits{Y.}}:
\batitle{Conversational question answering: A survey}.
\bjtitle{Knowledge and Information Systems}
\bvolume{64}(\bissue{12}),
\bfpage{3151}--\blpage{3195}
(\byear{2022})
\end{barticle}
\endbibitem

\bibitem{von2020combining}
\begin{bchapter}
\bauthor{\bparticle{von} \bsnm{Rueden}, \binits{L.}},
\bauthor{\bsnm{Mayer}, \binits{S.}},
\bauthor{\bsnm{Sifa}, \binits{R.}},
\bauthor{\bsnm{Bauckhage}, \binits{C.}},
\bauthor{\bsnm{Garcke}, \binits{J.}}:
\bctitle{Combining machine learning and simulation to a hybrid modelling
  approach: Current and future directions}.
In: \bbtitle{Advances in Intelligent Data Analysis XVIII: 18th International
  Symposium on Intelligent Data Analysis, IDA 2020, Konstanz, Germany, April
  27--29, 2020, Proceedings 18},
pp. \bfpage{548}--\blpage{560}
(\byear{2020}).
\bcomment{Springer}
\end{bchapter}
\endbibitem

\bibitem{yuan2020expert}
\begin{barticle}
\bauthor{\bsnm{Yuan}, \binits{S.}},
\bauthor{\bsnm{Zhang}, \binits{Y.}},
\bauthor{\bsnm{Tang}, \binits{J.}},
\bauthor{\bsnm{Hall}, \binits{W.}},
\bauthor{\bsnm{Cabot{\`a}}, \binits{J.B.}}:
\batitle{Expert finding in community question answering: a review}.
\bjtitle{Artificial Intelligence Review}
\bvolume{53},
\bfpage{843}--\blpage{874}
(\byear{2020})
\end{barticle}
\endbibitem

\bibitem{farshidi2020capturing}
\begin{barticle}
\bauthor{\bsnm{Farshidi}, \binits{S.}},
\bauthor{\bsnm{Jansen}, \binits{S.}},
\bauthor{\bparticle{van~der} \bsnm{Werf}, \binits{J.M.}}:
\batitle{Capturing software architecture knowledge for pattern-driven design}.
\bjtitle{Journal of Systems and Software}
\bvolume{169},
\bfpage{110714}
(\byear{2020})
\end{barticle}
\endbibitem

\bibitem{farshidi2020multi}
\begin{botherref}
\oauthor{\bsnm{Farshidi}, \binits{S.}}:
Multi-criteria decision-making in software production.
PhD thesis,
Utrecht University
(2020)
\end{botherref}
\endbibitem

\bibitem{jansen2009applied}
\begin{bchapter}
\bauthor{\bsnm{Jansen}, \binits{S.}}:
\bctitle{Applied multi-case research in a mixed-method research project:
  Customer configuration updating improvement}.
In: \bbtitle{Information Systems Research Methods, Epistemology, and
  Applications},
pp. \bfpage{120}--\blpage{139}.
\bpublisher{IGI Global}, \blocation{???}
(\byear{2009})
\end{bchapter}
\endbibitem

\bibitem{johnson2004mixed}
\begin{barticle}
\bauthor{\bsnm{Johnson}, \binits{R.B.}},
\bauthor{\bsnm{Onwuegbuzie}, \binits{A.J.}}:
\batitle{Mixed methods research: A research paradigm whose time has come}.
\bjtitle{Educational researcher}
\bvolume{33}(\bissue{7}),
\bfpage{14}--\blpage{26}
(\byear{2004})
\end{barticle}
\endbibitem

\bibitem{yin2017case}
\begin{bbook}
\bauthor{\bsnm{Yin}, \binits{R.K.}}:
\bbtitle{Case Study Research and Applications: Design and Methods}.
\bpublisher{Sage publications}, \blocation{???}
(\byear{2017})
\end{bbook}
\endbibitem

\bibitem{kilgarriff2014sketch}
\begin{barticle}
\bauthor{\bsnm{Kilgarriff}, \binits{A.}},
\bauthor{\bsnm{Baisa}, \binits{V.}},
\bauthor{\bsnm{Bu{\v{s}}ta}, \binits{J.}},
\bauthor{\bsnm{Jakub{\'\i}{\v{c}}ek}, \binits{M.}},
\bauthor{\bsnm{Kov{\'a}{\v{r}}}, \binits{V.}},
\bauthor{\bsnm{Michelfeit}, \binits{J.}},
\bauthor{\bsnm{Rychl{\`y}}, \binits{P.}},
\bauthor{\bsnm{Suchomel}, \binits{V.}}:
\batitle{The sketch engine: ten years on}.
\bjtitle{Lexicography}
\bvolume{1}(\bissue{1}),
\bfpage{7}--\blpage{36}
(\byear{2014})
\end{barticle}
\endbibitem

\bibitem{ni2021effective}
\begin{barticle}
\bauthor{\bsnm{Ni}, \binits{J.}},
\bauthor{\bsnm{Huang}, \binits{Z.}},
\bauthor{\bsnm{Cheng}, \binits{J.}},
\bauthor{\bsnm{Gao}, \binits{S.}}:
\batitle{An effective recommendation model based on deep representation
  learning}.
\bjtitle{Information Sciences}
\bvolume{542},
\bfpage{324}--\blpage{342}
(\byear{2021})
\end{barticle}
\endbibitem

\bibitem{wang2018streaming}
\begin{bchapter}
\bauthor{\bsnm{Wang}, \binits{W.}},
\bauthor{\bsnm{Yin}, \binits{H.}},
\bauthor{\bsnm{Huang}, \binits{Z.}},
\bauthor{\bsnm{Wang}, \binits{Q.}},
\bauthor{\bsnm{Du}, \binits{X.}},
\bauthor{\bsnm{Nguyen}, \binits{Q.V.H.}}:
\bctitle{Streaming ranking based recommender systems}.
In: \bbtitle{The 41st International ACM SIGIR Conference on Research \&
  Development in Information Retrieval},
pp. \bfpage{525}--\blpage{534}
(\byear{2018})
\end{bchapter}
\endbibitem

\bibitem{qu2019user}
\begin{bchapter}
\bauthor{\bsnm{Qu}, \binits{C.}},
\bauthor{\bsnm{Yang}, \binits{L.}},
\bauthor{\bsnm{Croft}, \binits{W.B.}},
\bauthor{\bsnm{Zhang}, \binits{Y.}},
\bauthor{\bsnm{Trippas}, \binits{J.R.}},
\bauthor{\bsnm{Qiu}, \binits{M.}}:
\bctitle{User intent prediction in information-seeking conversations}.
In: \bbtitle{Proceedings of the 2019 Conference on Human Information
  Interaction and Retrieval},
pp. \bfpage{25}--\blpage{33}
(\byear{2019})
\end{bchapter}
\endbibitem

\bibitem{zhang2021discovering}
\begin{bchapter}
\bauthor{\bsnm{Zhang}, \binits{H.}},
\bauthor{\bsnm{Xu}, \binits{H.}},
\bauthor{\bsnm{Lin}, \binits{T.-E.}},
\bauthor{\bsnm{Lyu}, \binits{R.}}:
\bctitle{Discovering new intents with deep aligned clustering}.
In: \bbtitle{Proceedings of the AAAI Conference on Artificial Intelligence},
vol. \bseriesno{35},
pp. \bfpage{14365}--\blpage{14373}
(\byear{2021})
\end{bchapter}
\endbibitem

\bibitem{agarwal2020evaluation}
\begin{barticle}
\bauthor{\bsnm{Agarwal}, \binits{N.}},
\bauthor{\bsnm{Sikka}, \binits{G.}},
\bauthor{\bsnm{Awasthi}, \binits{L.K.}}:
\batitle{Evaluation of web service clustering using dirichlet multinomial
  mixture model based approach for dimensionality reduction in service
  representation}.
\bjtitle{Information Processing \& Management}
\bvolume{57}(\bissue{4}),
\bfpage{102238}
(\byear{2020})
\end{barticle}
\endbibitem

\bibitem{zhang2018discrete}
\begin{bchapter}
\bauthor{\bsnm{Zhang}, \binits{Y.}},
\bauthor{\bsnm{Yin}, \binits{H.}},
\bauthor{\bsnm{Huang}, \binits{Z.}},
\bauthor{\bsnm{Du}, \binits{X.}},
\bauthor{\bsnm{Yang}, \binits{G.}},
\bauthor{\bsnm{Lian}, \binits{D.}}:
\bctitle{Discrete deep learning for fast content-aware recommendation}.
In: \bbtitle{Proceedings of the Eleventh ACM International Conference on Web
  Search and Data Mining},
pp. \bfpage{717}--\blpage{726}
(\byear{2018})
\end{bchapter}
\endbibitem

\bibitem{yu2022graph}
\begin{barticle}
\bauthor{\bsnm{Yu}, \binits{B.}},
\bauthor{\bsnm{Zhang}, \binits{R.}},
\bauthor{\bsnm{Chen}, \binits{W.}},
\bauthor{\bsnm{Fang}, \binits{J.}}:
\batitle{Graph neural network based model for multi-behavior session-based
  recommendation}.
\bjtitle{GeoInformatica}
\bvolume{26}(\bissue{2}),
\bfpage{429}--\blpage{447}
(\byear{2022})
\end{barticle}
\endbibitem

\bibitem{lin2021go}
\begin{barticle}
\bauthor{\bsnm{Lin}, \binits{H.}},
\bauthor{\bsnm{Liu}, \binits{G.}},
\bauthor{\bsnm{Li}, \binits{F.}},
\bauthor{\bsnm{Zuo}, \binits{Y.}}:
\batitle{Where to go? predicting next location in iot environment}.
\bjtitle{Frontiers of Computer Science}
\bvolume{15},
\bfpage{1}--\blpage{13}
(\byear{2021})
\end{barticle}
\endbibitem

\bibitem{Farshidi_Rezaee_2023}
\begin{botherref}
\oauthor{\bsnm{Farshidi}, \binits{S.}},
\oauthor{\bsnm{Rezaee}, \binits{K.}}:
Understanding User Intent: A Systematic Literature Review of Modeling
  Techniques.
Mendeley Data
(2023).
\url{http://dx.doi.org/10.17632/nw79y7mcvd.1}
\end{botherref}
\endbibitem

\bibitem{latifi2021session}
\begin{barticle}
\bauthor{\bsnm{Latifi}, \binits{S.}},
\bauthor{\bsnm{Mauro}, \binits{N.}},
\bauthor{\bsnm{Jannach}, \binits{D.}}:
\batitle{Session-aware recommendation: A surprising quest for the
  state-of-the-art}.
\bjtitle{Information Sciences}
\bvolume{573},
\bfpage{291}--\blpage{315}
(\byear{2021})
\end{barticle}
\endbibitem

\bibitem{park2020click}
\begin{barticle}
\bauthor{\bsnm{Park}, \binits{C.}},
\bauthor{\bsnm{Kim}, \binits{D.}},
\bauthor{\bsnm{Yang}, \binits{M.-C.}},
\bauthor{\bsnm{Lee}, \binits{J.-T.}},
\bauthor{\bsnm{Yu}, \binits{H.}}:
\batitle{Click-aware purchase prediction with push at the top}.
\bjtitle{Information Sciences}
\bvolume{521},
\bfpage{350}--\blpage{364}
(\byear{2020})
\end{barticle}
\endbibitem

\bibitem{ludewig2018evaluation}
\begin{barticle}
\bauthor{\bsnm{Ludewig}, \binits{M.}},
\bauthor{\bsnm{Jannach}, \binits{D.}}:
\batitle{Evaluation of session-based recommendation algorithms}.
\bjtitle{User Modeling and User-Adapted Interaction}
\bvolume{28},
\bfpage{331}--\blpage{390}
(\byear{2018})
\end{barticle}
\endbibitem

\bibitem{zhou2020leveraging}
\begin{bchapter}
\bauthor{\bsnm{Zhou}, \binits{K.}},
\bauthor{\bsnm{Zhao}, \binits{W.X.}},
\bauthor{\bsnm{Wang}, \binits{H.}},
\bauthor{\bsnm{Wang}, \binits{S.}},
\bauthor{\bsnm{Zhang}, \binits{F.}},
\bauthor{\bsnm{Wang}, \binits{Z.}},
\bauthor{\bsnm{Wen}, \binits{J.-R.}}:
\bctitle{Leveraging historical interaction data for improving conversational
  recommender system}.
In: \bbtitle{Proceedings of the 29th ACM International Conference on
  Information \& Knowledge Management},
pp. \bfpage{2349}--\blpage{2352}
(\byear{2020})
\end{bchapter}
\endbibitem

\bibitem{white2013enhancing}
\begin{bchapter}
\bauthor{\bsnm{White}, \binits{R.W.}},
\bauthor{\bsnm{Chu}, \binits{W.}},
\bauthor{\bsnm{Hassan}, \binits{A.}},
\bauthor{\bsnm{He}, \binits{X.}},
\bauthor{\bsnm{Song}, \binits{Y.}},
\bauthor{\bsnm{Wang}, \binits{H.}}:
\bctitle{Enhancing personalized search by mining and modeling task behavior}.
In: \bbtitle{Proceedings of the 22nd International Conference on World Wide
  Web},
pp. \bfpage{1411}--\blpage{1420}
(\byear{2013})
\end{bchapter}
\endbibitem

\bibitem{zou2022improving}
\begin{bchapter}
\bauthor{\bsnm{Zou}, \binits{J.}},
\bauthor{\bsnm{Kanoulas}, \binits{E.}},
\bauthor{\bsnm{Ren}, \binits{P.}},
\bauthor{\bsnm{Ren}, \binits{Z.}},
\bauthor{\bsnm{Sun}, \binits{A.}},
\bauthor{\bsnm{Long}, \binits{C.}}:
\bctitle{Improving conversational recommender systems via transformer-based
  sequential modelling}.
In: \bbtitle{Proceedings of the 45th International ACM SIGIR Conference on
  Research and Development in Information Retrieval},
pp. \bfpage{2319}--\blpage{2324}
(\byear{2022})
\end{bchapter}
\endbibitem

\bibitem{zhou2019real}
\begin{barticle}
\bauthor{\bsnm{Zhou}, \binits{X.}},
\bauthor{\bsnm{Qin}, \binits{D.}},
\bauthor{\bsnm{Chen}, \binits{L.}},
\bauthor{\bsnm{Zhang}, \binits{Y.}}:
\batitle{Real-time context-aware social media recommendation}.
\bjtitle{The VLDB Journal}
\bvolume{28},
\bfpage{197}--\blpage{219}
(\byear{2019})
\end{barticle}
\endbibitem

\bibitem{musto2019linked}
\begin{barticle}
\bauthor{\bsnm{Musto}, \binits{C.}},
\bauthor{\bsnm{Narducci}, \binits{F.}},
\bauthor{\bsnm{Lops}, \binits{P.}},
\bauthor{\bparticle{de} \bsnm{Gemmis}, \binits{M.}},
\bauthor{\bsnm{Semeraro}, \binits{G.}}:
\batitle{Linked open data-based explanations for transparent recommender
  systems}.
\bjtitle{International Journal of Human-Computer Studies}
\bvolume{121},
\bfpage{93}--\blpage{107}
(\byear{2019})
\end{barticle}
\endbibitem

\bibitem{mandayam2017intent}
\begin{bchapter}
\bauthor{\bsnm{Mandayam~Comar}, \binits{P.}},
\bauthor{\bsnm{Sengamedu}, \binits{S.H.}}:
\bctitle{Intent based relevance estimation from click logs}.
In: \bbtitle{Proceedings of the 2017 ACM on Conference on Information and
  Knowledge Management},
pp. \bfpage{59}--\blpage{66}
(\byear{2017})
\end{bchapter}
\endbibitem

\bibitem{ding2022tdtmf}
\begin{barticle}
\bauthor{\bsnm{Ding}, \binits{H.}},
\bauthor{\bsnm{Liu}, \binits{Q.}},
\bauthor{\bsnm{Hu}, \binits{G.}}:
\batitle{Tdtmf: A recommendation model based on user temporal interest drift
  and latent review topic evolution with regularization factor}.
\bjtitle{Information Processing \& Management}
\bvolume{59}(\bissue{5}),
\bfpage{103037}
(\byear{2022})
\end{barticle}
\endbibitem

\bibitem{pradhan2021claver}
\begin{barticle}
\bauthor{\bsnm{Pradhan}, \binits{T.}},
\bauthor{\bsnm{Kumar}, \binits{P.}},
\bauthor{\bsnm{Pal}, \binits{S.}}:
\batitle{Claver: An integrated framework of convolutional layer, bidirectional
  lstm with attention mechanism based scholarly venue recommendation}.
\bjtitle{Information Sciences}
\bvolume{559},
\bfpage{212}--\blpage{235}
(\byear{2021})
\end{barticle}
\endbibitem

\bibitem{yu2018pave}
\begin{barticle}
\bauthor{\bsnm{Yu}, \binits{S.}},
\bauthor{\bsnm{Liu}, \binits{J.}},
\bauthor{\bsnm{Yang}, \binits{Z.}},
\bauthor{\bsnm{Chen}, \binits{Z.}},
\bauthor{\bsnm{Jiang}, \binits{H.}},
\bauthor{\bsnm{Tolba}, \binits{A.}},
\bauthor{\bsnm{Xia}, \binits{F.}}:
\batitle{Pave: Personalized academic venue recommendation exploiting
  co-publication networks}.
\bjtitle{Journal of Network and Computer Applications}
\bvolume{104},
\bfpage{38}--\blpage{47}
(\byear{2018})
\end{barticle}
\endbibitem

\bibitem{schlaefer2011statistical}
\begin{bchapter}
\bauthor{\bsnm{Schlaefer}, \binits{N.}},
\bauthor{\bsnm{Chu-Carroll}, \binits{J.}},
\bauthor{\bsnm{Nyberg}, \binits{E.}},
\bauthor{\bsnm{Fan}, \binits{J.}},
\bauthor{\bsnm{Zadrozny}, \binits{W.}},
\bauthor{\bsnm{Ferrucci}, \binits{D.}}:
\bctitle{Statistical source expansion for question answering}.
In: \bbtitle{Proceedings of the 20th ACM International Conference on
  Information and Knowledge Management},
pp. \bfpage{345}--\blpage{354}
(\byear{2011})
\end{bchapter}
\endbibitem

\bibitem{kim2017deep}
\begin{barticle}
\bauthor{\bsnm{Kim}, \binits{D.}},
\bauthor{\bsnm{Park}, \binits{C.}},
\bauthor{\bsnm{Oh}, \binits{J.}},
\bauthor{\bsnm{Yu}, \binits{H.}}:
\batitle{Deep hybrid recommender systems via exploiting document context and
  statistics of items}.
\bjtitle{Information Sciences}
\bvolume{417},
\bfpage{72}--\blpage{87}
(\byear{2017})
\end{barticle}
\endbibitem

\bibitem{zhang2016improving}
\begin{barticle}
\bauthor{\bsnm{Zhang}, \binits{H.}},
\bauthor{\bsnm{Zhong}, \binits{G.}}:
\batitle{Improving short text classification by learning vector representations
  of both words and hidden topics}.
\bjtitle{Knowledge-Based Systems}
\bvolume{102},
\bfpage{76}--\blpage{86}
(\byear{2016})
\end{barticle}
\endbibitem

\bibitem{xu2015topic}
\begin{barticle}
\bauthor{\bsnm{Xu}, \binits{Z.}},
\bauthor{\bsnm{Chen}, \binits{L.}},
\bauthor{\bsnm{Chen}, \binits{G.}}:
\batitle{Topic based context-aware travel recommendation method exploiting
  geotagged photos}.
\bjtitle{Neurocomputing}
\bvolume{155},
\bfpage{99}--\blpage{107}
(\byear{2015})
\end{barticle}
\endbibitem

\bibitem{tang2010combination}
\begin{barticle}
\bauthor{\bsnm{Tang}, \binits{J.}},
\bauthor{\bsnm{Yao}, \binits{L.}},
\bauthor{\bsnm{Zhang}, \binits{D.}},
\bauthor{\bsnm{Zhang}, \binits{J.}}:
\batitle{A combination approach to web user profiling}.
\bjtitle{ACM Transactions on Knowledge Discovery from Data (TKDD)}
\bvolume{5}(\bissue{1}),
\bfpage{1}--\blpage{44}
(\byear{2010})
\end{barticle}
\endbibitem

\bibitem{li2014identifying}
\begin{bchapter}
\bauthor{\bsnm{Li}, \binits{L.}},
\bauthor{\bsnm{Deng}, \binits{H.}},
\bauthor{\bsnm{Dong}, \binits{A.}},
\bauthor{\bsnm{Chang}, \binits{Y.}},
\bauthor{\bsnm{Zha}, \binits{H.}}:
\bctitle{Identifying and labeling search tasks via query-based hawkes
  processes}.
In: \bbtitle{Proceedings of the 20th ACM SIGKDD International Conference on
  Knowledge Discovery and Data Mining},
pp. \bfpage{731}--\blpage{740}
(\byear{2014})
\end{bchapter}
\endbibitem

\bibitem{binkley2018need}
\begin{barticle}
\bauthor{\bsnm{Binkley}, \binits{D.}},
\bauthor{\bsnm{Lawrie}, \binits{D.}},
\bauthor{\bsnm{Morrell}, \binits{C.}}:
\batitle{The need for software specific natural language techniques}.
\bjtitle{Empirical Software Engineering}
\bvolume{23},
\bfpage{2398}--\blpage{2425}
(\byear{2018})
\end{barticle}
\endbibitem

\bibitem{izadi2022predicting}
\begin{barticle}
\bauthor{\bsnm{Izadi}, \binits{M.}},
\bauthor{\bsnm{Akbari}, \binits{K.}},
\bauthor{\bsnm{Heydarnoori}, \binits{A.}}:
\batitle{Predicting the objective and priority of issue reports in software
  repositories}.
\bjtitle{Empirical Software Engineering}
\bvolume{27}(\bissue{2}),
\bfpage{50}
(\byear{2022})
\end{barticle}
\endbibitem

\bibitem{venkateswara2022societal}
\begin{barticle}
\bauthor{\bsnm{Venkateswara~Rao}, \binits{P.}},
\bauthor{\bsnm{Kumar}, \binits{A.S.}}:
\batitle{The societal communication of the q\&a community on topic modeling}.
\bjtitle{The Journal of Supercomputing}
\bvolume{78}(\bissue{1}),
\bfpage{1117}--\blpage{1143}
(\byear{2022})
\end{barticle}
\endbibitem

\bibitem{yu2015combining}
\begin{barticle}
\bauthor{\bsnm{Yu}, \binits{J.}},
\bauthor{\bsnm{Zhu}, \binits{T.}}:
\batitle{Combining long-term and short-term user interest for personalized
  hashtag recommendation}.
\bjtitle{Frontiers of Computer Science}
\bvolume{9},
\bfpage{608}--\blpage{622}
(\byear{2015})
\end{barticle}
\endbibitem

\bibitem{yadav2022clus}
\begin{barticle}
\bauthor{\bsnm{Yadav}, \binits{N.}},
\bauthor{\bsnm{Pal}, \binits{S.}},
\bauthor{\bsnm{Singh}, \binits{A.K.}},
\bauthor{\bsnm{Singh}, \binits{K.}}:
\batitle{Clus-dr: Cluster-based pre-trained model for diverse recommendation
  generation}.
\bjtitle{Journal of King Saud University-Computer and Information Sciences}
\bvolume{34}(\bissue{8}),
\bfpage{6385}--\blpage{6399}
(\byear{2022})
\end{barticle}
\endbibitem

\bibitem{xu2022deep}
\begin{barticle}
\bauthor{\bsnm{Xu}, \binits{H.}},
\bauthor{\bsnm{Ding}, \binits{W.}},
\bauthor{\bsnm{Shen}, \binits{W.}},
\bauthor{\bsnm{Wang}, \binits{J.}},
\bauthor{\bsnm{Yang}, \binits{Z.}}:
\batitle{Deep convolutional recurrent model for region recommendation with
  spatial and temporal contexts}.
\bjtitle{Ad Hoc Networks}
\bvolume{129},
\bfpage{102545}
(\byear{2022})
\end{barticle}
\endbibitem

\bibitem{gozuacik2023technological}
\begin{barticle}
\bauthor{\bsnm{Gozuacik}, \binits{N.}},
\bauthor{\bsnm{Sakar}, \binits{C.O.}},
\bauthor{\bsnm{Ozcan}, \binits{S.}}:
\batitle{Technological forecasting based on estimation of word embedding matrix
  using lstm networks}.
\bjtitle{Technological Forecasting and Social Change}
\bvolume{191},
\bfpage{122520}
(\byear{2023})
\end{barticle}
\endbibitem

\bibitem{yengikand2023dhsirs}
\begin{botherref}
\oauthor{\bsnm{Yengikand}, \binits{A.K.}},
\oauthor{\bsnm{Meghdadi}, \binits{M.}},
\oauthor{\bsnm{Ahmadian}, \binits{S.}}:
Dhsirs: a novel deep hybrid side information-based recommender system.
Multimedia Tools and Applications,
1--27
(2023)
\end{botherref}
\endbibitem

\bibitem{chen2020handling}
\begin{bchapter}
\bauthor{\bsnm{Chen}, \binits{T.}},
\bauthor{\bsnm{Wong}, \binits{R.C.-W.}}:
\bctitle{Handling information loss of graph neural networks for session-based
  recommendation}.
In: \bbtitle{Proceedings of the 26th ACM SIGKDD International Conference on
  Knowledge Discovery \& Data Mining},
pp. \bfpage{1172}--\blpage{1180}
(\byear{2020})
\end{bchapter}
\endbibitem

\bibitem{elfaik2023leveraging}
\begin{barticle}
\bauthor{\bsnm{Elfaik}, \binits{H.}}, \betal:
\batitle{Leveraging feature-level fusion representations and attentional
  bidirectional rnn-cnn deep models for arabic affect analysis on twitter}.
\bjtitle{Journal of King Saud University-Computer and Information Sciences}
\bvolume{35}(\bissue{1}),
\bfpage{462}--\blpage{482}
(\byear{2023})
\end{barticle}
\endbibitem

\bibitem{ittoo2016text}
\begin{barticle}
\bauthor{\bsnm{Ittoo}, \binits{A.}},
\bauthor{\bparticle{van~den} \bsnm{Bosch}, \binits{A.}}, \betal:
\batitle{Text analytics in industry: Challenges, desiderata and trends}.
\bjtitle{Computers in Industry}
\bvolume{78},
\bfpage{96}--\blpage{107}
(\byear{2016})
\end{barticle}
\endbibitem

\bibitem{de2020intelligent}
\begin{barticle}
\bauthor{\bparticle{de} \bsnm{Barcelos~Silva}, \binits{A.}},
\bauthor{\bsnm{Gomes}, \binits{M.M.}},
\bauthor{\bparticle{da} \bsnm{Costa}, \binits{C.A.}},
\bauthor{\bparticle{da} \bsnm{Rosa~Righi}, \binits{R.}},
\bauthor{\bsnm{Barbosa}, \binits{J.L.V.}},
\bauthor{\bsnm{Pessin}, \binits{G.}},
\bauthor{\bsnm{De~Doncker}, \binits{G.}},
\bauthor{\bsnm{Federizzi}, \binits{G.}}:
\batitle{Intelligent personal assistants: A systematic literature review}.
\bjtitle{Expert Systems with Applications}
\bvolume{147},
\bfpage{113193}
(\byear{2020})
\end{barticle}
\endbibitem

\bibitem{pan2022test}
\begin{barticle}
\bauthor{\bsnm{Pan}, \binits{R.}},
\bauthor{\bsnm{Bagherzadeh}, \binits{M.}},
\bauthor{\bsnm{Ghaleb}, \binits{T.A.}},
\bauthor{\bsnm{Briand}, \binits{L.}}:
\batitle{Test case selection and prioritization using machine learning: a
  systematic literature review}.
\bjtitle{Empirical Software Engineering}
\bvolume{27}(\bissue{2}),
\bfpage{29}
(\byear{2022})
\end{barticle}
\endbibitem

\bibitem{pu2012evaluating}
\begin{barticle}
\bauthor{\bsnm{Pu}, \binits{P.}},
\bauthor{\bsnm{Chen}, \binits{L.}},
\bauthor{\bsnm{Hu}, \binits{R.}}:
\batitle{Evaluating recommender systems from the user's perspective: survey of
  the state of the art}.
\bjtitle{User Modeling and User-Adapted Interaction}
\bvolume{22}(\bissue{4}),
\bfpage{317}--\blpage{355}
(\byear{2012})
\end{barticle}
\endbibitem

\bibitem{salle2022cosearcher}
\begin{barticle}
\bauthor{\bsnm{Salle}, \binits{A.}},
\bauthor{\bsnm{Malmasi}, \binits{S.}},
\bauthor{\bsnm{Rokhlenko}, \binits{O.}},
\bauthor{\bsnm{Agichtein}, \binits{E.}}:
\batitle{Cosearcher: studying the effectiveness of conversational search
  refinement and clarification through user simulation}.
\bjtitle{Information Retrieval Journal}
\bvolume{25}(\bissue{2}),
\bfpage{209}--\blpage{238}
(\byear{2022})
\end{barticle}
\endbibitem

\bibitem{baykan2011comprehensive}
\begin{barticle}
\bauthor{\bsnm{Baykan}, \binits{E.}},
\bauthor{\bsnm{Henzinger}, \binits{M.}},
\bauthor{\bsnm{Marian}, \binits{L.}},
\bauthor{\bsnm{Weber}, \binits{I.}}:
\batitle{A comprehensive study of features and algorithms for url-based topic
  classification}.
\bjtitle{ACM Transactions on the Web (TWEB)}
\bvolume{5}(\bissue{3}),
\bfpage{1}--\blpage{29}
(\byear{2011})
\end{barticle}
\endbibitem

\bibitem{wang2022causal}
\begin{botherref}
\oauthor{\bsnm{Wang}, \binits{X.}},
\oauthor{\bsnm{Li}, \binits{Q.}},
\oauthor{\bsnm{Yu}, \binits{D.}},
\oauthor{\bsnm{Cui}, \binits{P.}},
\oauthor{\bsnm{Wang}, \binits{Z.}},
\oauthor{\bsnm{Xu}, \binits{G.}}:
Causal disentanglement for semantics-aware intent learning in recommendation.
IEEE Transactions on Knowledge and Data Engineering
(2022)
\end{botherref}
\endbibitem

\bibitem{phan2010hidden}
\begin{barticle}
\bauthor{\bsnm{Phan}, \binits{X.-H.}},
\bauthor{\bsnm{Nguyen}, \binits{C.-T.}},
\bauthor{\bsnm{Le}, \binits{D.-T.}},
\bauthor{\bsnm{Nguyen}, \binits{L.-M.}},
\bauthor{\bsnm{Horiguchi}, \binits{S.}},
\bauthor{\bsnm{Ha}, \binits{Q.-T.}}:
\batitle{A hidden topic-based framework toward building applications with short
  web documents}.
\bjtitle{IEEE Transactions on Knowledge and Data Engineering}
\bvolume{23}(\bissue{7}),
\bfpage{961}--\blpage{976}
(\byear{2010})
\end{barticle}
\endbibitem

\bibitem{yu2019adaptive}
\begin{bchapter}
\bauthor{\bsnm{Yu}, \binits{Z.}},
\bauthor{\bsnm{Lian}, \binits{J.}},
\bauthor{\bsnm{Mahmoody}, \binits{A.}},
\bauthor{\bsnm{Liu}, \binits{G.}},
\bauthor{\bsnm{Xie}, \binits{X.}}:
\bctitle{Adaptive user modeling with long and short-term preferences for
  personalized recommendation.}
In: \bbtitle{IJCAI},
pp. \bfpage{4213}--\blpage{4219}
(\byear{2019})
\end{bchapter}
\endbibitem

\bibitem{ashkan2009classifying}
\begin{bchapter}
\bauthor{\bsnm{Ashkan}, \binits{A.}},
\bauthor{\bsnm{Clarke}, \binits{C.L.}},
\bauthor{\bsnm{Agichtein}, \binits{E.}},
\bauthor{\bsnm{Guo}, \binits{Q.}}:
\bctitle{Classifying and characterizing query intent}.
In: \bbtitle{Advances in Information Retrieval: 31th European Conference on IR
  Research, ECIR 2009, Toulouse, France, April 6-9, 2009. Proceedings 31},
pp. \bfpage{578}--\blpage{586}
(\byear{2009}).
\bcomment{Springer}
\end{bchapter}
\endbibitem

\bibitem{xu2016spatio}
\begin{bchapter}
\bauthor{\bsnm{Xu}, \binits{P.}},
\bauthor{\bsnm{Sugano}, \binits{Y.}},
\bauthor{\bsnm{Bulling}, \binits{A.}}:
\bctitle{Spatio-temporal modeling and prediction of visual attention in
  graphical user interfaces}.
In: \bbtitle{Proceedings of the 2016 CHI Conference on Human Factors in
  Computing Systems},
pp. \bfpage{3299}--\blpage{3310}
(\byear{2016})
\end{bchapter}
\endbibitem

\bibitem{liu2022multi}
\begin{bchapter}
\bauthor{\bsnm{Liu}, \binits{P.}},
\bauthor{\bsnm{Liao}, \binits{D.}},
\bauthor{\bsnm{Wang}, \binits{J.}},
\bauthor{\bsnm{Wu}, \binits{Y.}},
\bauthor{\bsnm{Li}, \binits{G.}},
\bauthor{\bsnm{Xia}, \binits{S.-T.}},
\bauthor{\bsnm{Xu}, \binits{J.}}:
\bctitle{Multi-task ranking with user behaviors for text-video search}.
In: \bbtitle{Companion Proceedings of the Web Conference 2022},
pp. \bfpage{126}--\blpage{130}
(\byear{2022})
\end{bchapter}
\endbibitem

\bibitem{wu2019context}
\begin{barticle}
\bauthor{\bsnm{Wu}, \binits{L.}},
\bauthor{\bsnm{Quan}, \binits{C.}},
\bauthor{\bsnm{Li}, \binits{C.}},
\bauthor{\bsnm{Wang}, \binits{Q.}},
\bauthor{\bsnm{Zheng}, \binits{B.}},
\bauthor{\bsnm{Luo}, \binits{X.}}:
\batitle{A context-aware user-item representation learning for item
  recommendation}.
\bjtitle{ACM Transactions on Information Systems (TOIS)}
\bvolume{37}(\bissue{2}),
\bfpage{1}--\blpage{29}
(\byear{2019})
\end{barticle}
\endbibitem

\bibitem{mao2019multiobjective}
\begin{barticle}
\bauthor{\bsnm{Mao}, \binits{M.}},
\bauthor{\bsnm{Lu}, \binits{J.}},
\bauthor{\bsnm{Han}, \binits{J.}},
\bauthor{\bsnm{Zhang}, \binits{G.}}:
\batitle{Multiobjective e-commerce recommendations based on hypergraph
  ranking}.
\bjtitle{Information Sciences}
\bvolume{471},
\bfpage{269}--\blpage{287}
(\byear{2019})
\end{barticle}
\endbibitem

\bibitem{ni2012user}
\begin{barticle}
\bauthor{\bsnm{Ni}, \binits{X.}},
\bauthor{\bsnm{Lu}, \binits{Y.}},
\bauthor{\bsnm{Quan}, \binits{X.}},
\bauthor{\bsnm{Wenyin}, \binits{L.}},
\bauthor{\bsnm{Hua}, \binits{B.}}:
\batitle{User interest modeling and its application for question recommendation
  in user-interactive question answering systems}.
\bjtitle{Information Processing \& Management}
\bvolume{48}(\bissue{2}),
\bfpage{218}--\blpage{233}
(\byear{2012})
\end{barticle}
\endbibitem

\bibitem{liu2020dynamic}
\begin{barticle}
\bauthor{\bsnm{Liu}, \binits{P.}},
\bauthor{\bsnm{Zhang}, \binits{L.}},
\bauthor{\bsnm{Gulla}, \binits{J.A.}}:
\batitle{Dynamic attention-based explainable recommendation with textual and
  visual fusion}.
\bjtitle{Information Processing \& Management}
\bvolume{57}(\bissue{6}),
\bfpage{102099}
(\byear{2020})
\end{barticle}
\endbibitem

\bibitem{kaptein2013exploiting}
\begin{barticle}
\bauthor{\bsnm{Kaptein}, \binits{R.}},
\bauthor{\bsnm{Kamps}, \binits{J.}}:
\batitle{Exploiting the category structure of wikipedia for entity ranking}.
\bjtitle{Artificial Intelligence}
\bvolume{194},
\bfpage{111}--\blpage{129}
(\byear{2013})
\end{barticle}
\endbibitem

\bibitem{cai2014object}
\begin{barticle}
\bauthor{\bsnm{Cai}, \binits{Y.}},
\bauthor{\bsnm{Lau}, \binits{R.Y.}},
\bauthor{\bsnm{Liao}, \binits{S.S.}},
\bauthor{\bsnm{Li}, \binits{C.}},
\bauthor{\bsnm{Leung}, \binits{H.-F.}},
\bauthor{\bsnm{Ma}, \binits{L.C.}}:
\batitle{Object typicality for effective web of things recommendations}.
\bjtitle{Decision support systems}
\bvolume{63},
\bfpage{52}--\blpage{63}
(\byear{2014})
\end{barticle}
\endbibitem

\bibitem{colace2015collaborative}
\begin{barticle}
\bauthor{\bsnm{Colace}, \binits{F.}},
\bauthor{\bsnm{De~Santo}, \binits{M.}},
\bauthor{\bsnm{Greco}, \binits{L.}},
\bauthor{\bsnm{Moscato}, \binits{V.}},
\bauthor{\bsnm{Picariello}, \binits{A.}}:
\batitle{A collaborative user-centered framework for recommending items in
  online social networks}.
\bjtitle{Computers in Human Behavior}
\bvolume{51},
\bfpage{694}--\blpage{704}
(\byear{2015})
\end{barticle}
\endbibitem

\bibitem{yao2017version}
\begin{barticle}
\bauthor{\bsnm{Yao}, \binits{Y.}},
\bauthor{\bsnm{Zhao}, \binits{W.X.}},
\bauthor{\bsnm{Wang}, \binits{Y.}},
\bauthor{\bsnm{Tong}, \binits{H.}},
\bauthor{\bsnm{Xu}, \binits{F.}},
\bauthor{\bsnm{Lu}, \binits{J.}}:
\batitle{Version-aware rating prediction for mobile app recommendation}.
\bjtitle{ACM Transactions on Information Systems (TOIS)}
\bvolume{35}(\bissue{4}),
\bfpage{1}--\blpage{33}
(\byear{2017})
\end{barticle}
\endbibitem

\bibitem{teevan2008personalize}
\begin{bchapter}
\bauthor{\bsnm{Teevan}, \binits{J.}},
\bauthor{\bsnm{Dumais}, \binits{S.T.}},
\bauthor{\bsnm{Liebling}, \binits{D.J.}}:
\bctitle{To personalize or not to personalize: modeling queries with variation
  in user intent}.
In: \bbtitle{Proceedings of the 31st Annual International ACM SIGIR Conference
  on Research and Development in Information Retrieval},
pp. \bfpage{163}--\blpage{170}
(\byear{2008})
\end{bchapter}
\endbibitem

\bibitem{wang2021adapting}
\begin{barticle}
\bauthor{\bsnm{Wang}, \binits{H.-C.}},
\bauthor{\bsnm{Jhou}, \binits{H.-T.}},
\bauthor{\bsnm{Tsai}, \binits{Y.-S.}}:
\batitle{Adapting topic map and social influence to the personalized hybrid
  recommender system}.
\bjtitle{Information Sciences}
\bvolume{575},
\bfpage{762}--\blpage{778}
(\byear{2021})
\end{barticle}
\endbibitem

\bibitem{papadimitriou2012generalized}
\begin{barticle}
\bauthor{\bsnm{Papadimitriou}, \binits{A.}},
\bauthor{\bsnm{Symeonidis}, \binits{P.}},
\bauthor{\bsnm{Manolopoulos}, \binits{Y.}}:
\batitle{A generalized taxonomy of explanations styles for traditional and
  social recommender systems}.
\bjtitle{Data Mining and Knowledge Discovery}
\bvolume{24},
\bfpage{555}--\blpage{583}
(\byear{2012})
\end{barticle}
\endbibitem

\bibitem{fan2022modeling}
\begin{bchapter}
\bauthor{\bsnm{Fan}, \binits{L.}},
\bauthor{\bsnm{Li}, \binits{Q.}},
\bauthor{\bsnm{Liu}, \binits{B.}},
\bauthor{\bsnm{Wu}, \binits{X.-M.}},
\bauthor{\bsnm{Zhang}, \binits{X.}},
\bauthor{\bsnm{Lv}, \binits{F.}},
\bauthor{\bsnm{Lin}, \binits{G.}},
\bauthor{\bsnm{Li}, \binits{S.}},
\bauthor{\bsnm{Jin}, \binits{T.}},
\bauthor{\bsnm{Yang}, \binits{K.}}:
\bctitle{Modeling user behavior with graph convolution for personalized product
  search}.
In: \bbtitle{Proceedings of the ACM Web Conference 2022},
pp. \bfpage{203}--\blpage{212}
(\byear{2022})
\end{bchapter}
\endbibitem

\bibitem{liu2022category}
\begin{bchapter}
\bauthor{\bsnm{Liu}, \binits{J.}},
\bauthor{\bsnm{Dou}, \binits{Z.}},
\bauthor{\bsnm{Zhu}, \binits{Q.}},
\bauthor{\bsnm{Wen}, \binits{J.-R.}}:
\bctitle{A category-aware multi-interest model for personalized product
  search}.
In: \bbtitle{Proceedings of the ACM Web Conference 2022},
pp. \bfpage{360}--\blpage{368}
(\byear{2022})
\end{bchapter}
\endbibitem

\bibitem{iso420102011iec}
\begin{botherref}
\oauthor{\bsnm{{ISO}}}:
Iec/ieee systems and software engineering: Architecture description.
ISO/IEC/IEEE 42010: 2011 (E)(Revision of ISO/IEC 42010: 2007 and IEEE Std
  1471-2000)
(2011)
\end{botherref}
\endbibitem

\bibitem{garg2018madm}
\begin{barticle}
\bauthor{\bsnm{Garg}, \binits{R.}},
\bauthor{\bsnm{Kumar}, \binits{R.}},
\bauthor{\bsnm{Garg}, \binits{S.}}:
\batitle{Madm-based parametric selection and ranking of e-learning websites
  using fuzzy copras}.
\bjtitle{IEEE Transactions on Education}
\bvolume{62}(\bissue{1}),
\bfpage{11}--\blpage{18}
(\byear{2018})
\end{barticle}
\endbibitem

\bibitem{xu2007concepts}
\begin{barticle}
\bauthor{\bsnm{Xu}, \binits{L.}},
\bauthor{\bsnm{Brinkkemper}, \binits{S.}}:
\batitle{Concepts of product software}.
\bjtitle{European Journal of Information Systems}
\bvolume{16}(\bissue{5}),
\bfpage{531}--\blpage{541}
(\byear{2007})
\end{barticle}
\endbibitem

\bibitem{fitzgerald2014continuous}
\begin{bchapter}
\bauthor{\bsnm{Fitzgerald}, \binits{B.}},
\bauthor{\bsnm{Stol}, \binits{K.-J.}}:
\bctitle{Continuous software engineering and beyond: trends and challenges}.
In: \bbtitle{Proceedings of the 1st International Workshop on Rapid Continuous
  Software Engineering},
pp. \bfpage{1}--\blpage{9}
(\byear{2014})
\end{bchapter}
\endbibitem

\bibitem{rus2003supporting}
\begin{barticle}
\bauthor{\bsnm{Rus}, \binits{I.}},
\bauthor{\bsnm{Halling}, \binits{M.}},
\bauthor{\bsnm{Biffl}, \binits{S.}}:
\batitle{Supporting decision-making in software engineering with process
  simulation and empirical studies}.
\bjtitle{International Journal of Software Engineering and Knowledge
  Engineering}
\bvolume{13}(\bissue{05}),
\bfpage{531}--\blpage{545}
(\byear{2003})
\end{barticle}
\endbibitem

\bibitem{fitzgerald2017differences}
\begin{barticle}
\bauthor{\bsnm{Fitzgerald}, \binits{D.R.}},
\bauthor{\bsnm{Mohammed}, \binits{S.}},
\bauthor{\bsnm{Kremer}, \binits{G.O.}}:
\batitle{Differences in the way we decide: The effect of decision style
  diversity on process conflict in design teams}.
\bjtitle{Personality and Individual Differences}
\bvolume{104},
\bfpage{339}--\blpage{344}
(\byear{2017})
\end{barticle}
\endbibitem

\bibitem{kaufmann2012rationality}
\begin{barticle}
\bauthor{\bsnm{Kaufmann}, \binits{L.}},
\bauthor{\bsnm{Kreft}, \binits{S.}},
\bauthor{\bsnm{Ehrgott}, \binits{M.}},
\bauthor{\bsnm{Reimann}, \binits{F.}}:
\batitle{Rationality in supplier selection decisions: The effect of the buyer's
  national task environment}.
\bjtitle{Journal of Purchasing and Supply Management}
\bvolume{18}(\bissue{2}),
\bfpage{76}--\blpage{91}
(\byear{2012})
\end{barticle}
\endbibitem

\bibitem{garg2020mcdm}
\begin{botherref}
\oauthor{\bsnm{Garg}, \binits{R.}}:
Mcdm-based parametric selection of cloud deployment models for an academic
  organization.
IEEE Transactions on Cloud Computing
(2020)
\end{botherref}
\endbibitem

\bibitem{garg2017mcdm}
\begin{barticle}
\bauthor{\bsnm{Garg}, \binits{R.}},
\bauthor{\bsnm{Sharma}, \binits{R.}},
\bauthor{\bsnm{Sharma}, \binits{K.}}:
\batitle{Mcdm based evaluation and ranking of commercial off-the-shelf using
  fuzzy based matrix method}.
\bjtitle{Decision Science Letters}
\bvolume{6}(\bissue{2}),
\bfpage{117}--\blpage{136}
(\byear{2017})
\end{barticle}
\endbibitem

\bibitem{sandhya2018computational}
\begin{barticle}
\bauthor{\bsnm{Sandhya}},
\bauthor{\bsnm{Garg}, \binits{R.}},
\bauthor{\bsnm{Kumar}, \binits{R.}}:
\batitle{Computational madm evaluation and ranking of cloud service providers
  using distance-based approach}.
\bjtitle{International Journal of Information and Decision Sciences}
\bvolume{10}(\bissue{3}),
\bfpage{222}--\blpage{234}
(\byear{2018})
\end{barticle}
\endbibitem

\bibitem{garg2019parametric}
\begin{barticle}
\bauthor{\bsnm{Garg}, \binits{R.}}:
\batitle{Parametric selection of software reliability growth models using
  multi-criteria decision-making approach}.
\bjtitle{International Journal of Reliability and Safety}
\bvolume{13}(\bissue{4}),
\bfpage{291}--\blpage{309}
(\byear{2019})
\end{barticle}
\endbibitem

\bibitem{doumpos2013multicriteria}
\begin{botherref}
\oauthor{\bsnm{Doumpos}, \binits{M.}},
\oauthor{\bsnm{Grigoroudis}, \binits{E.}}:
Multicriteria decision aid and artificial intelligence.
Whiley (UK)
(2013)
\end{botherref}
\endbibitem

\bibitem{majumder2015multi}
\begin{bchapter}
\bauthor{\bsnm{Majumder}, \binits{M.}}:
\bctitle{Multi criteria decision making}.
In: \bbtitle{Impact of Urbanization on Water Shortage in Face of Climatic
  Aberrations},
pp. \bfpage{35}--\blpage{47}.
\bpublisher{Springer}, \blocation{???}
(\byear{2015})
\end{bchapter}
\endbibitem

\bibitem{caprara2000algorithms}
\begin{barticle}
\bauthor{\bsnm{Caprara}, \binits{A.}},
\bauthor{\bsnm{Toth}, \binits{P.}},
\bauthor{\bsnm{Fischetti}, \binits{M.}}:
\batitle{Algorithms for the set covering problem}.
\bjtitle{Annals of Operations Research}
\bvolume{98}(\bissue{1-4}),
\bfpage{353}--\blpage{371}
(\byear{2000})
\end{barticle}
\endbibitem

\bibitem{MANZOOR-CaseStudy}
\begin{barticle}
\bauthor{\bsnm{Manzoor}, \binits{A.}},
\bauthor{\bsnm{Jannach}, \binits{D.}}:
\batitle{Towards retrieval-based conversational recommendation}.
\bjtitle{Information Systems}
\bvolume{109},
\bfpage{102083}
(\byear{2022}).
\doiurl{10.1016/j.is.2022.102083}
\end{barticle}
\endbibitem

\bibitem{Mehrab-caseStudy}
\begin{bchapter}
\bauthor{\bsnm{Tanjim}, \binits{M.M.}},
\bauthor{\bsnm{Su}, \binits{C.}},
\bauthor{\bsnm{Benjamin}, \binits{E.}},
\bauthor{\bsnm{Hu}, \binits{D.}},
\bauthor{\bsnm{Hong}, \binits{L.}},
\bauthor{\bsnm{McAuley}, \binits{J.}}:
\bctitle{Attentive sequential models of latent intent for next item
  recommendation}.
In: \bbtitle{Proceedings of The Web Conference 2020}.
\bsertitle{WWW '20},
pp. \bfpage{2528}--\blpage{2534}.
\bpublisher{Association for Computing Machinery},
\blocation{New York, NY, USA}
(\byear{2020}).
\doiurl{10.1145/3366423.3380002}.
\burl{https://doi.org/10.1145/3366423.3380002}
\end{bchapter}
\endbibitem

\bibitem{haefliger2008code}
\begin{barticle}
\bauthor{\bsnm{Haefliger}, \binits{S.}},
\bauthor{\bsnm{Von~Krogh}, \binits{G.}},
\bauthor{\bsnm{Spaeth}, \binits{S.}}:
\batitle{Code reuse in open source software}.
\bjtitle{Management science}
\bvolume{54}(\bissue{1}),
\bfpage{180}--\blpage{193}
(\byear{2008})
\end{barticle}
\endbibitem

\bibitem{amershi2019software}
\begin{bchapter}
\bauthor{\bsnm{Amershi}, \binits{S.}},
\bauthor{\bsnm{Begel}, \binits{A.}},
\bauthor{\bsnm{Bird}, \binits{C.}},
\bauthor{\bsnm{DeLine}, \binits{R.}},
\bauthor{\bsnm{Gall}, \binits{H.}},
\bauthor{\bsnm{Kamar}, \binits{E.}},
\bauthor{\bsnm{Nagappan}, \binits{N.}},
\bauthor{\bsnm{Nushi}, \binits{B.}},
\bauthor{\bsnm{Zimmermann}, \binits{T.}}:
\bctitle{Software engineering for machine learning: A case study}.
In: \bbtitle{2019 IEEE/ACM 41st International Conference on Software
  Engineering: Software Engineering in Practice (ICSE-SEIP)},
pp. \bfpage{291}--\blpage{300}
(\byear{2019}).
\bcomment{IEEE}
\end{bchapter}
\endbibitem

\bibitem{kuwajima2020engineering}
\begin{barticle}
\bauthor{\bsnm{Kuwajima}, \binits{H.}},
\bauthor{\bsnm{Yasuoka}, \binits{H.}},
\bauthor{\bsnm{Nakae}, \binits{T.}}:
\batitle{Engineering problems in machine learning systems}.
\bjtitle{Machine Learning}
\bvolume{109}(\bissue{5}),
\bfpage{1103}--\blpage{1126}
(\byear{2020})
\end{barticle}
\endbibitem

\bibitem{chen2022intent}
\begin{bchapter}
\bauthor{\bsnm{Chen}, \binits{Y.}},
\bauthor{\bsnm{Liu}, \binits{Z.}},
\bauthor{\bsnm{Li}, \binits{J.}},
\bauthor{\bsnm{McAuley}, \binits{J.}},
\bauthor{\bsnm{Xiong}, \binits{C.}}:
\bctitle{Intent contrastive learning for sequential recommendation}.
In: \bbtitle{Proceedings of the ACM Web Conference 2022},
pp. \bfpage{2172}--\blpage{2182}
(\byear{2022})
\end{bchapter}
\endbibitem

\bibitem{garcia2021topic}
\begin{barticle}
\bauthor{\bsnm{Garcia}, \binits{K.}},
\bauthor{\bsnm{Berton}, \binits{L.}}:
\batitle{Topic detection and sentiment analysis in twitter content related to
  covid-19 from brazil and the usa}.
\bjtitle{Applied soft computing}
\bvolume{101},
\bfpage{107057}
(\byear{2021})
\end{barticle}
\endbibitem

\bibitem{hashemi2020guided}
\begin{bchapter}
\bauthor{\bsnm{Hashemi}, \binits{H.}},
\bauthor{\bsnm{Zamani}, \binits{H.}},
\bauthor{\bsnm{Croft}, \binits{W.B.}}:
\bctitle{Guided transformer: Leveraging multiple external sources for
  representation learning in conversational search}.
In: \bbtitle{Proceedings of the 43rd International Acm Sigir Conference on
  Research and Development in Information Retrieval},
pp. \bfpage{1131}--\blpage{1140}
(\byear{2020})
\end{bchapter}
\endbibitem

\bibitem{carvallo2020automatic}
\begin{barticle}
\bauthor{\bsnm{Carvallo}, \binits{A.}},
\bauthor{\bsnm{Parra}, \binits{D.}},
\bauthor{\bsnm{Lobel}, \binits{H.}},
\bauthor{\bsnm{Soto}, \binits{A.}}:
\batitle{Automatic document screening of medical literature using word and text
  embeddings in an active learning setting}.
\bjtitle{Scientometrics}
\bvolume{125},
\bfpage{3047}--\blpage{3084}
(\byear{2020})
\end{barticle}
\endbibitem

\bibitem{gao2022search}
\begin{bchapter}
\bauthor{\bsnm{Gao}, \binits{C.}},
\bauthor{\bsnm{Lam}, \binits{W.}}:
\bctitle{Search clarification selection via query-intent-clarification graph
  attention}.
In: \bbtitle{European Conference on Information Retrieval},
pp. \bfpage{230}--\blpage{243}
(\byear{2022}).
\bcomment{Springer}
\end{bchapter}
\endbibitem

\bibitem{wu2021exploration}
\begin{barticle}
\bauthor{\bsnm{Wu}, \binits{Z.}},
\bauthor{\bsnm{Liang}, \binits{J.}},
\bauthor{\bsnm{Zhang}, \binits{Z.}},
\bauthor{\bsnm{Lei}, \binits{J.}}:
\batitle{Exploration of text matching methods in chinese disease q\&a systems:
  A method using ensemble based on bert and boosted tree models}.
\bjtitle{Journal of biomedical informatics}
\bvolume{115},
\bfpage{103683}
(\byear{2021})
\end{barticle}
\endbibitem

\bibitem{devlin2018bert}
\begin{botherref}
\oauthor{\bsnm{Devlin}, \binits{J.}},
\oauthor{\bsnm{Chang}, \binits{M.-W.}},
\oauthor{\bsnm{Lee}, \binits{K.}},
\oauthor{\bsnm{Toutanova}, \binits{K.}}:
Bert: Pre-training of deep bidirectional transformers for language
  understanding.
arXiv preprint arXiv:1810.04805
(2018)
\end{botherref}
\endbibitem

\bibitem{sarker2021machine}
\begin{barticle}
\bauthor{\bsnm{Sarker}, \binits{I.H.}}:
\batitle{Machine learning: Algorithms, real-world applications and research
  directions}.
\bjtitle{SN computer science}
\bvolume{2}(\bissue{3}),
\bfpage{160}
(\byear{2021})
\end{barticle}
\endbibitem

\bibitem{blei2003latent}
\begin{barticle}
\bauthor{\bsnm{Blei}, \binits{D.M.}},
\bauthor{\bsnm{Ng}, \binits{A.Y.}},
\bauthor{\bsnm{Jordan}, \binits{M.I.}}:
\batitle{Latent dirichlet allocation}.
\bjtitle{Journal of machine Learning research}
\bvolume{3}(\bissue{Jan}),
\bfpage{993}--\blpage{1022}
(\byear{2003})
\end{barticle}
\endbibitem

\bibitem{raffel2020exploring}
\begin{barticle}
\bauthor{\bsnm{Raffel}, \binits{C.}},
\bauthor{\bsnm{Shazeer}, \binits{N.}},
\bauthor{\bsnm{Roberts}, \binits{A.}},
\bauthor{\bsnm{Lee}, \binits{K.}},
\bauthor{\bsnm{Narang}, \binits{S.}},
\bauthor{\bsnm{Matena}, \binits{M.}},
\bauthor{\bsnm{Zhou}, \binits{Y.}},
\bauthor{\bsnm{Li}, \binits{W.}},
\bauthor{\bsnm{Liu}, \binits{P.J.}}:
\batitle{Exploring the limits of transfer learning with a unified text-to-text
  transformer}.
\bjtitle{The Journal of Machine Learning Research}
\bvolume{21}(\bissue{1}),
\bfpage{5485}--\blpage{5551}
(\byear{2020})
\end{barticle}
\endbibitem

\bibitem{ribeiro2016should}
\begin{bchapter}
\bauthor{\bsnm{Ribeiro}, \binits{M.T.}},
\bauthor{\bsnm{Singh}, \binits{S.}},
\bauthor{\bsnm{Guestrin}, \binits{C.}}:
\bctitle{" why should i trust you?" explaining the predictions of any
  classifier}.
In: \bbtitle{Proceedings of the 22nd ACM SIGKDD International Conference on
  Knowledge Discovery and Data Mining},
pp. \bfpage{1135}--\blpage{1144}
(\byear{2016})
\end{bchapter}
\endbibitem

\bibitem{pujol2020fair}
\begin{bchapter}
\bauthor{\bsnm{Pujol}, \binits{D.}},
\bauthor{\bsnm{McKenna}, \binits{R.}},
\bauthor{\bsnm{Kuppam}, \binits{S.}},
\bauthor{\bsnm{Hay}, \binits{M.}},
\bauthor{\bsnm{Machanavajjhala}, \binits{A.}},
\bauthor{\bsnm{Miklau}, \binits{G.}}:
\bctitle{Fair decision making using privacy-protected data}.
In: \bbtitle{Proceedings of the 2020 Conference on Fairness, Accountability,
  and Transparency},
pp. \bfpage{189}--\blpage{199}
(\byear{2020})
\end{bchapter}
\endbibitem

\bibitem{bagdasaryan2019differential}
\begin{botherref}
\oauthor{\bsnm{Bagdasaryan}, \binits{E.}},
\oauthor{\bsnm{Poursaeed}, \binits{O.}},
\oauthor{\bsnm{Shmatikov}, \binits{V.}}:
Differential privacy has disparate impact on model accuracy.
Advances in neural information processing systems
\textbf{32}
(2019)
\end{botherref}
\endbibitem

\bibitem{zhou2016map}
\begin{bchapter}
\bauthor{\bsnm{Zhou}, \binits{X.}},
\bauthor{\bsnm{Jin}, \binits{Y.}},
\bauthor{\bsnm{Zhang}, \binits{H.}},
\bauthor{\bsnm{Li}, \binits{S.}},
\bauthor{\bsnm{Huang}, \binits{X.}}:
\bctitle{A map of threats to validity of systematic literature reviews in
  software engineering}.
In: \bbtitle{2016 23rd Asia-Pacific Software Engineering Conference (APSEC)},
pp. \bfpage{153}--\blpage{160}
(\byear{2016}).
\bcomment{IEEE}
\end{bchapter}
\endbibitem

\bibitem{zhang2011identifying}
\begin{barticle}
\bauthor{\bsnm{Zhang}, \binits{H.}},
\bauthor{\bsnm{Babar}, \binits{M.A.}},
\bauthor{\bsnm{Tell}, \binits{P.}}:
\batitle{Identifying relevant studies in software engineering}.
\bjtitle{Information and Software Technology}
\bvolume{53}(\bissue{6}),
\bfpage{625}--\blpage{637}
(\byear{2011})
\end{barticle}
\endbibitem

\bibitem{keyvan2022approach}
\begin{barticle}
\bauthor{\bsnm{Keyvan}, \binits{K.}},
\bauthor{\bsnm{Huang}, \binits{J.X.}}:
\batitle{How to approach ambiguous queries in conversational search: A survey
  of techniques, approaches, tools, and challenges}.
\bjtitle{ACM Computing Surveys}
\bvolume{55}(\bissue{6}),
\bfpage{1}--\blpage{40}
(\byear{2022})
\end{barticle}
\endbibitem

\bibitem{iovine2023virtual}
\begin{barticle}
\bauthor{\bsnm{Iovine}, \binits{A.}},
\bauthor{\bsnm{Narducci}, \binits{F.}},
\bauthor{\bsnm{Musto}, \binits{C.}},
\bauthor{\bparticle{de} \bsnm{Gemmis}, \binits{M.}},
\bauthor{\bsnm{Semeraro}, \binits{G.}}:
\batitle{Virtual customer assistants in finance: From state of the art and
  practices to design guidelines}.
\bjtitle{Computer Science Review}
\bvolume{47},
\bfpage{100534}
(\byear{2023})
\end{barticle}
\endbibitem

\bibitem{saka2023conversational}
\begin{barticle}
\bauthor{\bsnm{Saka}, \binits{A.B.}},
\bauthor{\bsnm{Oyedele}, \binits{L.O.}},
\bauthor{\bsnm{Akanbi}, \binits{L.A.}},
\bauthor{\bsnm{Ganiyu}, \binits{S.A.}},
\bauthor{\bsnm{Chan}, \binits{D.W.}},
\bauthor{\bsnm{Bello}, \binits{S.A.}}:
\batitle{Conversational artificial intelligence in the aec industry: A review
  of present status, challenges and opportunities}.
\bjtitle{Advanced Engineering Informatics}
\bvolume{55},
\bfpage{101869}
(\byear{2023})
\end{barticle}
\endbibitem

\bibitem{liu2022review}
\begin{barticle}
\bauthor{\bsnm{Liu}, \binits{T.}},
\bauthor{\bsnm{Wu}, \binits{Q.}},
\bauthor{\bsnm{Chang}, \binits{L.}},
\bauthor{\bsnm{Gu}, \binits{T.}}:
\batitle{A review of deep learning-based recommender system in e-learning
  environments}.
\bjtitle{Artificial Intelligence Review}
\bvolume{55}(\bissue{8}),
\bfpage{5953}--\blpage{5980}
(\byear{2022})
\end{barticle}
\endbibitem

\bibitem{tamine2010evaluation}
\begin{barticle}
\bauthor{\bsnm{Tamine-Lechani}, \binits{L.}},
\bauthor{\bsnm{Boughanem}, \binits{M.}},
\bauthor{\bsnm{Daoud}, \binits{M.}}:
\batitle{Evaluation of contextual information retrieval effectiveness: overview
  of issues and research}.
\bjtitle{Knowledge and Information Systems}
\bvolume{24},
\bfpage{1}--\blpage{34}
(\byear{2010})
\end{barticle}
\endbibitem

\bibitem{jiang2013mining}
\begin{barticle}
\bauthor{\bsnm{Jiang}, \binits{D.}},
\bauthor{\bsnm{Pei}, \binits{J.}},
\bauthor{\bsnm{Li}, \binits{H.}}:
\batitle{Mining search and browse logs for web search: A survey}.
\bjtitle{ACM Transactions on Intelligent Systems and Technology (TIST)}
\bvolume{4}(\bissue{4}),
\bfpage{1}--\blpage{37}
(\byear{2013})
\end{barticle}
\endbibitem

\bibitem{jindal2014review}
\begin{barticle}
\bauthor{\bsnm{Jindal}, \binits{V.}},
\bauthor{\bsnm{Bawa}, \binits{S.}},
\bauthor{\bsnm{Batra}, \binits{S.}}:
\batitle{A review of ranking approaches for semantic search on web}.
\bjtitle{Information Processing \& Management}
\bvolume{50}(\bissue{2}),
\bfpage{416}--\blpage{425}
(\byear{2014})
\end{barticle}
\endbibitem

\end{thebibliography}
\newpage
\appendix
\section{Models}\label{Appendix_Models}
\begin{table}[!h]
\scriptsize
\caption{Model definitions}
\centering
\includegraphics[trim=10 70 10 0  ,clip,width=1\textwidth]{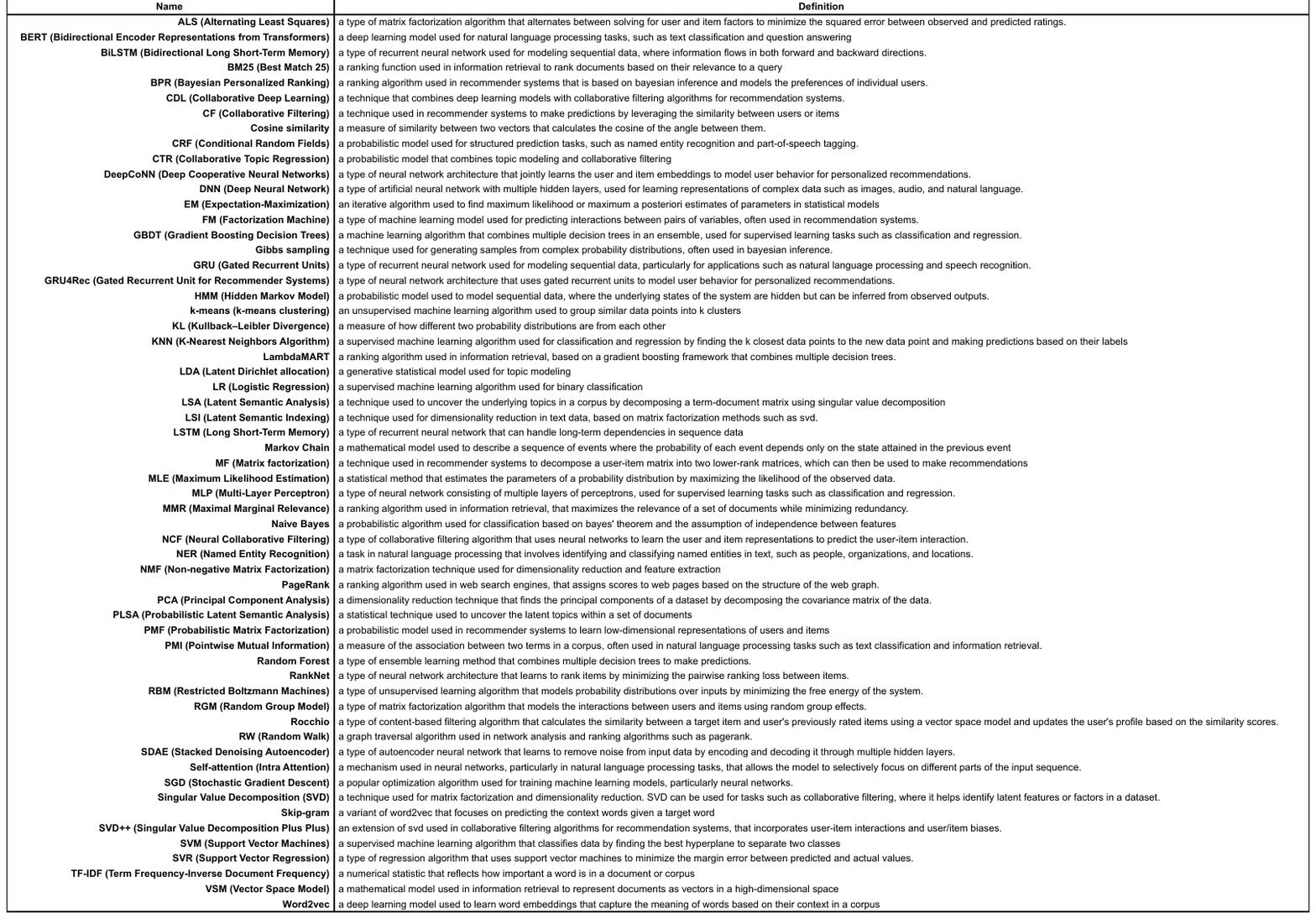}
\label{table:ModelDefinitions}
\end{table}
\section{Categories}\label{Appendix_CategoriesOfModels}
\begin{table}[!h]
\scriptsize
\caption{Categories}
\centering
\includegraphics[trim=10 350 10 0  ,clip,width=1\textwidth]{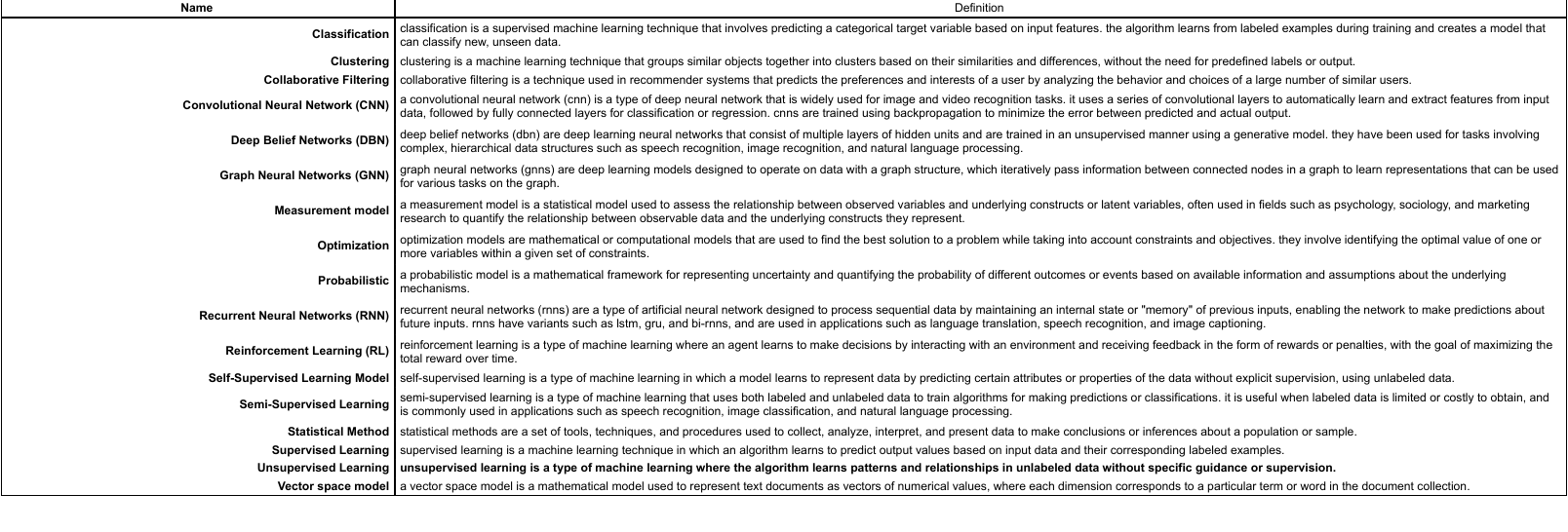}
\label{table:ModelCategoryDefinitions}
\end{table}
\newpage
\section{Features}\label{Appendix_Features}
\begin{table}[!h]
\scriptsize
\caption{Features}
\centering
\includegraphics[trim=00 50 00 0  ,clip,width=1\textwidth]{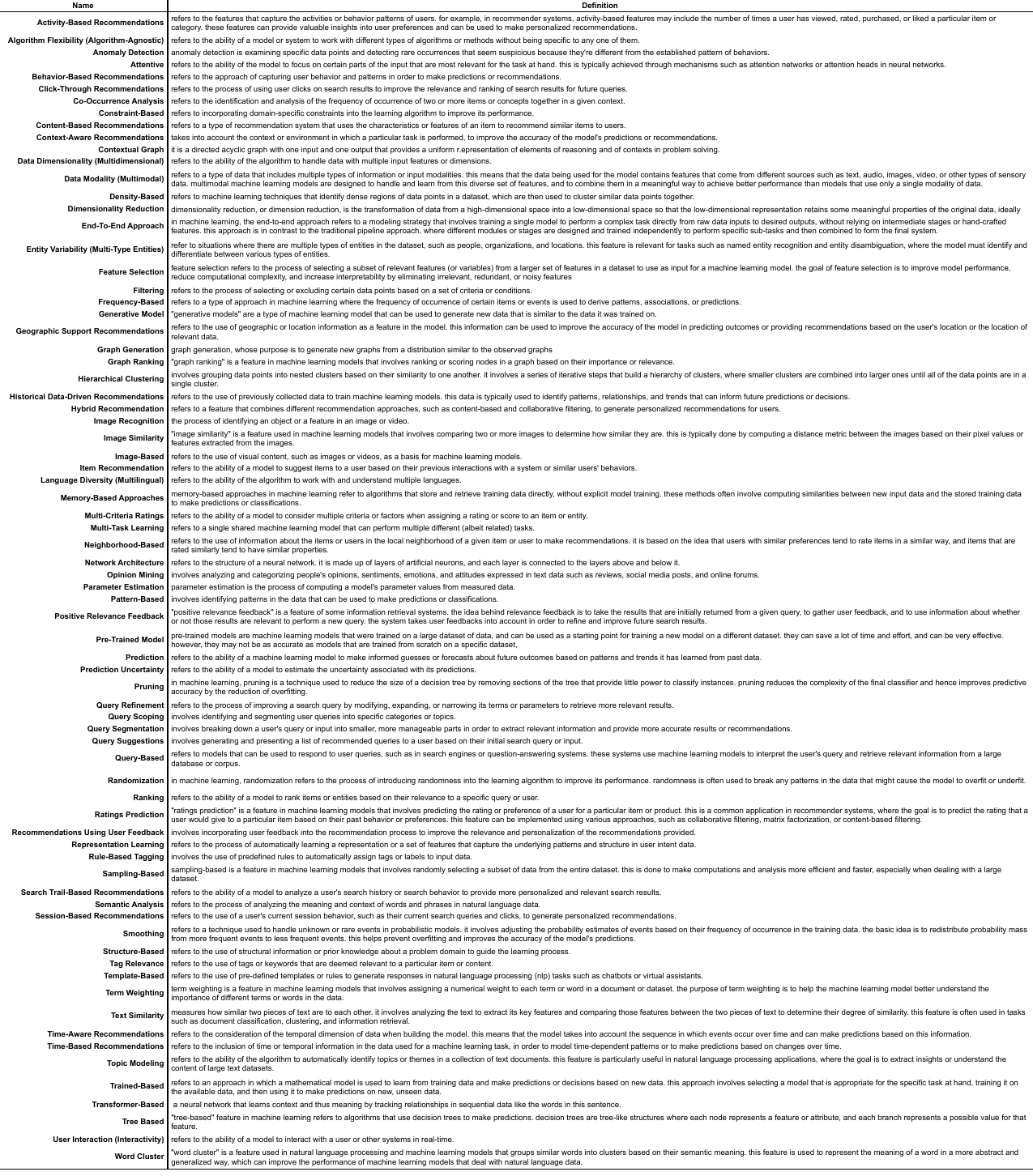}
\label{table:FeatureDefinition}
\end{table}

\newpage

\section{Quality attributes and evaluation measures}\label{Appendix_QA_EvaluationMeasures}
\begin{table}[!h]
\scriptsize
\caption{Quality attributes and evaluation measures}
\centering
\includegraphics[trim=00 150 00 0  ,clip,width=1\textwidth]{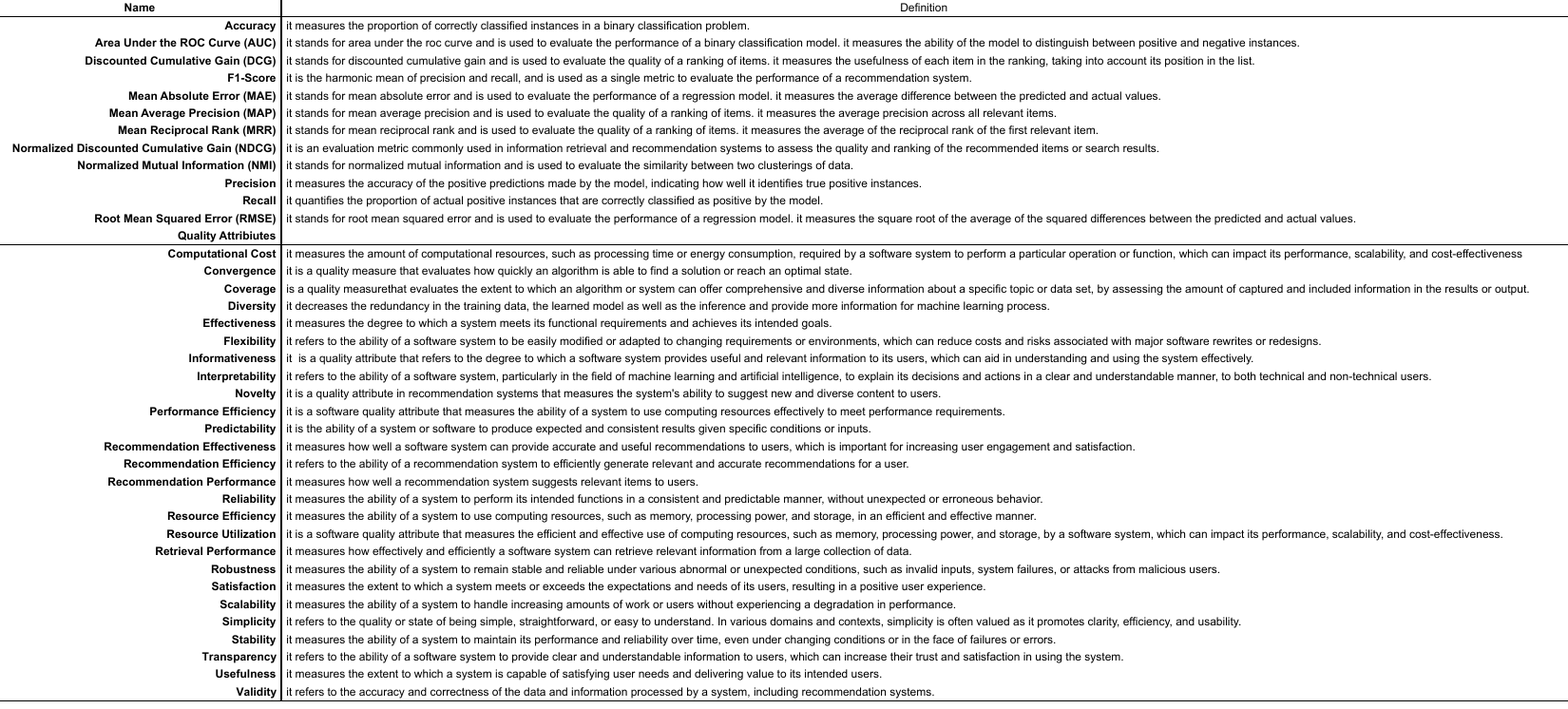}
\label{table:QAMeasureDefinition}
\end{table}
\end{document}